\documentclass{article}

\PassOptionsToPackage{numbers, compress}{natbib}
\usepackage[preprint]{neurips_2025}
\usepackage[utf8]{inputenc} % allow utf-8 input
\usepackage[T1]{fontenc}    % use 8-bit T1 fonts
\usepackage{hyperref}       % hyperlinks
\usepackage{url}            % simple URL typesetting
\usepackage{booktabs}       % professional-quality tables
\usepackage{amsfonts}       % blackboard math symbols
\usepackage{nicefrac}       % compact symbols for 1/2, etc.
\usepackage{microtype}      % microtypography
\usepackage{xcolor}         % colors

\usepackage{bbm}
\usepackage{algorithm}
\usepackage{subfigure}
\usepackage{subcaption}
\usepackage{multirow}
\usepackage{enumitem}
\usepackage{graphicx}
\usepackage{subcaption}
\usepackage{amsthm,amsmath,amssymb}
\usepackage{wrapfig}

\usepackage{algpseudocode}

\newcommand{\yw}[1]{\textcolor{black}{#1}}

\title{Generative Auto-Bidding in Large-Scale Competitive Auctions via Diffusion Completer-Aligner}

\author{
  Yewen Li$^{1}$\quad
  Jingtong Gao$^{3}$\quad
  Nan Jiang$^{1}$\quad
  Shuai Mao$^{4}$\quad 
  Ruyi An$^{5}$\quad \\
  \textbf{Fei Pan}$^{1}$\quad
  \textbf{Xiangyu Zhao}$^{3}$\quad
  \textbf{Bo An}$^{2}$\quad
  \textbf{Qingpeng Cai}$^{1}$\footnotemark[1]\thanks{Corresponding to: Qingpeng Cai <caiqingpeng@kuaishou.com>.}\quad
  \textbf{Peng Jiang}$^{1}$\quad
  \\
  $^1$Kuaishou Technology\quad
  $^2$Nanyang Technological University\quad \\
  $^3$City University of HongKong\quad
  $^4$Chinese University of HongKong\quad
  $^5$University of Texas at Austin
}

\begin{document}

\maketitle

\begin{abstract}
Auto-bidding is central to computational advertising, achieving notable commercial success by optimizing advertisers' bids within economic constraints.
Recently, large generative models show potential to revolutionize auto-bidding by generating bids that could flexibly adapt to complex, competitive environments. Among them, diffusers stand out for their ability to address sparse-reward challenges by focusing on trajectory-level accumulated rewards, as well as their explainable capability, \textit{i.e.}, planning a future trajectory of states and executing bids accordingly.
However, diffusers struggle with generation uncertainty, particularly regarding dynamic legitimacy between adjacent states, which can lead to poor bids and further cause significant loss of ad impression opportunities when competing with other advertisers in a highly competitive auction environment.
To address it, we propose a \textbf{C}ausal auto-\textbf{B}idding method based on a \textbf{D}iffusion completer-aligner framework, termed {\textbf{CBD}}.  
Firstly, we augment the diffusion training process with an extra random variable $t$, where the model observes $t$-length historical sequences with the goal of completing the remaining sequence, thereby enhancing the generated sequences' dynamic legitimacy. Then, we employ a trajectory-level return model to refine the generated trajectories, aligning more closely with advertisers' objectives.
% Experimental results with diverse settings demonstrate our approach not only achieves superior performance on large-scale auto-bidding benchmarks, \textit{e.g.}, 29.9$\%$ improvement of conversion value in the challenging sparse-reward auction setting, 
\yw{Experimental results across diverse settings demonstrate that our approach not only achieves superior performance on large-scale auto-bidding benchmarks, such as a 29.9\% improvement in conversion value in the challenging sparse-reward auction setting, but also delivers significant improvements on the Kuaishou online advertising platform, including a 2.0\% increase in target cost.}
% but also brings benefits for general offline RL tasks.
\end{abstract}

% \begin{figure}[t!]
%     \centering
%     \includegraphics[width=\linewidth]{figs/nips_cbd_main_fig.pdf}
%     % \caption{}
%     \label{fig:main}
%     \vspace{-5mm}
%     \caption{Overview of our causal auto-bidding method via a diffusion completer-aligner.}
% \end{figure}

\begin{figure}[t!]
    \centering
    \includegraphics[width=\linewidth]{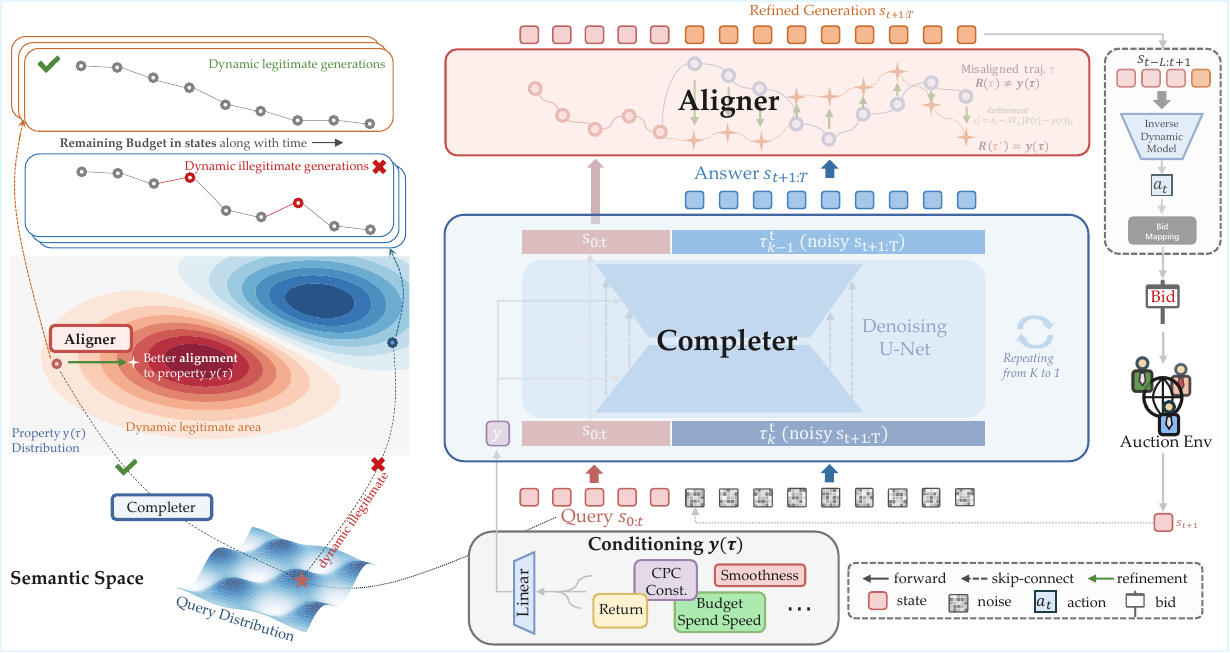}
    \caption{Overview of our CBD method: At each timestep $t$, a completer first receives a ``Query'' (observed states $s_{0:t}$ reflecting ad impression features and auction status) and outputs an ``Answer'' (generated future states $s_{t+1:T}$) within the dynamic legitimate area; The aligner then refines the generation to achieve closer alignment with the expected trajectory property; Finally, a bid is determined based on the refined trajectory and sent to the auction to compete with other advertisers.}
    \label{fig:main}
    \vspace{-5mm}
\end{figure}

\vspace{-2mm}
\section{Introduction}
\vspace{-2mm}
The rapid digital transformation of commerce has significantly expanded the reach of online advertising platforms, making them essential for engaging audiences and driving sales \cite{DBLP:conf/wsdm/WangY15, evans2009online}. 
With the continuous influx of ad impression opportunities on the platform, real-time auto-bidding for display advertising, which emerged in 2009, has become a dominant paradigm in computational advertising \cite{emerge}.
Specifically, when an ad impression opportunity is generated from a user's visits, a bid request for this opportunity, including its features like user, context, and auction status, is broadcast to numerous advertisers via hosting a real-time ad auction. Then, a specific auto-bidding algorithm is selected by the advertiser to determine a bid price in real-time \cite{BM}.
The goal of auto-bidding is to maximize the total accumulated impression value for an advertiser while adhering to some economic constraints, such as cost-per-conversion (CPC), throughout an entire bidding period, typically a day \cite{aigb}. 
Previous auto-bidding strategies can be categorized into rule-based methods \cite{OCPC, aiads, Branding, M-PID,Multi-KPI,Video,PID,PRUD}, prediction-based methods \cite{MEOW,DistributionShading,PointShading,LiftBidding,attribution_biding,causal_modeling,BGD}, and sequential decision-making methods \cite{maab,DCMAB,USCB,CBRL,RSDRL,ClusterA3C,RMDP,DRLB,RLB,LADDER} based on reinforcement learning (RL) \cite{sutton2018reinforcement}. 
Recently, generative models have achieved remarkable success in tasks such as NLP and CV \cite{ChatGPT, latent_diffusion}. Given the immense commercial value, it is worthwhile to explore generative models' potential to revolutionize auto-bidding by directly generating bids that can handle the complex advertising environment while adapting to the economic conditions of diverse advertisers.

In existing literature, there are two major paradigms employing generative models for sequential decision-making:
decision transformer (DT) \cite{dt, cdt} and  diffuser \cite{diffuser, aigb, diffuser_0, adapt_diffuser, online_diffuser}.
However, DT faces reliability as it directly outputs a bid without showcasing the reasoning process. Even when attempting long-term planning, it suffers from accumulated errors due to its auto-regressive generation.
Thus, DT complicates the diagnosis and rectification of errors in industrial auto-bidding services.
Moreover, DT relies on the single-step reward function, which can lead to poor performance in scenarios with sparse rewards \cite{DBLP:conf/ijcai/ZhuQZ023}, which is common in online advertising, such as in-app purchase (IAP).
Diffuser, instead, offers a promising alternative by generating an entire trajectory of states based on a specified expected property as a condition. Actions are then executed according to the trajectory, resembling a planning-based decision-making paradigm \cite{diffuser_0, diffuser}. 
This approach significantly enhances explainability by revealing the model's thought process, \textit{i.e.}, the planned trajectory.
% This transparency alleviates the reliability concerns associated with black-box learning-based models. 
Additionally, the diffuser learns a joint distribution between an entire trajectory and its accumulated rewards, which helps alleviate the sparse reward issues, leading to a more general usage than DT in auto-bidding.  
% Thus, diffusion models are a promising choice for achieving both reliability and effectiveness in auto-bidding.
A pioneering work, DiffBid \cite{aigb}, supports this by successfully deploying a decision diffuser for auto-bidding in a small private industry dataset, achieving superior performance than RL and DT.

However, directly applying diffusers could lead to poor performance in large-scale auctions with a large number of advertisers, millions of impression opportunities, and complex, uncertain impression distributions \cite{auctionNet}.
This could be due to diffusers' inherent generation uncertainty issue \cite{CleanDiffuser}, as shown in Fig. \ref{fig:causal}. 
Specifically, the planned next state may not be achievable from the current state, leading to a dynamic legitimacy issue \cite{metadiffuser}. 
Additionally, randomly sampling from prior distributions in the denoising process could result in multiple possible generations \cite{diffusion, ddim}, but not all align well with the advertiser's objectives.
The situation worsens in large-scale auctions, where the tolerance for bids is minimal due to high competitiveness among advertisers and complex ad impression distributions. Suboptimal bids can lead to significant loss of impression opportunities or spend waste.
Previous works have tried to address this issue in offline RL tasks, typically involving single-agent, non-competitive tasks like locomotion control \cite{d4rl}.
AdaptDiffuser \cite{adapt_diffuser} and MetaDiffuser \cite{metadiffuser} use a forward dynamic model (FDM) for enhancing legitimacy, but a poor base model could make FDM useless \cite{diffuser}.
DiffuserLite \cite{DiffuserLite} revises training process as generative interpolation by training multiple diffusers, but it could limit overall planning. 
Adopting causal network architectures for training a diffuser, such as DiT \cite{dit} with causal attention \cite{causal_atten}, is closely related to addressing this issue, but also raises concerns about accumulated errors and learning difficulty.

To tackle the generation uncertainty issue of diffusers in auto-bidding for large-scale competitive auctions, we draw inspiration from the training and alignment techniques used in LLMs \cite{ChatGPT, gpt-2, aligner, waswani2017attention}, which face a similar intrinsic dilemma of uncertainty, known as \textit{hallucination}.
Firstly, during training, we augment the denoising process by introducing an additional random variable $t$, reformulating model training as learning a completer to perform masked completion. 
This involves observing random $t$-length historical sequences as ``query'' and then completing the remaining sequence as ``answer'', thereby improving the dynamic legitimacy between randomly adjacent states.
However, a dynamic legitimate trajectory may still be misaligned with advertisers' objectives due to the randomness in denoising.
Thus, during inference, we employ a trajectory-level return model as an efficient aligner to refine the final generated trajectories, aligning more closely with advertisers’ objectives.
Combining them leads to our \textbf{C}ausal auto-\textbf{B}idding method based on \textbf{D}iffusion completer-aligner, termed \textbf{CBD}.
Overall, our contributions could be summarized as follows:
\begin{itemize}[leftmargin=1cm]
\item We highlight the crucial and valuable need for generative auto-bidding methods in large-scale auctions and analyze the factors hindering their development.
\item We propose a novel CBD method to improve the diffuser's decision-making performance in auto-bidding by addressing generation uncertainty in both training and inference.
\item Experiments show that our method excels in large-scale real-world auto-bidding benchmarks, like achieving a notable 29.9\% improvement in the challenging sparse-reward auction scenario. Given the sequential decision-making nature of our method, it could also provide insights and advantages for general offline RL tasks. 
\yw{More importantly, we successfully deploy our CBD method to the Kuaishou Online advertising system and achieve a significant improvement, \textit{e.g.}, a 2.0\% increase in the target cost. }
\end{itemize}
% Our method tackles the causality issue by reformulating the training process as completion to align with the inference process. 
% Specifically, we concentrate on generating only the unseen future trajectory based on the observed historical trajectory and conditions, promoting the learning of the causal relationship between them.
% Then, we implement a trajectory-level return model that updates the next state during the post-training phase, allowing the model to potentially perform beyond the underlying dataset policy.
% To provide a clearer overview of the development trend from traditional methods like RL to generative methods including DT and Diffuser, we present an illustration in Figure \ref{fig:main}.
% To summarize, our contributions are as follows:
% \vspace{-1mm}
% \begin{itemize}[leftmargin=1cm]
% \item This perspective paper highlights the need for next-generation real-time bidding display advertising based on a generative bidding strategy. We discuss and experimentally identify the potential direction for building such a strategy.
% \item We introduce a novel causal bidding strategy based on diffusion completion (CBD) that aligns training with inference by modeling bidding as a completion task. Additionally, we implement a trajectory-level return model to overcome dataset performance limitations.
% \item We demonstrate the effectiveness of CBD through superior performance and reliability on large-scale real-world bidding datasets, paving the way for the development of new generative bidding strategies.
% \end{itemize}

\vspace{-3mm}
\section{Preliminary and Related Work}
\vspace{-2mm}
\subsection{Problem Statement of Auto-Bidding}\label{sec:statement}
\vspace{-2mm}
Consider a scenario where there are \(H\) impression opportunities arriving sequentially, each labeled by an index \(i\). 
% On an advertising platform, advertisers place bids to compete for these impression opportunities. 
An advertiser wins an impression if its bid \(b_i\) surpasses those advertisers and incurs a cost \(c_i\).
% , which is usually the highest bid among the other advertisers in a second-price auction format. 
% Throughout this period, the advertiser aims 
The goal is to maximize the total value of the impressions won, represented by \(\sum_i o_i v_i\), where \(v_i\) denotes the impression value and \(o_i\) is a binary variable indicating whether the advertiser wins impression \(i\). 
Additionally, it is essential to consider the budget and various KPI constraints, such as limiting the unit cost of specific advertising events like CPC and CPA.\cite{USCB}.
% The budget constraint is defined as \(\sum_i o_i c_i \leq B\), where \(B\) is the budget. 
% Other KPI constraints are more intricate, such as limiting the unit cost of specific advertising events like CPC and CPA.
Therefore, the auto-bidding problem could be formulated as:
% \[
% \frac{\sum_i c_{ij} o_i}{\sum_i p_{ij} o_i} \leq C_j,
% \]
% where \(C_j\) is the upper bound of the \(j\)-th constraint provided by the advertiser, \(p_{ij}\) can be any performance indicator, such as conventions, and \(c_{ij}\) is the cost of constraint \(j\). Given \(J\) constraints, we have the multi-constraints bidding problem:
{\small
\begin{equation}
\begin{aligned}
\text{maximize} \quad & \sum\nolimits_i o_i v_i \\
\text{s.t.} \quad & \sum\nolimits_i o_i c_i \leq B  \\
& \frac{\sum_i c_{ij} o_i}{\sum_i p_{ij} o_i} \leq C_j, \quad \forall j  \\
& o_i \in \{0, 1\}, \quad \forall i
\label{eq:goal}
\end{aligned}
\end{equation}}

\vspace{-2mm}
where $B$ is the budget, \(C_j\) is the upper bound of the \(j\)-th constraint provided by the advertiser. \(p_{ij}\) can be any performance indicator, such as conventions, and \(c_{ij}\) is the cost of constraint \(j\).
A previous study \cite{USCB} has shown the optimal solution:

\vspace{-6mm}
{\small
\begin{equation}
b_i^* = \lambda_0 v_i + \sum\nolimits_{j=1}^{J} \lambda_j p_{ij}{C_j}, 
\label{eq:action_bid}
\end{equation}
}

\vspace{-2mm}
where \(b_i^*\) is the optimal bid for impression \(i\). 
The parameters \(\lambda_j\) represent the optimal bidding parameters. 
\yw{Please note that in industrial ad platforms, auto-bidding methods typically update bidding parameters at a relatively low frequency, such as every 30 minutes, rather than for every impression opportunity.}
However, given the uncertain and ever-changing advertising environment, calculating these optimal bidding parameters directly is impossible. 
Therefore, it stimulates research into advanced techniques for addressing this multi-constraint bidding problem in large-scale auctions, where \textbf{a detailed literature review is in Appendix \ref{app:related}}.
% We discuss the related work below.
% , where we discuss the \textbf{related work} in details in Appendix \ref{app:related}.
% Advertisers of different types may express their preferences through various combinations of constraints. For instance, advertisers focused on maximizing returns typically only consider the budget constraint, whereas those using target-CPA bidding take into account both the budget constraint and the CPA constraint.

\vspace{-2mm}
\subsection{Auto-Bidding as Decision-making}
\vspace{-2mm}
% Given the highly dynamic nature of the advertising environment, it is essential to regularly adjust the optimal bidding parameters to maximize the total received value over the entire period. 
% This adjustment transforms the bidding task into a sequential decision-making problem. 
We consider a sequential decision-making framework where an auto-bidding agent interacts with the advertising environment \(\mathcal{E}\) at discrete timesteps. At each timestep \(t\), the agent receives a state \(s_t \in \mathcal{S}\) that describes the ad impression features and auction status, and then outputs an action \(a_t \in \mathcal{A}\).
The action is the adjustment to the bidding parameters $\lambda_j$, \textit{i.e.}, $a_t:=\{a_t^{\lambda_0},...,a_t^{\lambda_J} \}$, which is then mapped to the final bid using Eq.~(\ref{eq:action_bid}). 
% Assuming a Markov Decision Process (MDP), the transition dynamic \(\mathcal{T}\) can be expressed as \(\mathcal{T}: s_t \times a_t \rightarrow s_{t+1}\), indicating that the next state \(s_{t+1} \in \mathcal{S}\) is determined by the current state \(s_t\) and action \(a_t\). 
In this context, the agent's policy is represented as \(\pi(a_t|s_t)\). 
The advertising environment has an underlying unknown state transition dynamic \(\mathcal{T}\).
Without assuming an MDP, the next state could be influenced by both the historical trajectory \(\tau\) and the current state. 
After transitioning to the next state, the env \(\mathcal{E}\) emits a reward \(r_t\), which represents the value contributed to the objective within the time period \(t\). This process repeats until the end of a bidding period, such as one day.
A detailed description of these elements can be seen in Appendix \ref{app:dataset}.
To achieve the goal in Eq.~(\ref{eq:goal}), we discuss several representative related works below.

\textbf{Related Work.} As the complexity of online ad environments increased, RL became a major approach for auto-bidding. 
USCB \cite{USCB} employs RL to maximize the value under multiple constraints.
MAAB \cite{maab} considers the multi-agent RL to maximize value and social welfare.
SORL \cite{sorl} proposes a sustainable online RL framework to explore in the online ad system. 
Recently, DiffBid \cite{aigb}, a pioneering generative auto-bidding method, utilized a decision diffuser to create a planning and control scheme, demonstrating the remarkable potential of generative models over reinforcement learning (RL).
Despite the promising properties, some methods, such as GAS \cite{gas} and GAVE \cite{gave}, have found that DiffBid is ineffective in large-scale auctions.
Therefore, this paper deeply investigates this issue and explores the potential of diffusers in large-scale auctions.

% \begin{itemize}[leftmargin=0.5cm]
%     \item \textbf{$s_t$}: The state is a collection of information describing the advertising impression feature and auction status, such as impression value prediction, remaining time, remaining budget, and current KPI ratio for constraints.
%     \item \(a_t\): Adjustment to the bidding parameters \(\lambda_j\), \(j = 0, \ldots, J\), at time period \(t\), modeled as \((a_t^{\lambda_0}, \ldots, a_t^{\lambda_J})\).
%     \item $r_t$: The value contributed to the objective obtained by the candidate impression set between step \(t\) and \(t + 1\)
% \end{itemize}

\vspace{-2mm}
\subsection{Diffuser for Auto-Bidding}
\vspace{-2mm}
The diffuser is developed for decision-making based on a diffusion model \cite{old_diffuser,diffuser}.
For simplicity, we introduce the basic elements of the decision diffuser (DD) employed by DiffBid.
DD aims to maximize likelihood estimation (MLE) of a trajectory $x_0(\tau)$ given a condition $y(\tau)$ \cite{aigb}:
{\small
\begin{align}
    \max_\theta \mathbb{E}_{\tau\sim\mathcal{D}}[\log p_\theta (\boldsymbol{x}_0(\tau) | \boldsymbol{y}(\tau))],
    \label{eq:objective_diffusion}
\end{align}
}

\vspace{-4mm}
where $\tau$ is the trajectory index and $\boldsymbol{y}(\tau)$ can be the property of the trajectory $\boldsymbol{x}_0(\tau)$, such as the accumulated reward of the trajectory.
For brevity, we index the states from 0 to  $T$, \textit{i.e.},
$
    \boldsymbol{{x}}_0 = \{\boldsymbol{{s}}_0, \boldsymbol{{s}}_1, ..., \boldsymbol{{s}}_T\}.
$
Specifically, DD is built with a predefined forward noising process
$
    q(\boldsymbol{x}_{k+1}(\tau) | \boldsymbol{x}_k(\tau)) := \mathcal{N}(\boldsymbol{x}_{k+1}(\tau)| \sqrt{\alpha_k} \boldsymbol{x}_k(\tau), (1-\alpha_k)\bf{I})
$
and a trainable reverse denoising process 
 $
{\small
    p_{\theta} (\boldsymbol{x}_{k-1}(\tau) | \boldsymbol{x}_k(\tau), \boldsymbol{y}(\tau)):=\mathcal{N}(\boldsymbol{x}_{k-1}(\tau)| \mu_\theta(\boldsymbol{x}_k(\tau), k, \boldsymbol{y}(\tau)), \sigma^2_k\bf{I}),
    % \label{eq:denoise}
    }
$
where $\mathcal{N}$ denotes a Gaussian distribution,
% with mean $\boldsymbol{\mu}$ and variance $\sigma^2$, $\alpha_k\in\mathbb{R}$ determines the variance schedule, 
and 
$
    \boldsymbol{x}_{k}(\tau):= \{\boldsymbol{s}_0^k, \boldsymbol{s}_1^k, ..., \boldsymbol{s}_T^k\}
$ 
denotes the noisy trajectory at step $k$.
% In this paper, we follow the cosine schedule for setting the value of $\alpha_k$ that smoothly increases the diffusion noise level using a cosine function.
We could get such a DD by training on a tractable variational evidence lower bound of \(\log p_\theta(\boldsymbol{x}_0(\tau)|\boldsymbol{y}(\tau))\) with a surrogate loss \cite{diffusion}:

\vspace{-6mm}
{\small
\begin{equation}
\small{
    \mathcal{L}_{\text{denoise}}(\theta) := \mathbb{E}_{k \sim [1,K], \boldsymbol{x}_0 \sim \mathcal{D}, \boldsymbol{\epsilon} \sim \mathcal{N}(\mathbf{0}, \mathbf{I})} \left\| \boldsymbol{\epsilon} - \epsilon_\theta(\boldsymbol{x}_k(\tau), k, \boldsymbol{y}(\tau)) \right\|^2,
    \label{eq:loss_denoise}
    }
\end{equation}}

\vspace{-2mm}
where \(\boldsymbol{\epsilon} \sim \mathcal{N}(\mathbf{0}, \mathbf{I})\) is the noise added to the data sample \(\boldsymbol{x}_0\) to produce \(\boldsymbol{x}_k\), and we use a deep neural network \(\epsilon_\theta(\boldsymbol{x}_k, k, \boldsymbol{y}(\tau))\) to estimate the noise \(\boldsymbol{\epsilon}\). 
% Although we are estimating noise, it is actually equivalent to predicting the mean of \(p_\theta(\boldsymbol{x}_{k-1}(\tau)|\boldsymbol{x}_k(\tau), \boldsymbol{y}(\tau))\), as \(\mu_\theta(\boldsymbol{x}_k, k, \boldsymbol{y}(\tau))\) is parameterized with the guidance \(\boldsymbol{\hat{\epsilon}}_k\).
% This is why it is also called denoising, \textit{i.e.}, predicting a clean data sample $\boldsymbol{x}_{k-1}$ from a noisy one $\boldsymbol{x}_{k}$.
Note that the condition \(\boldsymbol{y}(\tau)\) could be randomly dropped during training to form the classifier-free generation guidance \cite{diffuser}.
% \(\boldsymbol{\hat{\epsilon}}_k\) inEq. (\ref{eq:epsilon_k}). 

\vspace{-1mm}
At a timestep $t$ during auto-bidding, we could do planning by generating trajectories $\boldsymbol{\tilde{x}}_0$ that consist of a sequence of multi-step states with expected properties and then executing actions to get bids according to the selected trajectory, leading to an explainable and reliable bidding paradigm.
Specifically, an action $\boldsymbol{a}_t$ is typically obtained by inputting the observed states $\boldsymbol{s}_{t-L:t}$ with a horizon $L$ and the generated next state $\boldsymbol{\tilde{s}}_{t+1}$ into an inverse dynamics model $f_\phi$, \textit{i.e.}, 
$
    \boldsymbol{a}_t = f_\phi(\boldsymbol{s}_{t-L:t}, \boldsymbol{\tilde{s}}_{t+1}).
$
The inverse dynamic model could be independently trained using the training data, \textit{i.e.},

\vspace{-5mm}
{\small
\begin{equation}
    \mathcal{L}_a(\phi)=\mathbb{E}_{\{\boldsymbol{s}_{t-L:t+1},\boldsymbol{a}_t\}\sim\mathcal{D}}||\boldsymbol{a}_t - f_\phi(\boldsymbol{s}_{t-L:t}, \boldsymbol{\tilde{s}}_{t+1})||^2.
    \label{eq:loss_a}
\end{equation}
}

\vspace{-2mm}
By combining the diffusion model with the inverse dynamic model, a decision-making policy of DiffBid is formed, \textit{i.e.},
$
    \boldsymbol{a}_t \sim \pi_{\theta,\phi}(\boldsymbol{a}_t|\boldsymbol{s}_{t-L:t}, \boldsymbol{y}(\tau)).
$

Obviously, the quality of the generation of $\boldsymbol{\tilde{x}}_0$ directly determines the auto-bidding performance. However, we found the DD could suffer from the generation uncertainty issue due to the randomness in the denosing process. Besides, the training objective in Eq.~(\ref{eq:loss_denoise}) lacks explicit guidance on dynamic legitimacy, resulting in the generation of next states that cannot be achieved from the previous observed states, as illustrated in Fig. \ref{fig:main} and experiments in Fig. \ref{fig:causal}, leading to inverse dynamic model outputting bad actions with these anomaly transitions. Below are some related works for addressing this issue, though not specifically for auto-bidding tasks.

\textbf{Related Work.} AdaptDiffuser (AD) \cite{adapt_diffuser} and MetaDiffuser \cite{metadiffuser} use a forward dynamic model for guiding the generation to pursue consistent trajectories, but poor base model generation can hinder effectiveness and necessitate training revision \cite{diffuser}.
DiffuserLite (DL) \cite{DiffuserLite} reformulates the training process as generative interpolation by training multiple diffusers to generate segment trajectories, but it could limit overall planning ability. 
Adopting causal network architectures for training diffusers, such as diffusers based on RNN \cite{DiffusionForcing} or DiT \cite{dit} with causal attention \cite{causal_atten}, is closely related to addressing this issue by enforcing the generation of next states based on historical states. However, they raise concerns about accumulated errors and learning difficulty.

\section{Causal Auto-Bidding via Diffusion Completer-Aligner}
\vspace{-2mm}
We propose a novel causal auto-bidding method, termed CBD, to address the generation uncertainty issue for enhancing bidding performance in large-scale auctions by \textit{\textbf{i}}) reformulating training a diffuser as learning a completer to do completion task and \textit{\textbf{ii}}) aligning the generated states to advertisers' objectives represented as the condition $y(\tau)$ in the inference process by refining the generated states via a trajectory-level return model as an aligner. A detailed illustration of our CBD method could be seen in Fig. \ref{fig:main}.

\subsection{\textbf{Completer: Learning to Complete in Training}}
% \vspace{-2mm}
Reflecting the large language model (LLM), when a user provides a query, the LLM completes the conversation session by generating an answer, which has demonstrated remarkable instruction-following ability in NLP tasks \cite{guo2025deepseek, qwen} and inspires us to address the dynamic legitimacy issue of diffusers. 
Similarly, we here use the diffuser to play such a role as a completer to fulfill an ``observe and plan'' session by only generating the remaining sequence of the trajectory, but in an efficient single forward generation instead of auto-regressive next-token generation of LLM, as an ``answer'' after receiving the observations as a ``query''.

The ``query'' in auto-bidding could be constructed by combining the historical observations with the current observation, \textit{i.e.}, $s_{0:t}$. 
Specifically, at a timestep $t$, we start generation by sampling $\boldsymbol{x}_K(\tau)=\{\boldsymbol{s}_0^K,\boldsymbol{s}_1^K,...,\boldsymbol{s}_T^K\}$ from $\mathcal{N}(\bf{0}, \bf{I})$ and assign the observed states $\boldsymbol{s}_{0:t}$ to the corresponding places to get the input of the diffuser, \textit{i.e.},
% \begin{equation} 
%     \boldsymbol{\tilde{x}}_K(\tau, t)= \{\boldsymbol{s}_0, ..., \boldsymbol{s}_t, \boldsymbol{s}_{t+1}^K,...,\boldsymbol{s}_{T}^K\}.
%     \label{eq:x_k_generate}
% \end{equation}
\begin{equation} 
    \boldsymbol{\tilde{x}}_K(\tau, t)= \{\underbrace{\boldsymbol{s}_0, \ldots, \boldsymbol{s}_t}_{\text{Query}}, \underbrace{\boldsymbol{s}_{t+1}^K, \ldots, \boldsymbol{s}_{T}^K}_{\text{padding with noise}}\}.
    \label{eq:x_k_generate}
\end{equation}
Note that in some offline RL tasks, previous methods always set the first position as $s_t$ and let the diffuser generate the remaining sequence \cite{CleanDiffuser}, which is not necessary for auto-bidding as the trajectory typically has a fixed length and historical information is crucial for bid assessment \cite{aigb}.
Then, according to denoising process $p_{\theta} (\boldsymbol{x}_{k-1}(\tau) | \boldsymbol{x}_k(\tau), \boldsymbol{y}(\tau))$, we could sample the predicted next step's trajectory by
\begin{equation}
    \boldsymbol{\tilde{x}}_{k-1}(\tau, t) = \mu_\theta(\boldsymbol{\tilde{x}}_k(\tau, t), k, \boldsymbol{y}(\tau)) + \sigma_k \boldsymbol{z}, 
    \label{eq:recursive_generate}
\end{equation}
where $\bf{z} \sim \mathcal{N}(\bf{0}, \bf{I})$.  
Specifically, the $\sigma_k$ could be typically set as $\sigma_k:= 1-\alpha_k$ and the $\mu_\theta$ could be parameterized by 
$
    \mu_\theta(\boldsymbol{\tilde{x}}_k(\tau, t), k, \boldsymbol{y}(\tau))=\frac{1}{\sqrt{\alpha_k}}(\boldsymbol{\tilde{x}}_k(\tau, t) - \frac{1-\alpha_k}{\sqrt{1-\bar{\alpha}_k}}\boldsymbol{\hat{\epsilon}}_k),
    % \label{eq:mu}
$
where $\bar{\alpha}_k := \prod_{i=1}^{k}\alpha_k$ and the $\boldsymbol{\hat{\epsilon}}_k$ could be modeled with a conditional model $\epsilon_\theta(\boldsymbol{\tilde{x}}_k(\tau, t), k, \boldsymbol{y}(\tau))$ by randomly dropping the condition $\boldsymbol{y}(\tau)$ in a classifier-free guidance scheme \cite{class-free}, \textit{i.e.},
\begin{equation}
    \boldsymbol{\hat{\epsilon}}_k := {\epsilon}_{\theta} (\boldsymbol{\tilde{x}}_k(\tau, t), k) + \omega (\epsilon_{\theta} (\boldsymbol{\tilde{x}}_k(\tau, t), \boldsymbol{y}(\tau), k) - \epsilon_{\theta} (\boldsymbol{\tilde{x}}_k(\tau, t), k)),
    \label{eq:epsilon_k}
\end{equation}
where the scalar $\omega$  affects the diversity of generation. 
Due to the low tolerance of actions in auto-bidding, we could set $\omega$ as 1 during inference to alleviate the generation uncertainty and enhance the property-aligned generation.
Though some works claim a larger guidance scalar $\omega$ may decrease the legitimacy \cite{DiffuserLite, aligndiff}, we enhance it by our proposed augmented training method later.
By recursively applying the above procedure and assigning the first $t$-length of the generation with the query, we could get the final generated trajectory
\begin{equation}
        \boldsymbol{\tilde{x}}_0(\tau, t)= \{\underbrace{\boldsymbol{s}_0, \ldots, \boldsymbol{s}_t}_{\text{Query}}, \underbrace{\boldsymbol{\tilde{s}}_{t+1}, \ldots, \boldsymbol{\tilde{s}}_{T}}_{\text{Answer}}\}.
\end{equation}
Additionally, as the dynamic legitimacy issue may occur in every position of the generated trajectory, we augment the previous diffuser training procedure by setting $t$ as an extra random variable to construct the observations query and reformulate training as learning to complete.
Specifically, with the augmented random variable $t$, the completer $p_\theta$ should capture the causality distribution between the observed part $\{\boldsymbol{s}_0, ..., \boldsymbol{s}_t\}$ and the remaining part $\{\boldsymbol{s}_{t+1}, ..., \boldsymbol{s}_T\}$ with the causal root $\boldsymbol{y}(\tau)$, thus the training objective is changed into
% \begin{equation}
%     \min_\psi ||\boldsymbol{s}_{t+1:T} - f_\psi(\boldsymbol{s}_{0:t}, \boldsymbol{y}(\tau))||^2.
%     \label{eq:objective_new}
% \end{equation}
\begin{align}
    \max_\theta \mathbb{E}_{\tau\sim\mathcal{D}}[\log p_\theta (\boldsymbol{s}_{t+1:T}|\boldsymbol{s}_{0:t},\boldsymbol{y}(\tau))].
    \label{eq:augment_diffusion}
\end{align}
% Therefore, the objective is significantly different from the objective of previous diffusion-based bidding methods inEq. (\ref{eq:objective_diffusion}), highlighting an urgent need for correct planning-based bidding methods based on deep generative models.
This objective could be seen as an augmented version of the vanilla objective of diffuser, \textit{i.e.}, replacing $s_{0:t}$ with $\emptyset$ to be $p_\theta (\boldsymbol{s}_{0:T}|\emptyset,\boldsymbol{y}(\tau))$ degenerates the above augmented objective to the vanilla one in Eq. (\ref{eq:objective_diffusion}).
To achieve this objective, we can make a small modification by introducing an augmented new completion training loss $\mathcal{L}_{c}$, while still utilizing the powerful generation ability of diffusers, \textit{i.e.}, 
\begin{equation}
{\small
    \mathcal{L}_{c}(\theta) :=\mathbb{E}_{k\sim[1,K], \boldsymbol{x}_0\sim\mathcal{D}, {t\sim[0,T-1]}, \boldsymbol{\epsilon} \sim \mathcal{N}(\bf{0}, \bf{I})}||\boldsymbol{\epsilon} - \epsilon_\theta( {\boldsymbol{\tilde{x}}_k(\tau, t)}, k, \boldsymbol{y}(\tau))||_{t+1:T}^2
    \label{eq:loss_comp}
}
\end{equation}
\noindent{where 
$||\cdot||_{t+1:T}^2$ means we only compute the squared loss on the position index $t+1$ to $T$ in a trajectory.}
With such an additional variable $t\sim[0,T-1]$ and replacing $\boldsymbol{x}_k(\tau)$ to $\boldsymbol{\tilde{x}}_k(\tau, t)$ of Eq. (\ref{eq:x_k_generate}), the model is learned to be aware of the dynamic legitimacy of adjacent states in every positions of the trajectory and thus enhancing the dynamic legitimacy of the whole trajectory.

\subsection{\textbf{Aligner: Refining Generations in Inference}} \label{sec:grad}
% Although the new objective in Eq. (\ref{eq:objective_new}) can compel the diffusion model to generate states similar to the offline dataset, this alone is insufficient for the bidding task.
After training, the learned policy $\pi^*$ consisting the diffuser $p_\theta$ and the inverse dynamic model $f_\phi$ minimizes the distribution matching error to the underlying training data distribution $\pi_\beta$ that was used to collect the data, \textit{i.e.},
% \vspace{-1mm}
$
    \min_{\theta,\phi} \text{D}_{\text{KL}}[\pi^*(\boldsymbol{a}_t|\boldsymbol{s}_{t-L:t}, \boldsymbol{y}(\tau)) || \pi_\beta(\boldsymbol{a}_t|\boldsymbol{s}_{t-L:t}, \boldsymbol{y}(\tau))].
$
% \vspace{-1mm}
However, there are still some concerns. First, the underlying data distribution $\pi_\beta$ could be biased to the property $\boldsymbol{y}(\tau)$ in some cases, leading to the bias of the learned model; Second, the generation results can be diverse but suboptimal due to the stochastic nature of sampling-based generation, as random noise $\bf{z}$ are recursively employed in the generation process in Eq. (\ref{eq:recursive_generate}).
\textit{i.e.}, the $a_t$ follows a complex and untractable policy distribution $\boldsymbol{a}_t \sim f(\boldsymbol{Z}|\boldsymbol{s}_{t-L:t}, \boldsymbol{y}(\tau)), \boldsymbol{Z}:=\{\boldsymbol{z^k}\sim \mathcal{N}(\boldsymbol{0}, \boldsymbol{I})\}_{k=1:K}$. It is still unclear how to explicitly control the sampling process in denoising to precisely get the optimal $a_t^*$, even with some advanced ODE solver methods \cite{ode, DPM-Solver-v3}.
Thus, due to these concerns, the generated trajectories may raise misalignment issues in satisfying the advertiser's expected property as shown in Fig. \ref{fig:causal}.

% Besides, the upper-bound performance of the trained model $\pi^*(\boldsymbol{a}_t|\boldsymbol{s}_{0:t}, \boldsymbol{y}(\tau))$ may be constrained by the underlying policy $\pi_\beta$ that collected the dataset, \textit{i.e.}, the diffusion-based decision-making policy is actually approaching $\pi_\beta$ instead of surpassing it that could be expressed as
% \begin{equation}
%     \min_\theta \text{D}_{\text{KL}}[\pi^*(\boldsymbol{a}_t|\boldsymbol{s}_{t-L:t}, \boldsymbol{y}(\tau)) || \pi_\beta(\boldsymbol{a}_t|\boldsymbol{s}_{t-L:t}, \boldsymbol{y}(\tau))].
% \end{equation}
To improve the quality of the generated trajectory for enhancing the following auto-bidding performance, we introduce an aligner $R_\varphi$ during the inference stage.
This aligner directly refines the generation by performing a gradient update on the $\boldsymbol{\tilde{s}}_{t+1:T}$ after the generation,  utilizing a trajectory-level return model $R_\varphi(\boldsymbol{\tilde{x}}_0(\tau, t))$, which could be expressed as
\begin{equation}
    \boldsymbol{\tilde{s}}_{t+1:T}' \leftarrow \boldsymbol{\tilde{s}}_{t+1:T} - \lambda \nabla_{\boldsymbol{\tilde{s}}_{t+1:T}} ||R_\varphi(\boldsymbol{\tilde{x}}_0(\tau, t))-\boldsymbol{y}(\tau)||^2.
\end{equation}
% Thus, the updated new next-state $\boldsymbol{\tilde{s}}_{t+1}'$ could further improve the quality of following actions according to this refined trajectory with a higher return.
When the generated trajectory has a higher estimated property value $R_\varphi(\boldsymbol{\tilde{x}}_0(\tau, t))$ than the expected property $\boldsymbol{y}(\tau)$ provided by the advertiser, the refinement operation would adjust the trajectory to a new one with a lower property value, and vice versa.
The return model could be independently trained to regress on the conditions $\boldsymbol{y}(\tau)$ in the datasets expressed as
\begin{equation}
    \mathcal{L}_r(\varphi) := \mathbb{E}_{\{\boldsymbol{x}_0, \boldsymbol{y}\}\sim\mathcal{D}}||R_\varphi(\boldsymbol{x}_0(\tau)) - \boldsymbol{y}(\tau)||^2.
\end{equation}
This regression objective could largely alleviate the issue of sparse rewards. Although the immediate reward \( r_t \) for most steps may be zero, leading to a sparse reward challenge, the accumulated rewards \(\sum\nolimits_{t=0:T} r_{t}\) are not zero and can be easy to learn.
Note that the aligner differs from the classifier-guidance denoising technique used in diffusers, which also relies on a trained return model \( \mathcal{J}(\boldsymbol{x}) \) \cite{old_diffuser}. 
This technique, briefly expressed as \( \boldsymbol{x}_{k-1}\sim\mathcal{N}(\boldsymbol{\mu}_k+\alpha\nabla_\theta \mathcal{J}(\boldsymbol{\mu}_k),\sigma_k^2\boldsymbol{I}) \), only aims to maximize the return but can reduce generation controllablity. In auto-bidding, advertisers may prefer the strategy to be conservative sometimes to avoid winning unnecessary ad impressions at unsuitable times or with inappropriate audiences. 
Moreover, the $\mathcal{J}$ must iteratively interact with each denoising step, creating challenges in extending it to various denoising sampling methods such as DDPM \cite{diffusion}, DDIM \cite{ddim}, and ODE \cite{ode} methods.
In contrast, we refine only the final output to maintain computational efficiency and enable flexible future development, regardless of the training and inference techniques employed by the diffuser.
For instance, future directions could include multi-round refinement or exploring non-gradient methods, such as directly adding a residual to the output like LoRA \cite{lora} for optimal alignment.

Finally, we feed the updated $\boldsymbol{\tilde{s}}_{t+1}'$ into the inverse dynamic model to obtain the final action.
For a better understanding of our CBD method, we provide the training and inference procedure in Algorithm \ref{algo:train} and \ref{algo:infer}.

\begin{figure}[t] % 使用 [t] 选项使其浮动到页面顶部
\centering
\begin{minipage}{0.48\textwidth}
\begin{algorithm}[H]
\caption{\textbf{Training Procedure of CBD}}
\begin{algorithmic}[1]
\scriptsize
\State Initialize model parameters of completer $\epsilon_\theta$, aligner $R_\varphi$, and inverse dynamic model $f_\phi$
\For{each iteration}
    \State Sample $k \sim [1, K]$, $x_0 \sim \mathcal{D}$, $t \sim [0, T-1]$, $\boldsymbol{\epsilon} \sim \mathcal{N}(\mathbf{0}, \mathbf{I})$
    \State Compute the losses:
    \State \quad\quad $\mathcal{L}_{c}(\theta) = \mathbb{E} ||\boldsymbol{\epsilon} - \epsilon_\theta(\boldsymbol{\tilde{x}}_k(\tau, t), k, \boldsymbol{y}(\tau))||_{t+1:T}^2$
    \State \quad\quad $\mathcal{L}_r(\varphi) = ||R_\varphi(\boldsymbol{x}_0(\tau)) - \boldsymbol{y}(\tau)||^2$
    \State \quad\quad $\mathcal{L}_a(\phi)= ||\boldsymbol{a}_t - f_\phi(\boldsymbol{s}_{t-L:t, \boldsymbol{s}_{t+1}})||^2$
    \State Update $\theta$, $\varphi$, $\phi$ using gradient descent on $\mathcal{L}_{c}(\theta)$, $\mathcal{L}_r(\varphi)$, and $\mathcal{L}_a(\phi)$, respectively
\EndFor
\end{algorithmic}
\label{algo:train}
\end{algorithm}
\end{minipage}
\hfill
\begin{minipage}{0.48\textwidth}
\begin{algorithm}[H]
\caption{\textbf{Inference Procedure of CBD}}
\begin{algorithmic}[1]
\scriptsize
\State Initialize: initial state $\{\boldsymbol{s}_0\}$, completer $\epsilon_\theta$, aligner $R_\varphi$, inverse dynamic model  $f_\phi$
\For{bidding timestep $t = 0$ to $T-1$}
    \State Generate future states $\{\boldsymbol{\tilde{s}}_{t+1}, ..., \boldsymbol{\tilde{s}}_T\}$ using completer $\epsilon_\theta$ by Eq. \ref{eq:recursive_generate}
    \State Do refinement by aligner: $\boldsymbol{\tilde{s}}'_{t+1} \leftarrow \boldsymbol{\tilde{s}}_{t+1:T} - \lambda \nabla_{\boldsymbol{\tilde{s}}_{t+1:T}} ||R_\varphi(\boldsymbol{\tilde{x}}_0(\tau, t))||^2$
    \State Get action: $\boldsymbol{a}_t = f_\varphi(\boldsymbol{s}_{t-L:t}, \boldsymbol{\tilde{s}}'_{t+1})$
    \State Execute bids and get next state $\boldsymbol{s}_{t+1}$ from the ad system
    \State Update query: $\{\boldsymbol{s}_0, ..., \boldsymbol{s}_t, \boldsymbol{s}_{t+1}\} \leftarrow \{\boldsymbol{s}_0, ..., \boldsymbol{s}_t\} \oplus {\boldsymbol{s}}_{t+1}$
\EndFor
\end{algorithmic}
\label{algo:infer}
\end{algorithm}
\end{minipage}
\end{figure}

% \begin{algorithm}[H]
% \caption{Inference Procedure of CBD}
% \begin{algorithmic}[1]
% \State \textbf{Input:} Observed states $\{s_0, ..., s_t\}$, condition $y(\tau)$, diffusion model $\epsilon_\theta$, return model $R_\varphi$, inverse dynamic model $f_\phi$
% \State \textbf{Output:} Final action for bidding
% \State Generate future states $\{\tilde{s}_{t+1}, ..., \tilde{s}_T\}$ using the trained diffusion model $\epsilon_\theta$ by  Eq. \ref{eq:recursive_generate}
% \State Do gradient update on $\tilde{s}_{t+1}$:  $\tilde{s}_{t+1}' \leftarrow \tilde{s}_{t+1} + \lambda \nabla_{\tilde{s}_{t+1}} R_\varphi(\tilde{x}_0(\tau, t))$
% \State Get the action by $a_t = f_\phi(s_{t-L:t}, \tilde{s}_{t+1}')$
% \State \textbf{Return:} Final action for bidding
% \end{algorithmic}
% \label{algo:infer}
% \end{algorithm}

% \section{Experiment}
\section{Experiment}

\subsection{Setup}
\noindent\textbf{Datasets.} 
To evaluate the performance of bidding strategy decision-making in large-scale ad auctions, we employ AuctionNet, a publicly available benchmark from Alibaba designed for ad auctions, based on a real-world online advertising platform \cite{auctionNet}. AuctionNet effectively simulates the complexity and integrity of real-world ad auctions through the interaction of several key modules: the ad opportunity generation module, the auto-bidding module, and the auction mechanism module. Additionally, AuctionNet provides a more challenging variant called Auction-Sparse, which offers a sparser dataset. Each dataset comprises 21 advertising delivery periods, with each period containing approximately 500,000 impression opportunities, divided into 48 intervals. 
% The auto-bidding strategy adopted by the 
A total of 48 advertisers of different settings participate in bidding, competing for every impression. 
The auto-bidding strategy is iteratively adopted by each advertiser in every period to evaluate its performance under diverse advertiser settings.
Details of the AuctionNet benchmark are shown in Table \ref{tab:dataset_setting} of Appendix \ref{app:dataset}.

% {More details are in Appendix \ref{parameter_setting}.}

\noindent\textbf{Evaluation Metrics.}
We test the baselines and our method on the maximize conversions bidding (MCB) task\footnote{https://support.google.com/google-ads/answer/7381968?hl=en}, 
% where advertisers specify a daily budget, 
% and the bidding strategy aims to maximize the total number
% of conversions. 
% In contrast, the Costcap setting involves advertisers setting both a budget and a CPA limit or Return On Investment (ROI) limit. 
and we adopt three metrics to evaluate the performance: 

\vspace{-3mm}
\begin{itemize}[leftmargin=0.5cm]
    \item  \text{\textbf{Value}}: the total received value, \textit{i.e.}, number of conventions $\sum_i o_i v_i$, during the bidding period;
    \item \textbf{ER} (\text{\text{Exceeding rate}} of the KPI constraints): 
    In a specific cost-cap setting, advertisers are concerned with how closely the strategy's actual KPI performance matches the target KPI limit, such as CPA. To assess this, we introduce the ER metric for a delivery period, defined by
    $
        \text{ER} = \frac{1}{J}\sum_jC_{j}^{real} / C_{j} = \frac{1}{J} \sum_j (\sum_i c_{ij} o_i / \sum_i p_{ij} o_i)/ C_{j}.
    $ 
    
    \item \textbf{Score}: by introducing a penalty term 
    $
        penaty_j=min\{(\frac{C_j}{\Sigma_i c_{ij} o_i / \Sigma_i p_{ij} o_i})^\beta, 1\},$
    where $\beta=2$,
    the score is a balance between the value and the KPI constraint, \textit{i.e.}, 
    $
        score = (\sum_i o_i v_i) \times min\{penaty_j\}_{j=1\sim J}.
    $
\end{itemize}

\vspace{-2mm}
\noindent\textbf{Baselines.} We compare our method to a wide range of baselines involving RL, decision transformer, and diffuser methods: 
For RL methods, \textbf{USCB} \cite{USCB}, an online RL approach that adjusts parameters dynamically to the optimal bids; \textbf{BCQ} \cite{bcq}, a typical offline RL method that updates policies  with only access to a fixed dataset; \textbf{CQL} \cite{cql}, an offline RL method for learning conservative value function by regularizing Q-values; \textbf{IQL} \cite{iql}, an offline RL method that enables multi-step dynamic programming updates without querying out-of-sample actions;
For transformer-based methods, \textbf{DT} \cite{dt}, a sequential decision-making generative method based on the transformer; \textbf{CDT} \cite{cdt}, a DT-based method that considers a vector of multiple constraints; {\textbf{DT}-\text{S}}: a DT-based method that utilizes the reward function considering both the winning value and the KPI constraints, an effective variant found in previous works including GAS \cite{gas} and GAVE \cite{gave};
For diffuser-based methods, \textbf{DiffBid} \cite{aigb}, a generative method based on diffusers to generate bidding trajectories given conditions;
\textbf{DiT} \cite{dit} builds upon DiffBid by replacing its backbone from U-Net \cite{unet} with a diffusion transformer; 
\textbf{DiT-causal} further modifies DiT by substituting its attention mechanism with a causal-attention mechanism \cite{waswani2017attention, causal_atten}.

\noindent\textbf{Implementation Details.}
For implementing the baselines, we use the default hyperparameters suggested in their respective papers and further fine-tune them to the best of our ability. For our CBD method, the diffuser backbone is based on Diffbid \cite{aigb}. The number of diffusion steps is set to 100, the horizon length of historical states is 3, and the batch size is 512 with a total of 500 training epochs. For the backbone of the diffuser, we adopt a temporal U-Net \cite{unet} with hidden sizes of [128, 256, 512]. The inverse dynamic model is implemented as an MLP with hidden sizes of [1024, 1024, 512]. 
The gradient guidance scale $\lambda$ is set as 0.1 following the previous typical setting of diffusers \cite{old_diffuser} for all tasks.
We train the model using the Adam optimizer \cite{adamw} with a learning rate of 1e-4 on an H100 GPU, applying momentum updates over 4 steps.

\begin{table}[t!]
\caption{Comparison with baselines under the metric of Value in different budget settings in the MCB task. The best results are bolded and the base policy model's results are underlined to demonstrate the performance improvement of our CBD method in the last column (improvements are statistically significant, \textit{i.e.}, two-sided t-test with $p<0.05$).}
% \vspace{-3mm}
\setlength\tabcolsep{8pt}
\renewcommand{\arraystretch}{1.5}
\resizebox{\linewidth}{!}{
\begin{tabular}{c|c|cccc|ccc|cccccc}
\hline
\hline
\multirow{2}{*}{ \textbf{Dataset} } & \multirow{2}{*}{ \textbf{Budget} }   & \multicolumn{4}{c}{\textbf{RL}} &  \multicolumn{3}{c}{\textbf{Transformer}}  &  \multicolumn{5}{c}{\textbf{Diffuser}}  \\
    & & {\textbf{USCB}} & {\textbf{BCQ}} & {\textbf{CQL}} & \multicolumn{1}{c}{\textbf{IQL}}  & {\textbf{DT}} & {\textbf{CDT}} & \multicolumn{1}{c}{ \textbf{DT}-S} & {\textbf{DiffBid}} & \textbf{DiT} & \textbf{DiT}-causal 
 & \textbf{CBD}-Completer & \textbf{CBD}  &\textit{improve} \\ \hline
          & 50\%   &  133.1      & 184.2                                 & 180.1                                 & 185.0                                 &  186.2                                  & 190.3                                   & 196.8                                    &  \underline{161.9}                                 & 156.4                  &  160.0              & 191.9 & \textbf{198.6} & +22.3\% \\
          & 75\%   & 201.9                                  & 260.1                                 & 256.6                                 &  259.9                                &  234.1                                  & 290.8                                   & 298.0                                    &     \underline{233.7}                              & 228.3                  &   229.1             & 280.3 & \textbf{298.3} & +27.8\%\\
AuctionNet & 100\%  & 267.4                                  &  270.6                                &  336.6                                &  322.2                                & 364.0                                   & 366.5                                   & 373.1                                    &  \underline{316.5}                                 & 307.0    &  311.8            &  370.4              & \textbf{374.0}  & +18.3\% \\
          & 125\%  & 335.0                                  &   333.0                               &  400.2                                & 398.6                                 & 393.0                                   & 400.2                                   &  418.6                                   &  \underline{384.0}                                 &  375.1                 & 381.8               & 420.5 & \textbf{427.2} &  +11.1\%\\
          & 150\%  & 415.9                                  &  388.5                                &  465.3                                &  457.3                                & 445.2                                   &   429.5                                 &   437.6                                  & \underline{452.5}                              & 435.1                  &  440.6              & 473.7 & \textbf{480.5}  & +6.19\%  \\ \hline
          & 50\%  & 15.42                                  &  18.10                                & 16.79                                 &  20.21                                & 18.20                                   & 18.01                                   & 19.62                                    &  \underline{14.54}                                 & 14.12                  & 14.33               & 20.08  & \textbf{20.56}  & +41.4\% \\ 
          & 75\%   & 21.63                                  & 27.00                                 &   21.80                               &  26.90                                &  27.50                                  & 27.54                                   &  28.70                                   & \underline{23.25}                                  & 22.58                  &   22.60             & 29.90 & \textbf{29.97} & +28.9\%
          \\
AuctionNet-sparse & 100\%  & 28.48                                  & 30.54                                 & 30.15                                 & 36.06                                 & 31.30                                   & 34.31                                   & 36.82                                    &  \underline{31.33}                                 &  31.02                 &  31.27              & 40.40  & \textbf{40.71}  & +29.9\%  \\
          & 125\%  & 34.88                                  &   35.19                               &  37.67                                & 43.92                                 & 43.51                                   & 42.52                                   &  43.91                                   &  \underline{40.54}                                 & 39.40                  &  39.14              & 45.54    & \textbf{46.04}  & +13.5\%\\
          & 150\%  & 43.08                                  &  36.97                                & 44.43                                 &   49.27                               &  50.04                                  &  50.93                                  &   49.65                                  &  \underline{47.18}                                &  45.81                 & 46.07               & 51.10   & \textbf{51.35}   & +8.91\%  \\ \hline
          \hline
\end{tabular}}
\label{tab:main}
\vspace{-5mm}
\end{table}

\vspace{-3mm}
\subsection{Performance Comparison with Baselines}

In this experiment, we conducted a performance comparison of various baseline methods under different settings, including varying datasets and budget constraints in the MCB task. The dataset used for this evaluation is AuctionNet and its sparse version, AuctionNet-sparse, and the budgets are set at 50\%, 75\%, 100\%, 125\%, and 150\% of the maximum allowable budget.
The results are indicated by the value as a comprehensive assessment of the performance.
The results are presented in Table \ref{tab:main}.
The CBD method consistently outperforms other models across all budget levels, with the most notable improvements in lower budget and sparse reward scenarios. 
Even without employing a trajectory-level return model for updating the predicted next state (referred to as CBD-Completer), its performance still significantly exceeds the baseline Diffbid, highlighting the benefits of employing a completer to address the dynamic legitimacy issue. 
\begin{table}[]
% \vspace{-3mm}
\caption{Comparison under more metrics of the diffuser-based methods and CBD variants, where arrow directions indicate performance improvement direction.}
\setlength\tabcolsep{8pt}
\renewcommand{\arraystretch}{1.2}
% \vspace{-2mm}
\resizebox{1.\linewidth}{!}{
\begin{tabular}{c|c|ccccccccc}
\hline
\hline
Dataset            & Metrics  & Diffbid  & DiT & DiT-causal &CBD-Transformer &CBD-CVAE &CBD-$\mathcal{A}$ & CBD-Distillation & CBD-Completer & CBD \\ \hline
                  & Value $\uparrow$   & 316    &  307 & 311 & 313  & 314  &354 & 365 & 370          & \textbf{374}    \\
AuctionNet        & ER $\downarrow$      & 1.37       & 1.42 & 1.40 &1.35 & \textbf{0.90} &1.13 & 1.13 & 1.12          & {1.10}    \\
                  & Score $\uparrow$   & 214   & 196     & 205  & 210 & 284  & 294 & 285 &   292        &  \textbf{298}   \\ \hline
                  & Value $\uparrow$  & 31.3     & 31.0   & 31.2 & 31.2 & 31.2 &32.0  & 39.8 &  40.4         &  \textbf{40.7}  \\
AuctionNet-sparse & ER $\downarrow$    & 1.23      & 1.32 & 1.30 & 1.21 & 1.25 &1.24 & 0.95 & 0.90           & \textbf{0.86}     \\
                  & Score $\uparrow$  & 23.1     & 22.0  &22.5 & 21.9 & 29.2 &29.5  & 34.0 &  35.5         & \textbf{37.0}     \\ \hline\hline
\end{tabular}
}
\label{tab:diffusion}
\end{table}
% \vspace{-3mm}
% \end{wraptable}

Additionally, the diffusion transformer (DiT) and its variant based on a causal-attention mechanism (DiT-causal) implementation do not enhance the diffuser's performance, underscoring the effectiveness of our training objective, \textit{i.e.}, learning to perform auto-bidding as learning to complete. This insight could be inspiring for future research.

\begin{figure}
    \centering
    \includegraphics[width=1.\linewidth]{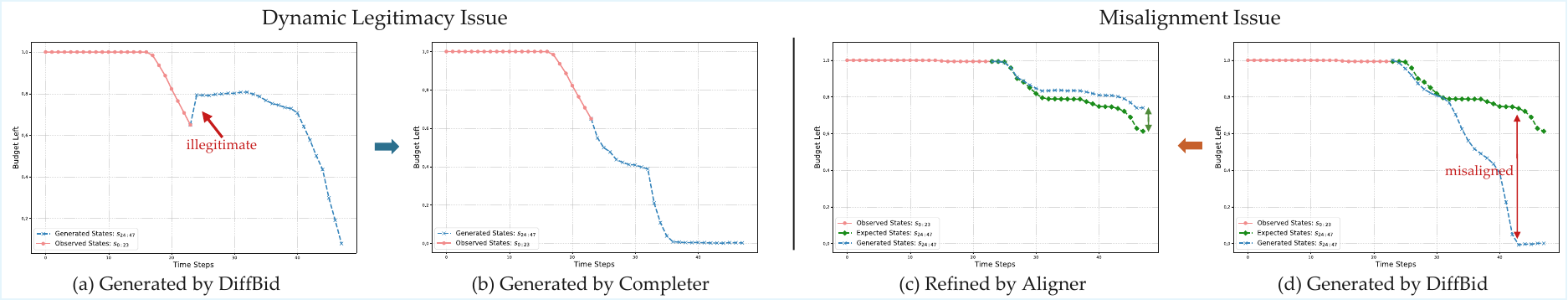}
    \vspace{-3mm}
    \caption{visualization of the generation uncertainty issue using an example case. In all subfigures, the red and green lines depict the ground-truth state information of the remaining budget from the selected trajectory in the dataset. The blue lines represent the generation results of DiffBid and our Diffusion Completer and Aligner. Observing from $t = 0$ to $23$, DiffBid generates states where the remaining budget at $t = 24$ is greater than at $t = 23$, indicating a \textbf{dynamic legitimacy issue}. Additionally, DiffBid could have significant deviations of the generated trajectories from the expected trajectory, even if satisfying dynamic legitimacy, demonstrating its poor ability to capture the relationship between the trajectory and its conditioning property $\boldsymbol{y}(\tau)$, highlighting a \textbf{misalignment issue}.}
    \label{fig:causal}
\end{figure}

To better understand the generation uncertainty issue, we provide a visualization in Fig. \ref{fig:causal} by randomly sampling trajectories from the training set and comparing the generated trajectories between DiffBid and our methods.
Visualization of more generations are in Appendix \ref{app:vis_gen}.
To enable a more comprehensive comparison with diffusion-based methods across a wider range of metrics, specifically considering both the number of conversion values and KPI constraints common in the cost-cap bidding mode, we present the results in Table \ref{tab:diffusion}. The CBD method generally demonstrates superior performance in ER, indicating a more cost-efficient bidding strategy, and it performs particularly well in sparse scenarios.

\subsection{Property Alignment Performance}
\begin{figure}[t!]
    \centering
    \subfigure[Selection on Smoothness]{
    \includegraphics[width=0.26\linewidth]{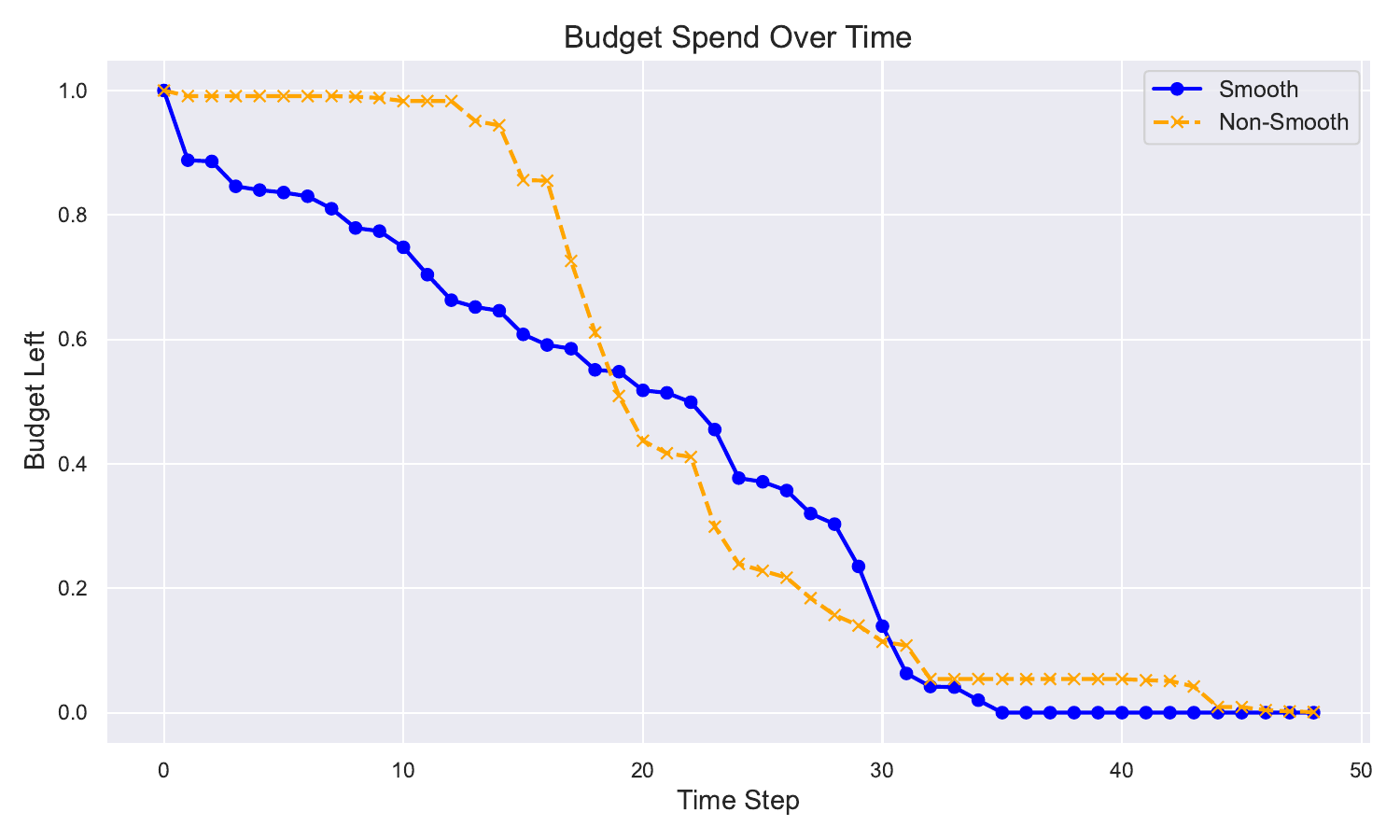}
    % \caption{}
    \label{fig:smooth_over_time}
    }
    \subfigure[Selection on Early-Spend]{
    \includegraphics[width=0.26\linewidth]{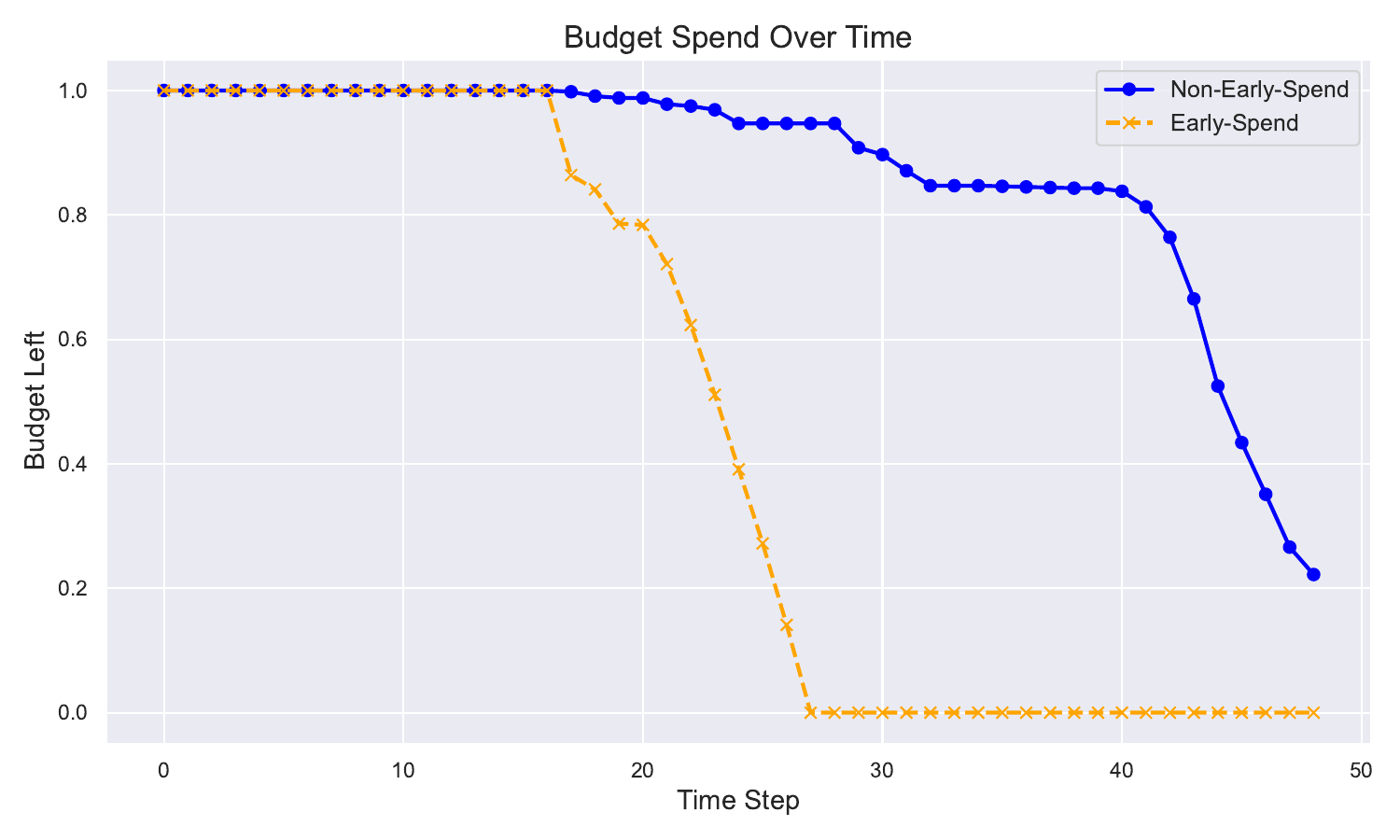}
    % \caption{}
    \label{fig:early_over_time}
    }
    \subfigure[Smoothness Dist.]{
    \includegraphics[width=0.205\linewidth]{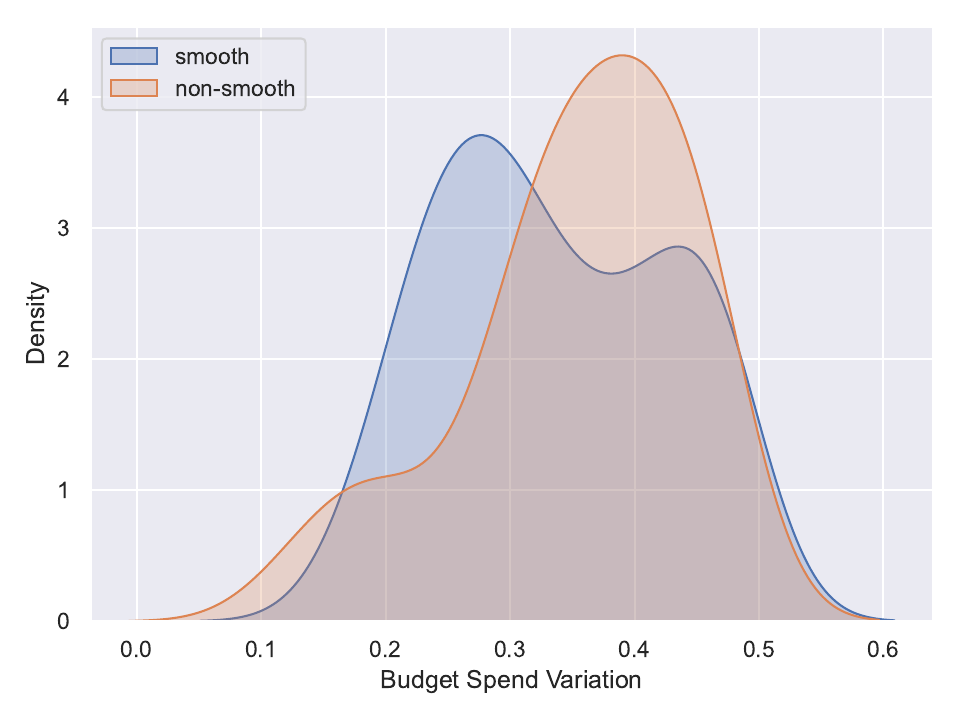}
    % \caption{}
    \label{fig:smooth}
    }
    \subfigure[Early-Spend Dist.]{
    \includegraphics[width=0.205\linewidth]{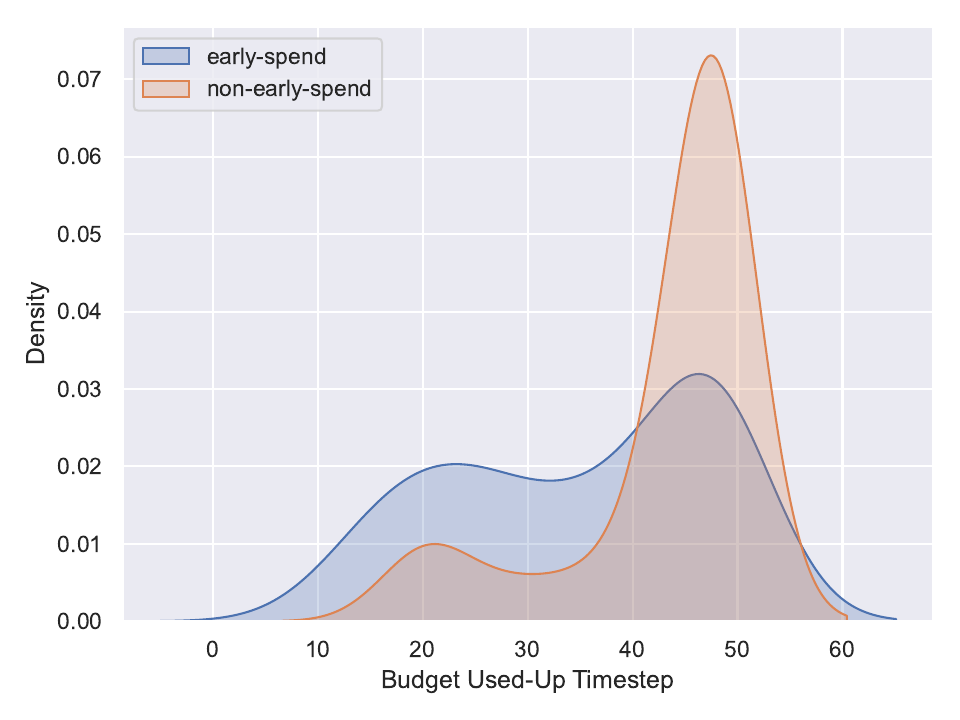}
    % \caption{}
    \label{fig:early}
    }
    \vspace{-4mm}
    \caption{Results for property alignment. (a-b) show the generated trajectories with different properties, including smoothness and early spending; (c-d) illustrate the property distribution after aligning the generated trajectories to the desired properties at each bidding timestep.}
    \label{fig:control}
    \vspace{-5mm}
\end{figure}

As illustrated in Figures \ref{fig:smooth_over_time} and \ref{fig:early_over_time}, we can generate diverse trajectories and employ the aligner trained on specific criteria to align the trajectory with desired characteristics, such as aligning the generation for a smoother trajectory. 
This flexibility is a major advantage of the planning-based auto-bidding strategy, as it enhances the reliability of the bidding process and facilitates the diagnosis and correction of errors or biases.

\noindent\textbf{Smoothness.}
Figure \ref{fig:smooth} illustrates the distribution of budget spending variation representing smoothness, \textit{i.e.}, the lower the variation, the smoother the strategy. It is evident that the smooth budget spend (depicted in blue) and the non-smooth budget spend (depicted in orange) exhibit distinct differences in their distributions, indicating the effectiveness of the aligner in refining the trajectory. The smooth budget spend is more concentrated and low in variation, indicating smaller variations, whereas the non-smooth budget spend is more spread out and higher, indicating larger variations.

\noindent\textbf{Early Spend.}
Figure \ref{fig:early} shows the distribution of budget used-up timesteps to present the property of early spending, where a property criterion for training the aligner could be $y(\tau)=\sum_{i=0}^{T/2}cost_i/\sum_{i=0}^{T}cost_i$. In real-world bidding, we always hope the bidding strategy does not use the budget too early, avoiding missing the impression opportunities of high value in the later phase of the bidding period. The early-spend (depicted in blue) and non-early-spend (depicted in orange) also differ in their timestep distributions. The early-spend distribution is concentrated at smaller timesteps, indicating that the budget is used up earlier before alignment.

\subsection{Ablation Study}
\begin{wrapfigure}{r}{0.3\textwidth}
    \centering
    \includegraphics[width=0.3\textwidth]{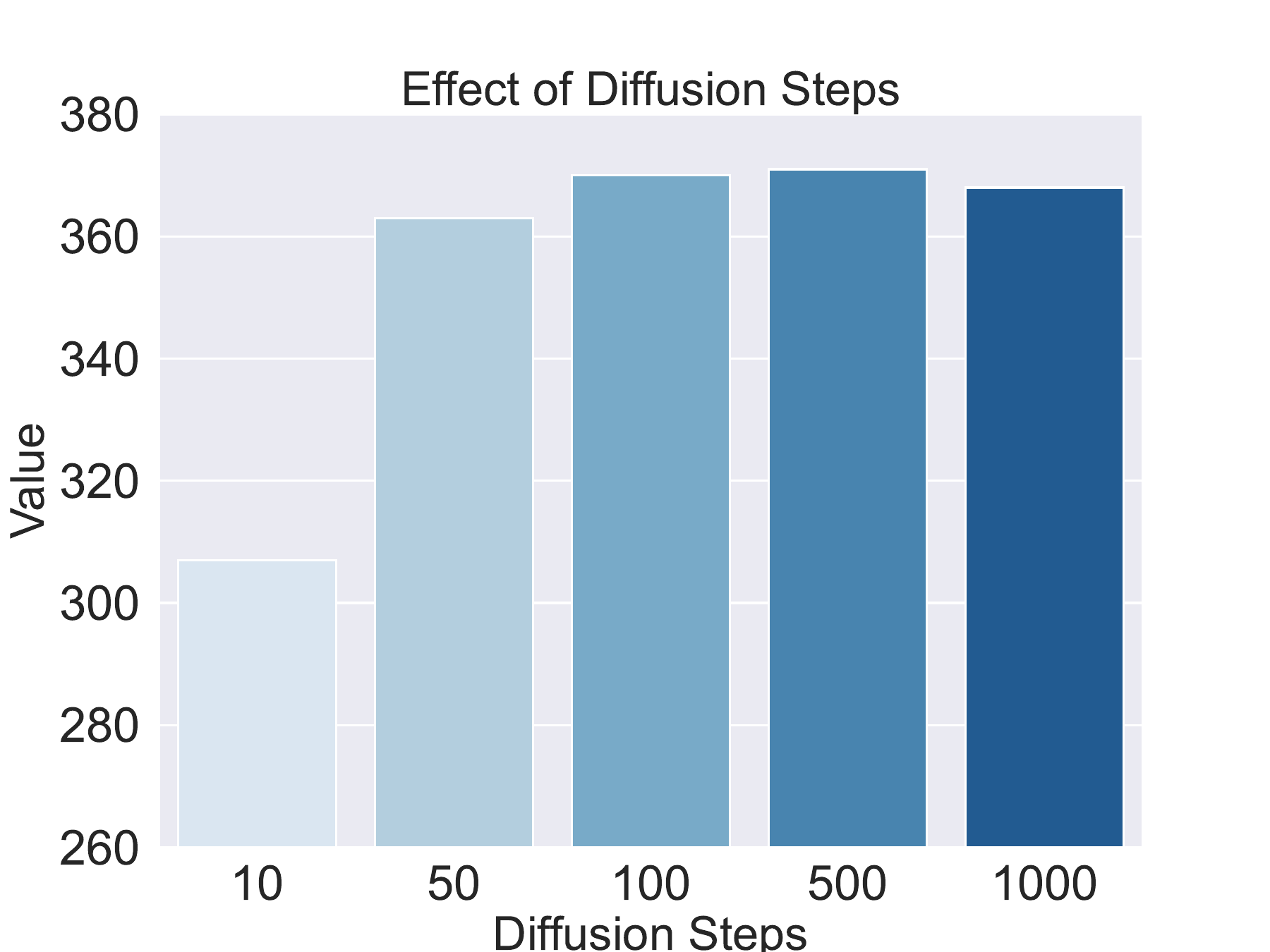}
    \vspace{-2mm}
    \caption{Effect of the diffusion steps.}
    \label{fig:steps}
\end{wrapfigure}

\noindent\textbf{Diffusion Steps.} In certain high-intensity bidding scenarios, 
where online auctions occur every minute, the computational efficiency, particularly the model's inference speed, is crucial. 
The diffusion model typically requires repeating the recursive denoising step in Eq. (\ref{eq:recursive_generate}) over multiple iterations, 
often up to 1000 steps. 
Therefore, we examined how the number of denoising steps affects performance to determine if a larger number of steps could improve outcomes.
% \vspace{-3mm}
% \vspace{-5mm}
% \begin{wraptable}{r}{0.4\textwidth}
% \caption{Effect of Diffusion Steps.}
% \vspace{-2mm}
% \centering
% \setlength\tabcolsep{2pt}
% \renewcommand{\arraystretch}{1.2}
% \footnotesize % 使用 \small 命令来缩小字体
% \begin{tabular}{c|ccccc}
% \bottomrule
% AuctionNet & 10 & 50 & 100 & 500 & 1000 \\ \hline
% Value & 307 & 363 & 370 & 371 & 368 \\ \toprule
% \end{tabular}
% \label{tab:diff_step}
% \end{wraptable}
As shown in Fig. \ref{fig:steps}, we trained the diffusion model with varying denoising steps, including 10, 50, 100, 500, and 1000. 
Our findings indicate that 100 denoising steps are sufficient for achieving good performance, resulting in a computationally efficient approach.
{Though industrial auto-bidding services have sufficient time for model's response as stated in Section \ref{sec:statement}, we also introduce a distillation method as a post-training trick, named SiD \cite{sid}, to distill our trained diffusion model into a one-step generator while maintaining similar performance as the results of \textbf{CBD-Distillation} shown in Table \ref{tab:diffusion}. As these acceleration techniques can be directly applied to our method, the inference latency can be significantly reduced. Therefore, we highlight the importance of identifying the root causes of failures and developing advanced methods for training the original diffusion models for auto-bidding, which is the primary contribution of our work and also the urgent need of this field.}

% For other acceleration methods for diffusion models}
% Other methods, such as DDIM \cite{ddim}, could further accelerate the inference speed. We consider this an avenue for future research.

% \noindent\textbf{Grad Scale.} 1.0, 0.1, 0.01, 0.001
% \begin{table}[h!]
% \caption{Grad Scale on the RM gradient update.}
% \centering
% \vspace{-3mm}
% \begin{tabular}{c|cccccc}
% \hline
% AuctionNet            & 1.  & 0.1 & 0.01 & 0.001  \\ \hline
% Value &   &     &      &        \\ \hline
% \end{tabular}
% \label{tab:grad_scale}
% \end{table}

\noindent\textbf{Necessity of Inverse Dynamic Model.}
Current diffuser-based methods, such as Diffbid and our CBD method, generate states first and then derive an action, which is not straightforward since our
major interest is in the action itself. 
% \begin{wraptable}{r}{0.4\textwidth}
% % \begin{wraptable}{r}{0.3\textwidth}
% \caption{Comparison of performance between our method and a variant CBD-$\mathcal{A}$.}
% \vspace{-2mm}
% \centering
% \setlength\tabcolsep{3pt}
% \renewcommand{\arraystretch}{1.0}
% \footnotesize % 使用 \small 命令来缩小字体
% % \vspace{-3mm}
% \begin{tabular}{c|ccc}
% \bottomrule
% AuctionNet            & Value $\uparrow$ & ER $\downarrow$ & Score $\uparrow$ \\ \hline
% CBD-$\mathcal{A}$ & 354  & 1.13  & 294     \\ 
% CBD & 374 & 1.10 & 298  \\
% \toprule
% \end{tabular}
% \label{tab:dp}
% % \end{wraptable}
% \end{wraptable}
This raises the question of whether we can directly generate the actions together with states, thereby avoiding the need for an additional inverse dynamic model. However, as shown in Table \ref{tab:diffusion}, our method still outperforms this variant (\textbf{CBD-$\mathcal{A}$}), empirically demonstrating the effectiveness and stability of training the model to learn planning first, resembling a chain-of-thought process \cite{cot} before determining the final action.

\yw{
\noindent\textbf{Alternative Backbones compared to Diffusion Model.}
To showcase the superior performance of using diffusion models as the backbone of CBD, we conducted an additional comparison experiment with other simpler generative models, including a Transformer-based CBD method and a Conditional VAE (CVAE)-based CBD method, noted as \textbf{CBD-Transformer} and \textbf{CBD-CVAE}, respectively.
Specifically, the CBD-Transformer utilizes a transformer where a masked sequence of states serves as the input, and the return of the trajectory acts as a condition interacting with the transformer block via AdaLN-zero. The CBD-CVAE employs a Temporal-UNet as the encoder and decoder, using the returns as the condition. 
The results in Table \ref{tab:diffusion} clearly demonstrate that employing diffusion models can achieve a better balance between maximizing impression value and meeting the KPI constraints, demonstrating the necessity of adopting diffusion models for CBD.
}

\yw{
\noindent\textbf{Robustness of the Aligner.}
For the principle of choosing a proper step size, as the offline data collected from online systems can be assumed to have a similar distribution to the online data, it is straightforward to use a portion of the offline data to evaluate the validity of different step sizes and employ the trajectory-level return model to assess alignment improvements. 
In this work, we choose the gradient step size of the aligner as 0.1, which generally brings improvement on various tasks.
Furthermore, we have included experimental results demonstrating the robustness of the gradient step size in Table \ref{tab:robust}. The validity is measured by the generated remaining budget's legitimacy. The GS-Align is an alternative alignment technique without gradient refinement, which is based on greedy selection of the best-aligned generated trajectory among parallelly sampled 10 generations.
}
\vspace{-4mm}
\begin{table}[h]
    \centering
    \setlength\tabcolsep{8pt}
    \renewcommand{\arraystretch}{1.2}
    \caption{Robustness and efficiency of the aligner's gradient step size.}
    \resizebox{1.\linewidth}{!}{\begin{tabular}{lcccccccc}
        \toprule
        \textbf{AuctionNet-Sparse} &DiffBid & $\lambda=0.001$ & $\lambda=0.01$ & $\lambda=0.05$ & $\lambda=0.1$ & $\lambda=1.0$ & $\lambda=5.0$ & {GS-Align} \\
        \midrule
        Value $\uparrow$ &31.3 & 40.4 & 40.4 & 40.6 & 40.7 & 40.6 & 39.4 & 40.6 \\
        ER $\downarrow$ &1.23 & 0.91 & 0.91 & 0.87 & 0.86 & 0.91 & 0.87 & 0.88 \\
        Score $\uparrow$ &23.1 & 35.5 & 35.5 & 36.2 & 37.0 & 35.7 & 34.2 & 35.8 \\
        Validity $\uparrow$ &77.4\% & 98.7\% & 98.7\% & 98.7\% & 98.7\% & 98.7\% & 96.0\% & 98.7\% \\
        \bottomrule
    \end{tabular}
    }
    \label{tab:robust}
\end{table}
\vspace{-3mm}

\vspace{-2mm}
\subsection{Extension to Offline RL Tasks}
\vspace{-2mm}
Our CBD method could also bearing insights and benefit for general offline RL tasks like the locomotion tasks.
We compare to the existing non-diffuser and diffuser-based methods together in Table \ref{tab:offlineRL}, and a detailed introduction to these baselines can be seen in Appendix \ref{app:dataset}.
% Note that in some offline RL tasks, we follow the setting of previous methods that always set the first position of a trajectory as $s_t$ and let the diffuser generate the remaining sequence \cite{CleanDiffuser}, though it is not necessary for auto-bidding as the trajectory typically has a fixed length and historical information is crucial for bid assessment \cite{aigb}.
Since these tasks have no fixed length, we follow the inference setting as DD, \textit{i.e.}, the current state is always assigned to the first place of the trajectory. 
As shown in Table \ref{tab:offlineRL}, the CBD method achieves state-of-the-art performance, demonstrating its effectiveness and generalizations in offline RL tasks. 
Notably, our method performs slightly better in the \textit{Medium-Expert} setting than in others, suggesting that transition quality also impacts performance, \textit{i.e.}, the \textit{Medium-Expert} data trains the model to generate both dynamic legitimate and expert-level transitions, whereas other settings only teach the model about dynamic legitimacy.
This situation differs from auto-bidding tasks, where the offline training log is derived from real advertisers' behaviors, which inherently have a basic quality guarantee. 
Poor strategies will be immediately discontinued to prevent significant economic losses. 
However, this also presents unique challenges in developing new methods that can outperform existing competitive strategies in large-scale auctions.

\vspace{-3mm}
\begin{table}[h]
\centering
\caption{Performance comparison across different datasets and environments of the D4RL benchmark. Results correspond to the mean and
standard error over 150 episode seeds, and the best scores of diffuser-based methods are emphasized in \textbf{bold}.}
\resizebox{1.\linewidth}{!}{
\begin{tabular}{ll|cccccc|cccccc}
\toprule
\multirow{2}{*}{ \textbf{Dataset} } & \multirow{2}{*}{ \textbf{Environment} }   & \multicolumn{6}{c|}{\textbf{non-Diffuser-based}} &  \multicolumn{5}{c}{\textbf{Diffuser-based}}   \\
 &  & \textbf{BC} & \textbf{CQL} & \textbf{IQL} & \textbf{DT} & \textbf{TT} & \textbf{MOReL} & \textbf{Diffuser} & \textbf{DD} &\textbf{AD} &\textbf{DL} &\textbf{CBD}\\
\midrule
            & HalfCheetah & 55.2 & 91.6 & 86.7 & 86.8 & {95} & 53.3 & 79.8 & ${90.6}$ & 90.4 & 88.5 & \textbf{93.5$\pm$0.83} \\
Medium-Expert  & Hopper & 52.5 & 105.4 & 91.5 & 107.6 & {110.0} & 108.7 & 107.2 & ${{111.8}}$ & 109.3 & 111.6 & \textbf{113.8$\pm$0.13} \\
            & Walker2d & {107.5} & 108.8 & 109.6 & 108.1 & 101.9 & 95.6 & 108.4 & ${{108.8}}$ & 108.2  & 107.1 & \textbf{109.1$\pm$0.33} \\
\midrule
        & HalfCheetah & 42.6 & 44.0 & 47.4 & 42.6 & 46.9 & 42.1 & 44.2 & ${{49.1}}$ & 44.3 &48.9 & \textbf{49.4$\pm$0.20} \\
Medium & Hopper & 52.9 & 58.5 & 66.3 & 67.6 & 61.1 & {95.4} & 58.5 & ${79.3}$ & 96.6 & \textbf{100.9} & 97.7$\pm$1.38 \\
        & Walker2d & 75.3 & 72.5 & 78.3 & 74.0 & 79.0 & 77.8 & 79.7 & ${82.5}$ & 84.4 & \textbf{88.8}  & 83.0$\pm$0.42 \\
\midrule
            & HalfCheetah & 36.6 & {45.5} & 44.2 & 36.6 & 41.9 & 40.2 & {42.2} & ${39.3}$ & 38.3 & 41.6 & \textbf{42.8$\pm$0.31}  \\
Medium-Replay  & Hopper & 18.1 & 95 & 94.7 & 82.7 & 91.5 & 93.6 & 96.8 & $\textbf{{100.0}}$ &92.2 & 96.6 & 96.5$\pm$0.18  \\
            & Walker2d & 26.0 & 77.2 & 73.9 & 66.6 & {82.6} & 49.8 & 61.2 & ${75.0}$ & 84.7 & \textbf{90.2} & 75.2$\pm$0.26  \\
\bottomrule
% \textbf{Average} & & 51.9 & 77.6 & 77.4 & 74.7 & 78.9 & 72.9 & 75.3 & {81.8} & 83.4 & 86.0  & 84.6  \\
% \bottomrule
\end{tabular}
\label{tab:offlineRL}
}
\end{table}

\vspace{-3mm}
\yw{
\section{Online A/B Test}}
\vspace{-2mm}
To verify the effectiveness of our proposed CBD, we have deployed it on the MCB production of Kuaishou’s advertising system. Under the MCB setting, the advertisers set the budget with or without the CPA/ROI constraint, and the bidding strategy aims to get as many conversions as possible under these constraints. Due to the limited online testing resources and the potential impact on the advertisers’ value, we
only compared CBD with the in-production DT-based method. 
% The experimental setup is as follows:
% \begin{itemize}
%     \item \textbf{State}: Budget, cost, time-based budget allocation, time-based cost speed, predicted conversion rates, real CPA or ROI states, etc.
%     \item \textbf{Action}: Adjustment to the last timestep's bidding parameter.
%     \item \textbf{Reward}: The acquired conversion value of each timestep.
% \end{itemize}
Over 7 days A/B test, there have been an average of 14700 campaigns per day. With the same budget allocated for each method's experiment, each method achieved a daily average of approximately 80,000,000 impressions, which is substantial enough to demonstrate the effectiveness of the methods. Both methods utilized 360 CPUs for computation, and inference latency was measured on it. 
\begin{wraptable}{r}{0.3\textwidth}
    \centering
    \setlength\tabcolsep{1pt}
    \renewcommand{\arraystretch}{1.5}
    % \footnotesize % 使用 \small 命令来缩小字体
    \caption{A/B test results.}
    \vspace{-2mm}
    \resizebox{1.\linewidth}{!}{
    \begin{tabular}{lc}
        \toprule
         Metric & -\textit{Compare}  \\
        \midrule
        Inference Latency & +6 ms \\
        Cost & +0.0\% \\
        Target Cost &  +2.0\% \\
        CPA Valid Ratio & +1.85\%\\
        \bottomrule
    \end{tabular}
    }
    \label{tab:ab}
\end{wraptable}
\vspace{-3mm}

As shown in Table \ref{tab:ab}, the average achieved conversions (Target Cost) for advertisers are improved by an average of +2.0\% while spending a similar budget (Cost), and the CPA valid ratio for costcap campaigns is also improved by +1.85\%, which are significant improvements for such a large-scale auto-bidding production. Though our method incurs an additional 6ms compared to DT (6ms), it is still acceptable and cost-effective without needing additional computational resources, which could support a maximum of 26ms at the same frequency (around 20s) of calling the service to update the bidding parameters.
Given the substantial commercial value achieved, the additional inference latency is both worthwhile and justifiable.

\section{Limitations and Conclusion}
\vspace{-2mm}
There are several \textbf{limitations} to our approach. 
This paper focuses exclusively on the auto-bidding strategy, which represents only one aspect of the real-time bidding system.
Therefore, future research should also consider other components, such as auction mechanism design, and their influence on the auto-bidding strategy.
CBD only considers the competition relationship between advertisers, but there may exist cooperation between them, \textit{e.g.}, advertisers may engage in private negotiations to reach a consensus and collectively lower their bids.
% Additionally, CBD uses the prediction of impression features, such as pCVR prediction, as inputs. 
% However, these predictions can sometimes be inaccurate and uncertain, which may introduce bias into the strategy.
% The aligner receives a property value rather than a language prompt, which may be more convenient for advertisers.

% The effectiveness of the CBD strategy is still dependent on the quality and scope of the training data, and further exploration regarding the data is needed to fully realize its potential. 
% Additionally, both the dataset and model sizes are relatively small compared to existing large language models (LLMs), indicating that the scaling laws in bidding warrant future exploration. Moreover, this paper focuses exclusively on the bidding strategy, which represents only one aspect of the real-time bidding system. Therefore, future research should also consider other components, such as auction mechanism design and user feedback prediction, and their influence on the bidding strategy.

In conclusion, this paper explores the potential of generative models, particularly diffusers, for real-time auto-bidding strategies in computational advertising. After examining previous failures in applying diffusers, we identified generation uncertainty as a key root cause and introduced a causal auto-bidding method based on a novel diffusion completer-aligner framework. Our method addresses the dynamic legitimacy issue through the completer and refines the generations to align closely with the advertiser's expected properties using an aligner. Experimental results support our analysis and demonstrate the effectiveness of our method, offering insights for general offline RL tasks as well.

\clearpage
\newpage
% \section{Appendices}

\bibliographystyle{unsrtnat}
\bibliography{reference}

%%%%%%%%%%%%%%%%%%%%%%%%%%%%%%%%%%%%%%%%%%%%%%%%%%%%%%%%%%%%
\clearpage
\newpage
\appendix

\section*{Technical Appendices and Supplementary Material}
% Technical appendices with additional results, figures, graphs and proofs may be submitted with the paper submission before the full submission deadline (see above), or as a separate PDF in the ZIP file below before the supplementary material deadline. There is no page limit for the technical appendices.

\section{Detailed Background}\label{app:related}
\subsection{Auto-Bidding in Computational Advertising}
\begin{figure}[h!]
    \centering
    \includegraphics[width=\linewidth]{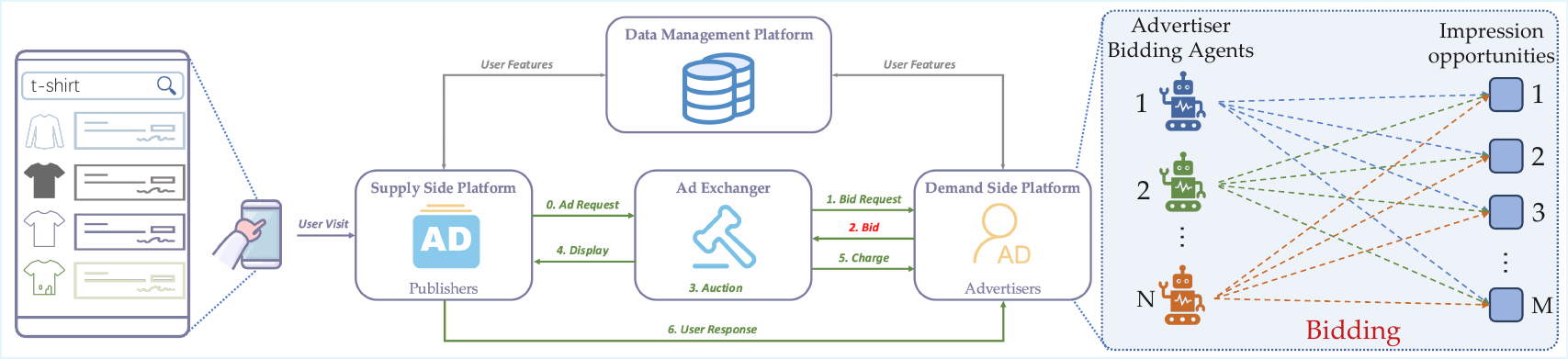}
    \vspace{-3mm}
    \caption{Real-time bidding system.}
    \label{fig:rtb}
\end{figure}

The rapid digital transformation of commerce has significantly expanded the reach of online advertising platforms, making them essential for engaging audiences and driving sales \cite{DBLP:conf/wsdm/WangY15, evans2009online}. 
% At the core of these platforms lies computational advertising, a scientific discipline that combines information retrieval, statistical modeling, machine learning, and optimization \cite{computation_ads}. 
As reported by the International Advertising Bureau, U.S. digital ad revenue from computational advertising revenue has reached \$114.2 billion \cite{IAB2024}, significantly contributing to the success and sustainability of information systems and technologies.
The primary goal of computational advertising is to effectively match ads with their relevant contexts on the web, such as the content where the ad is displayed and users' search queries or browsing behavior \cite{computation_ads,display_ads}.
With the continuous influx of ad impression opportunities to the platform, real-time bidding (RTB) based display advertising,  which emerged in 2009 \cite{emerge}, has become a major paradigm of computational advertising, 
which enables online advertising platforms to sell individual ad impressions via hosting a real-time auction and facilitates advertisers to bid for the impression opportunity based on estimated value.
Specifically, in real-time auto-bidding \cite{BM}, when an impression opportunity is generated from a user's visits, a bid request for the ad display impression opportunity, including its features like user, context, and auction status, is sent to numerous advertisers via an ad exchange. Then, advertisers employ specific auto-bidding algorithms to assess the bid request's potential value and determine a bid price in real-time on demand-side platforms (DSPs). 
The ad exchanger then selects the highest bidder to display the ad, charging them the market prices, typically the second-highest bid price in a second-price auction setting \cite{second_price}. 
A detailed illustration can be seen in Figure \ref{fig:rtb}.

To maximize the total value of impressions for an advertiser while adhering to certain economic constraints, research into advanced auto-bidding strategies has progressed from rule-based approaches to prediction-based methods, and more recently, to reinforcement learning-based techniques \cite{survey}.
For rule-based methods, the proportional-integral-differential (PID) is the most widely used controller method \cite{pid_old}. The control signal consists of proportional, integral, and differential terms of the error, accounting for current, past, and future trends, resulting in comprehensive control of the variable.
A typical example of applying PID in online advertising is controlling the campaign's Cost-Per-Action (CPA) to a reference level \cite{PID}.
The controller could also be used to control the budget pacing
to spend the budget smoothly \cite{rnn_control, DBLP:journals/mor/Myerson81}.
In addition to the PID controller, custom designs exist to adjust bids based on feedback data by classifying states as comfortable or risky and updating controlled variables accordingly \cite{Video}, while Multi-KPI \cite{Multi-KPI} controlled multiple KPIs simultaneously based on observed and target errors.
For prediction-based techniques integrating auto-bidding strategies with upstream prediction tasks, such as CTR prediction, CVR prediction, and market price prediction, bid optimization is no longer an independent task but is correlated with these prediction tasks and jointly optimized.
It is because a previous work, Bidding Machine \cite{BM}, argues that interconnected tasks and sequential optimization lead to suboptimal solutions, and joint optimization allows the learning process to focus on valuable and competitive impressions.
Some methods have tried this view, such as an EM-like (Expectation-Maximization) algorithm proposed to iteratively optimize the CTR learning and bid optimization \cite{em} and its extension incorporates market price estimation into the framework \cite{BM}.
Besides joint optimization, some methods tried introducing the lift/attribution prediction into the bid optimization, which is to predict the difference value before and after the ad is displayed \cite{lift_1, lift_2, lift_3, LiftBidding}. 
For the reinforcement learning based methods, they repeatedly interact with the environment, allowing bidding agents to gain value from winning ad impression opportunities and update their subsequent strategies. The entire bidding process for advertisers is essentially a sequential decision-making process, which has recently demonstrated superior effectiveness.

\subsection{Sequential Decision-Making for Auto-Bidding}
As the complexity of online bidding environments increased, reinforcement learning (RL) algorithms like USCB \cite{USCB}, SORL \cite{sorl}, and MAAB \cite{maab} became essential for bidding. 
USCB \cite{USCB} maximizes the value under multiple constraints and discovers a recursive structure to accelerate convergence.
MAAB \cite{maab} considers the multi-agent modeling to maximize value and social welfare,
proposing to mix cooperation and competition among multiple advertisers and improving the platform's revenue.
SORL \cite{sorl} proposes a sustainable online RL framework that trains the auto-bidding policy by directly interacting with the RTB system, using safe and efficient online exploration instead of a simulated environment. 
However, online RL poses risks to the RTB system and cannot fully leverage the extensive historical bidding logs, resulting in a preference for bidding methods based on offline RL.
Offline RL methods, which derive policies from existing datasets without requiring online interaction, have demonstrated considerable success. Noteworthy approaches include BCQ \cite{bcq}, which constrains the action space to encourage on-policy behavior; CQL \cite{cql}, which regularizes Q-values for conservative estimates; and IQL \cite{iql}, which facilitates multi-step dynamic programming updates without querying Q-values of out-of-sample actions during training. 

Despite their effectiveness, these methods are limited by the Markov Decision Process (MDP) assumption, whereas generative models \cite{vae, gan, gflownet, diffusion, dt} offer greater potential for bidding.
Some methods try to apply decision transformers (DTs) \cite{dt,liu2024sequential,gao2025future}  for auto-bidding.
GAS \cite{gas} employs MCTS-inspired post-training search methods to enhance the auto-bidding performance during inference;
GAVE \cite{gave} modifies the training of DT to enlarge its exploration space for finding better policies beyond the underlying dataset policy.
However, DT may suffer from poor explanation ability and sparse-reward challenges, making them limited in the wider auto-bidding tasks.
In contrast, DiffBid \cite{aigb} employs a decision diffuser to generate a long trajectory based on conditions like return, and then uses an inverse dynamic model to output the final action.
However, in large-scale auctions, DiffBid faces challenges related to generation uncertainty, which is the primary focus of our work.

\section{Details of Experiments}\label{app:dataset}
\textbf{Datasets.} The AuctionNet benchmark comprises two datasets: AuctionNet and AuctionNet-sparse. AuctionNet-sparse is a sparser version of AuctionNet, featuring fewer conversions. Each dataset includes 21 advertising delivery periods, with each period containing approximately 500,000 impression opportunities, divided into 48 intervals. 
The models are trained on these two datasets, while 5,000 randomly selected trajectories from each dataset are used exclusively for evaluating the visualization of the generation.
Detailed parameters are provided in Table \ref{tab:dataset_setting}.
Additionally, the state at each time step includes the following information:
\begin{itemize}
     \item time\_left: The remaining time steps left in the current advertising period.
     \item budget\_left: The remaining budget that the advertiser has available to spend in the current advertising period. 
     
     \item historical\_bid\_mean: The average values of bids made by the advertiser over past time steps.
    
     \item last\_three\_bid\_mean: The average values of bids over the last three time steps. 
    
     \item historical\_LeastWinningCost\_mean: The average of the least cost required to win an impression over previous time steps.
    
     \item historical\_pValues\_mean: The average of historical p-values over past time steps.
    
     \item historical\_conversion\_mean: The average number of conversions (e.g., sales, clicks, etc.) the advertiser achieved in previous time steps. 
    
     \item historical\_xi\_mean: The average winning status of advertisers in impression opportunities, where 1 represents winning and 0 represents not winning.
    
     \item last\_three\_LeastWinningCost\_mean: The average of the least winning costs over the last three time steps. 
    
     \item last\_three\_pValues\_mean: The average of conversion probability of advertising exposure to users over the last three time steps.
    
     \item last\_three\_conversion\_mean: The average number of conversions over the last three time steps.

     \item last\_three\_xi\_mean: The average winning status of advertisers over the last three time steps.
    
     \item current\_pValues\_mean: The mean of p-values at the current time step. 
    
     \item current\_pv\_num: The number of impressions served at the current time step
    
     \item last\_three\_pv\_num\_total: The total number of impressions served over the last three time steps. 
     
     \item historical\_pv\_num\_total: The total number of impressions served over past time steps. 
\end{itemize}

\begin{table}[h!]
  \centering
  \caption{The parameters of AuctionNet and AuctionNet-sparse.}
  \scalebox{0.9}{
    \begin{tabular}{ccc}
    \toprule
    Params & AuctionNet & AuctionNet-Sparse \\
    \midrule
    Trajectories & 479,376 & 479,376 \\
    Delivery Periods & 9,987 & 9,987 \\
    Time steps in a trajectory & 48    & 48 \\
    State dimension & 16    & 16 \\
    Action dimension & 1     & 1 \\
    % Return-To-Go Dimension & 1     & 1 \\
    Action range & [0, 493] & [0, 8178] \\
    Impression's value range & [0, 1] & [0, 1] \\
    CPA range & [6, 12] & [60, 130] \\
    Total conversion range & [0, 1512] & [0, 57] \\
    \bottomrule
    \end{tabular}%
    }
  \label{tab:dataset_setting}
\end{table}%

\noindent\textbf{Evaluation Details.} 
To assess the auto-bidding performance, we perform experiments in a simulated environment that mirrors a real-world advertising system, as provided by Alibaba \cite{auctionNet}. During the evaluation, an episode—also known as an advertising delivery period—consists of a day divided into 48 intervals, each lasting 30 minutes. Each episode includes approximately 500,000 impression opportunities that occur sequentially. An advertising delivery period features 48 advertisers from various categories, each with distinct budgets and CPAs, competing for all impression opportunities during the period. In each evaluation, our well-trained model acts as a specific advertiser, participating in bidding with a given budget and CPA. To thoroughly evaluate the model's performance across different advertisers, we employ various advertiser configurations and advertising periods, conducting multiple evaluations in the simulated environment and averaging the results to obtain the evaluation score.

\noindent\textbf{Details of Extension to Offline RL Tasks.}
The offline reinforcement learning (RL) tasks under evaluation include three widely recognized locomotion challenges: HalfCheetah, Hopper, and Walker2d. These tasks involve maneuvering three distinct Mujoco robots to achieve optimal speed while ensuring energy efficiency and maintaining stability. The D4RL \cite{d4rl} benchmark offers offline datasets at three quality tiers: "medium," which includes demonstrations of moderate proficiency; "medium-replay," which encompasses all replay buffer recordings observed during training up to the point where the policy attains medium-level performance; and "medium-expert," which equally blends medium and expert-level performances.
We also compared some additional non-diffuser-based methods on this benchmark, with details as follows: MOReL \cite{morel} addresses common pitfalls of model-based reinforcement learning, such as model exploitation, by learning a pessimistic Markov Decision Process (P-MDP) and training a near-optimal policy within this P-MDP. On the other hand, TT \cite{tt} employs a Transformer architecture to model distributions over trajectories and utilizes beam search as a planning algorithm.

\section{More Visualization of Generated Trajectories}\label{app:vis_gen}
\begin{figure}[htbp]
    \centering
    \subfigure{
        \includegraphics[width=0.31\linewidth]{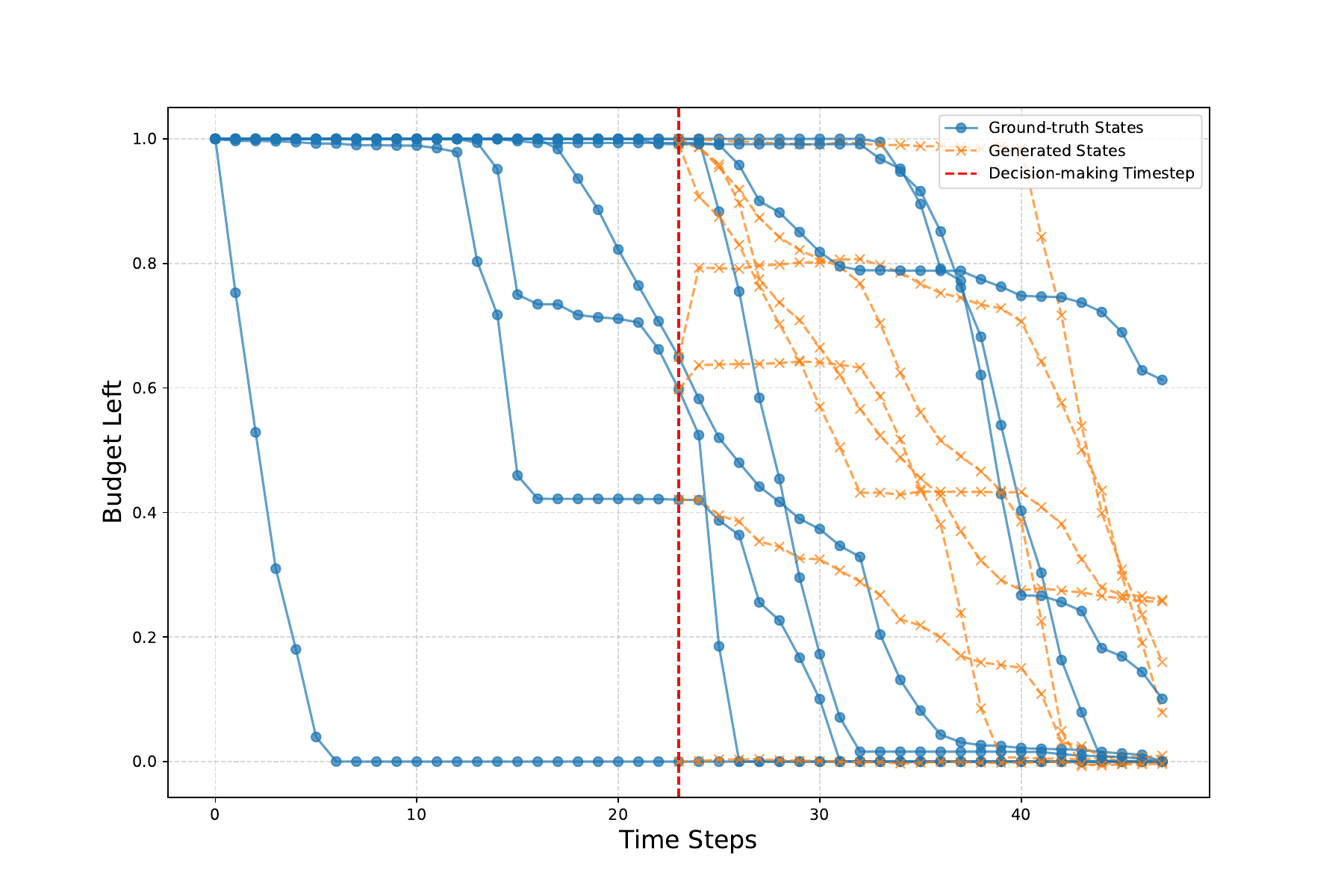}
    }
    \subfigure{
        \includegraphics[width=0.31\linewidth]{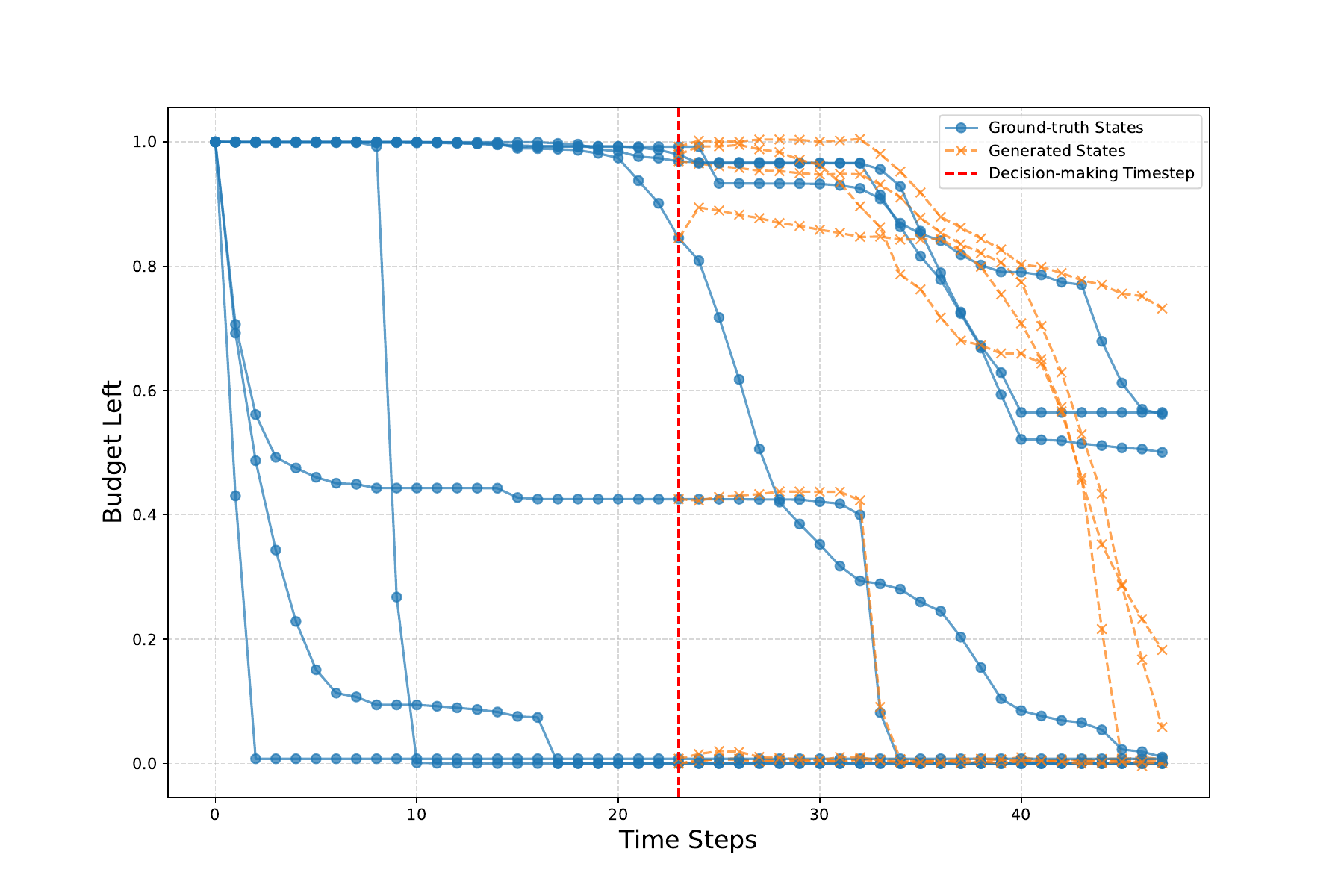}
    }
    \subfigure{
        \includegraphics[width=0.31\linewidth]{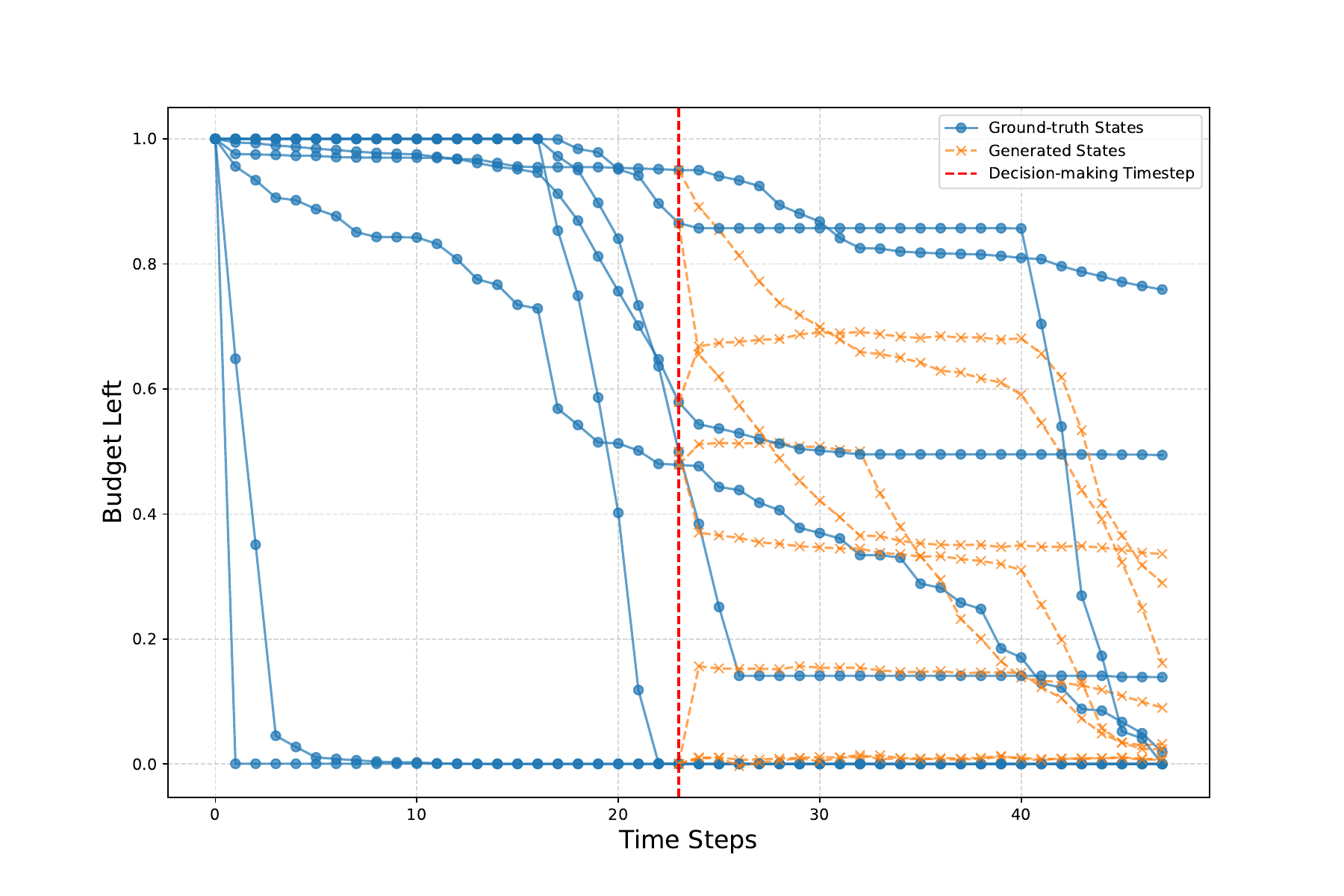}
    }
    \subfigure{
        \includegraphics[width=0.31\linewidth]{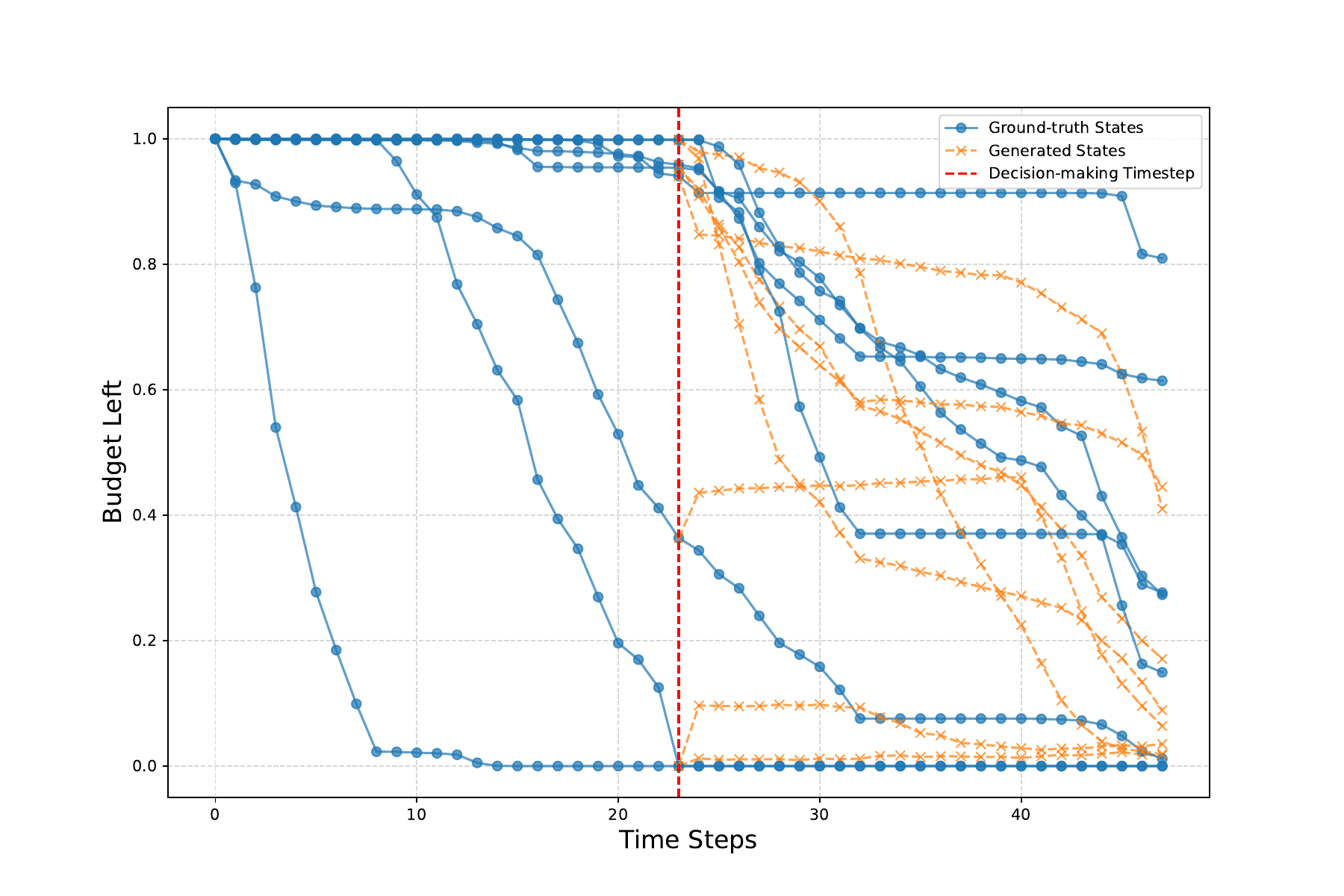}
    }
    \subfigure{
        \includegraphics[width=0.31\linewidth]{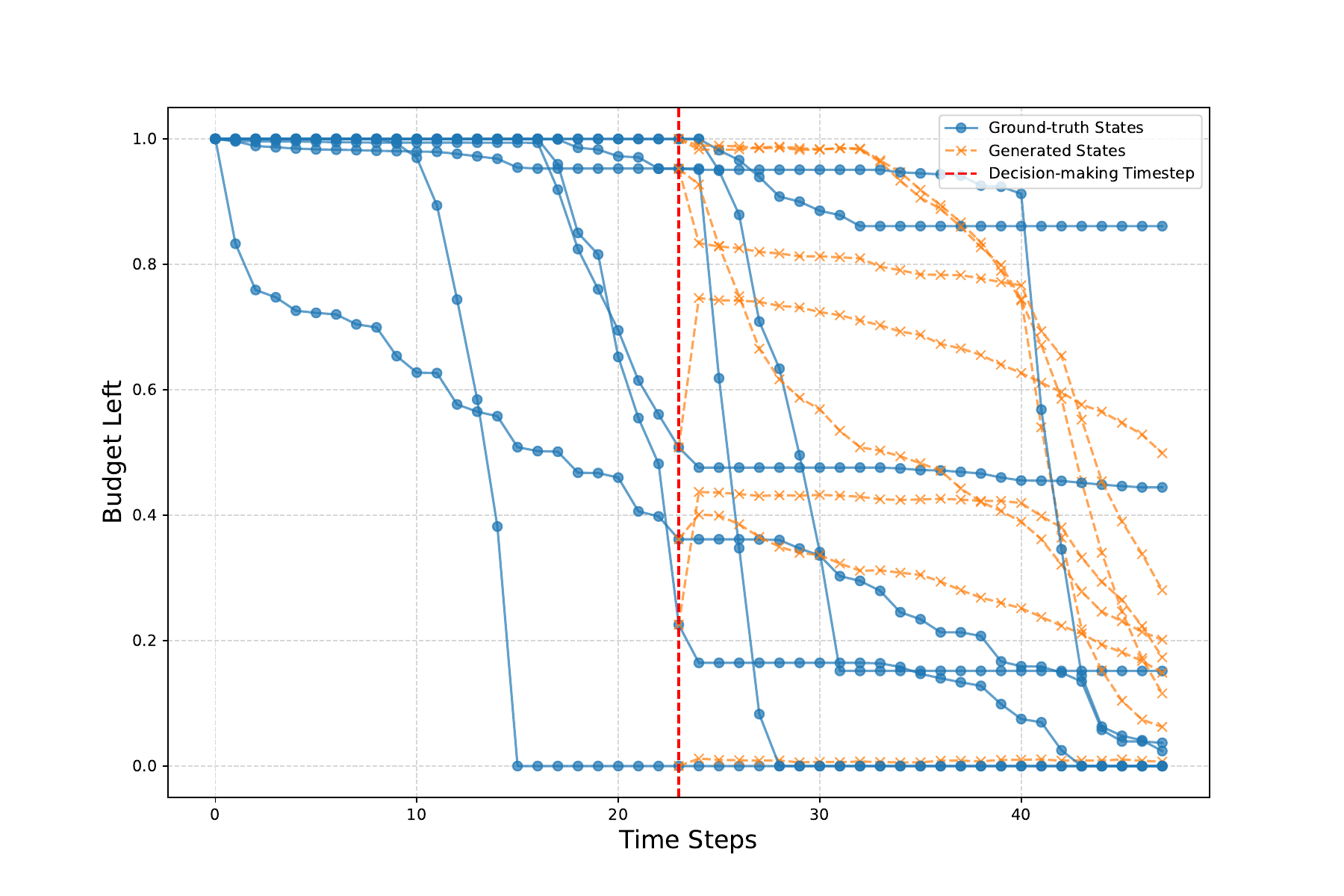}
    }
    \subfigure{
        \includegraphics[width=0.31\linewidth]{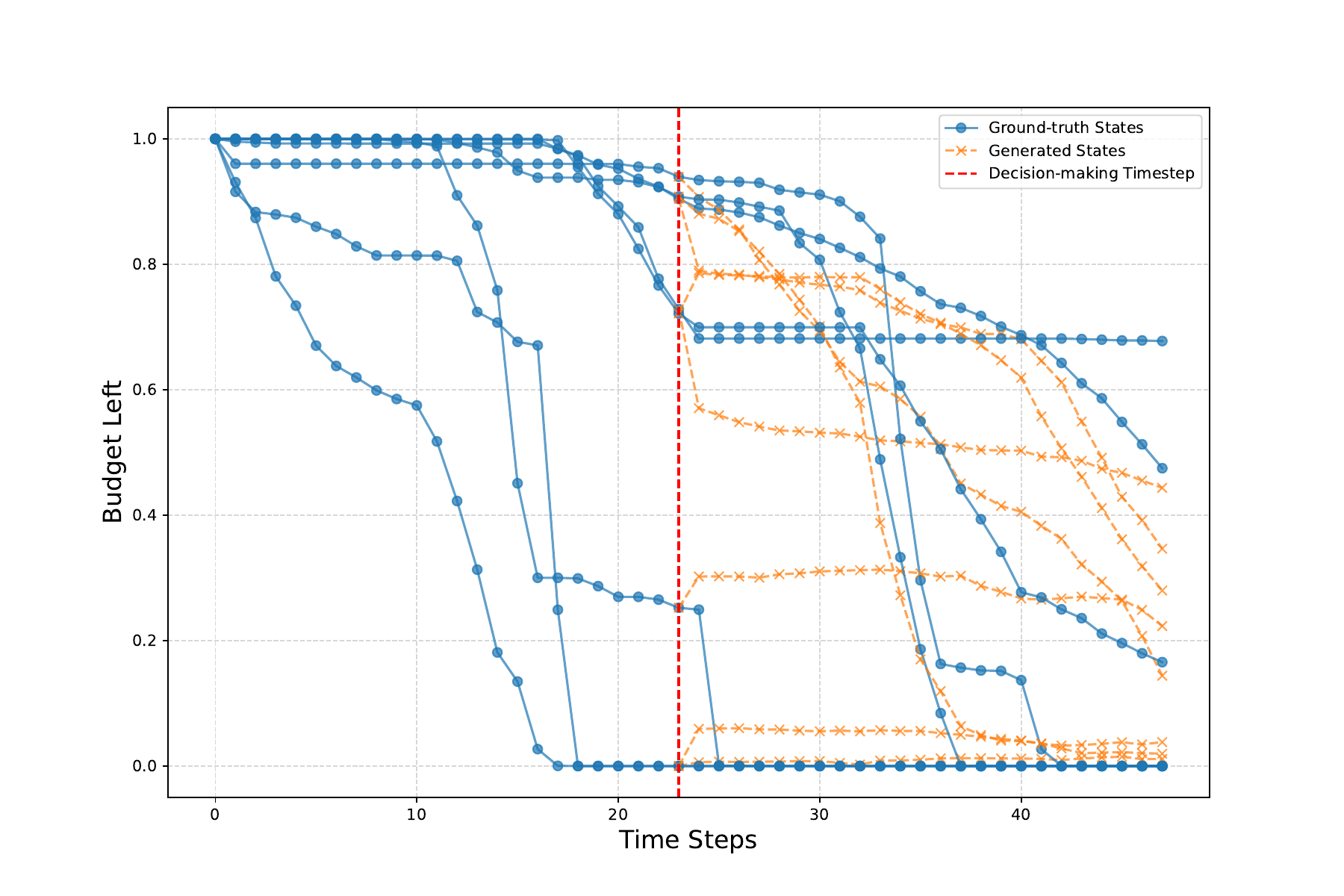}
    }
        \subfigure{
        \includegraphics[width=0.31\linewidth]{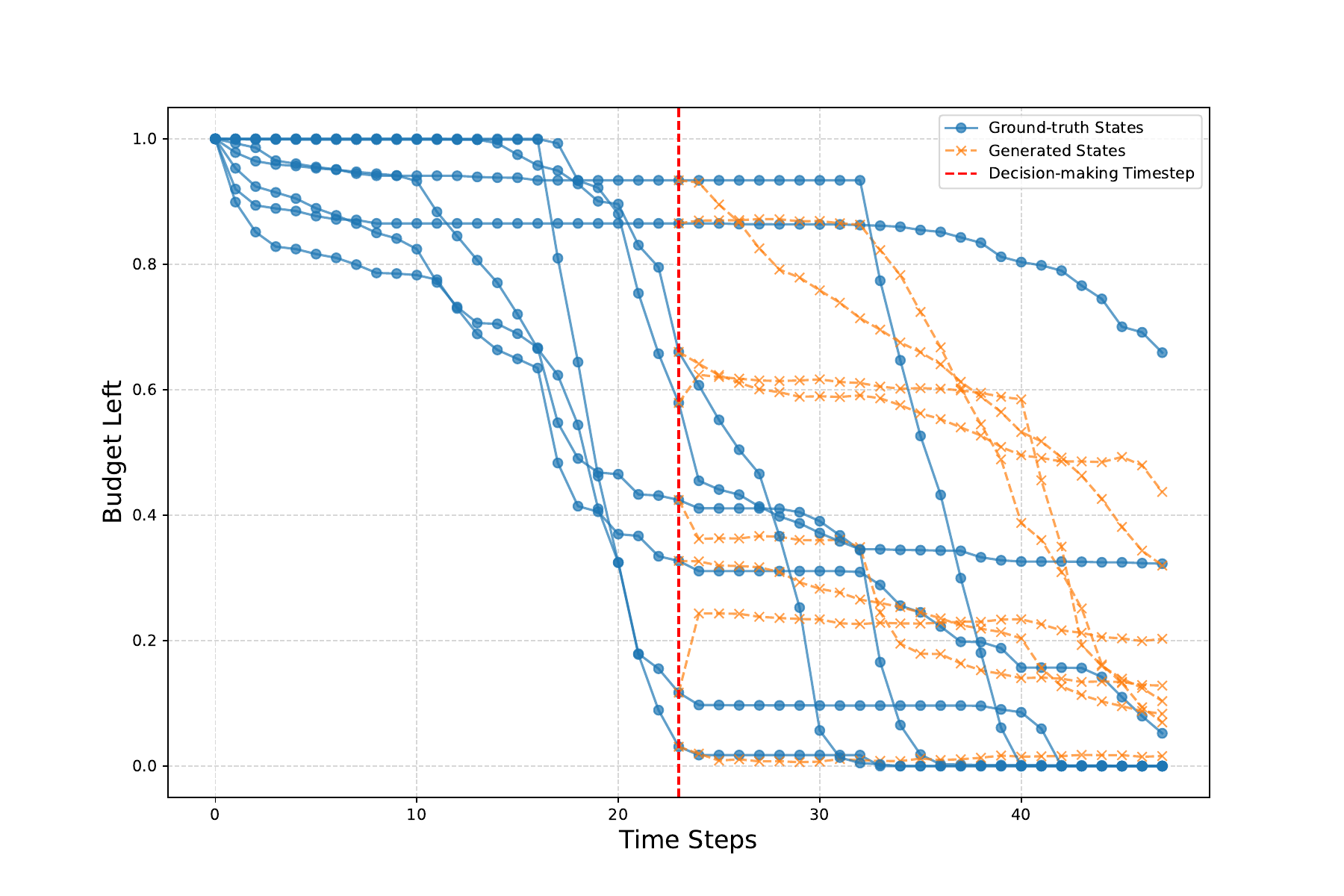}
    }
    \subfigure{
        \includegraphics[width=0.31\linewidth]{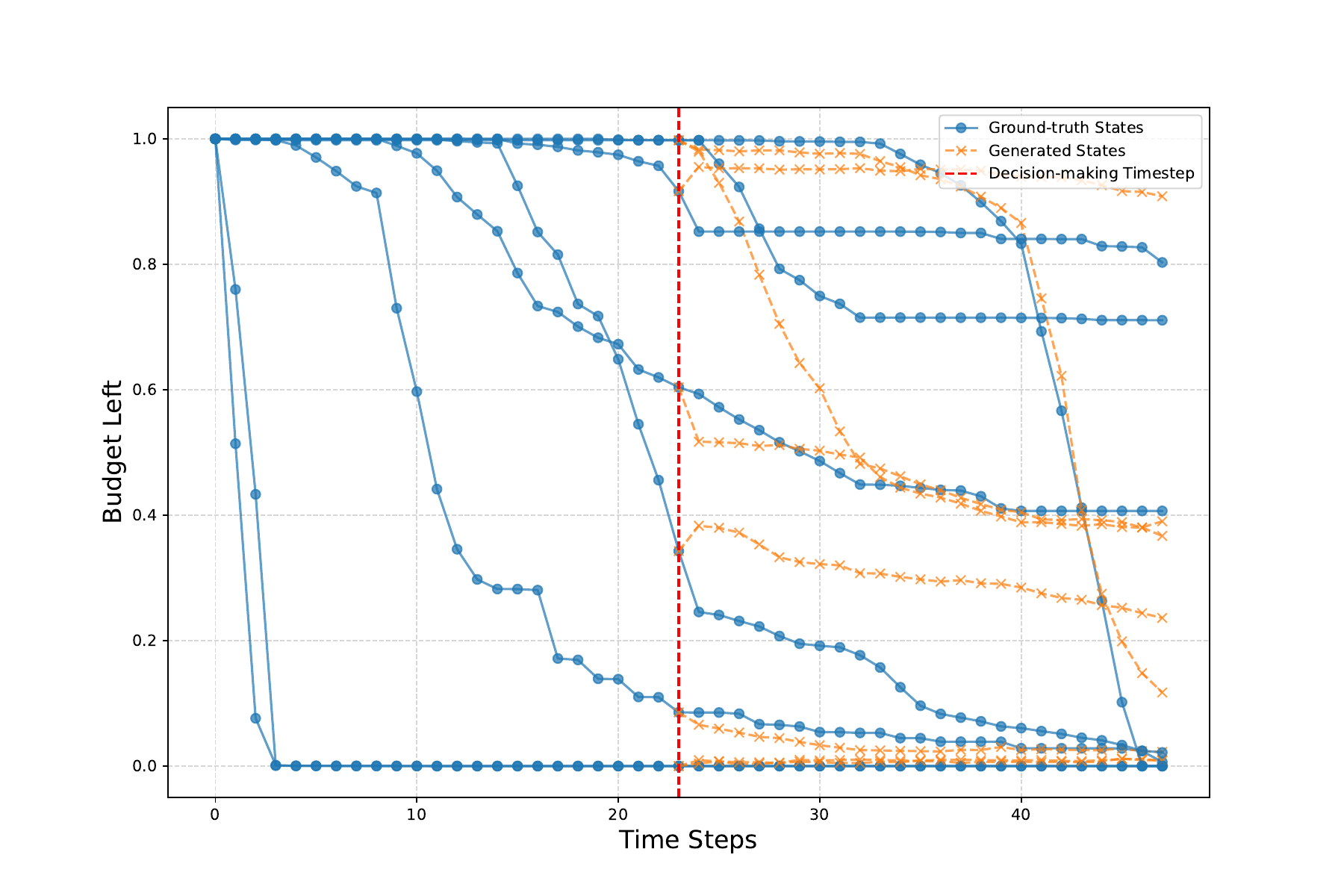}
    }
    \subfigure{
        \includegraphics[width=0.31\linewidth]{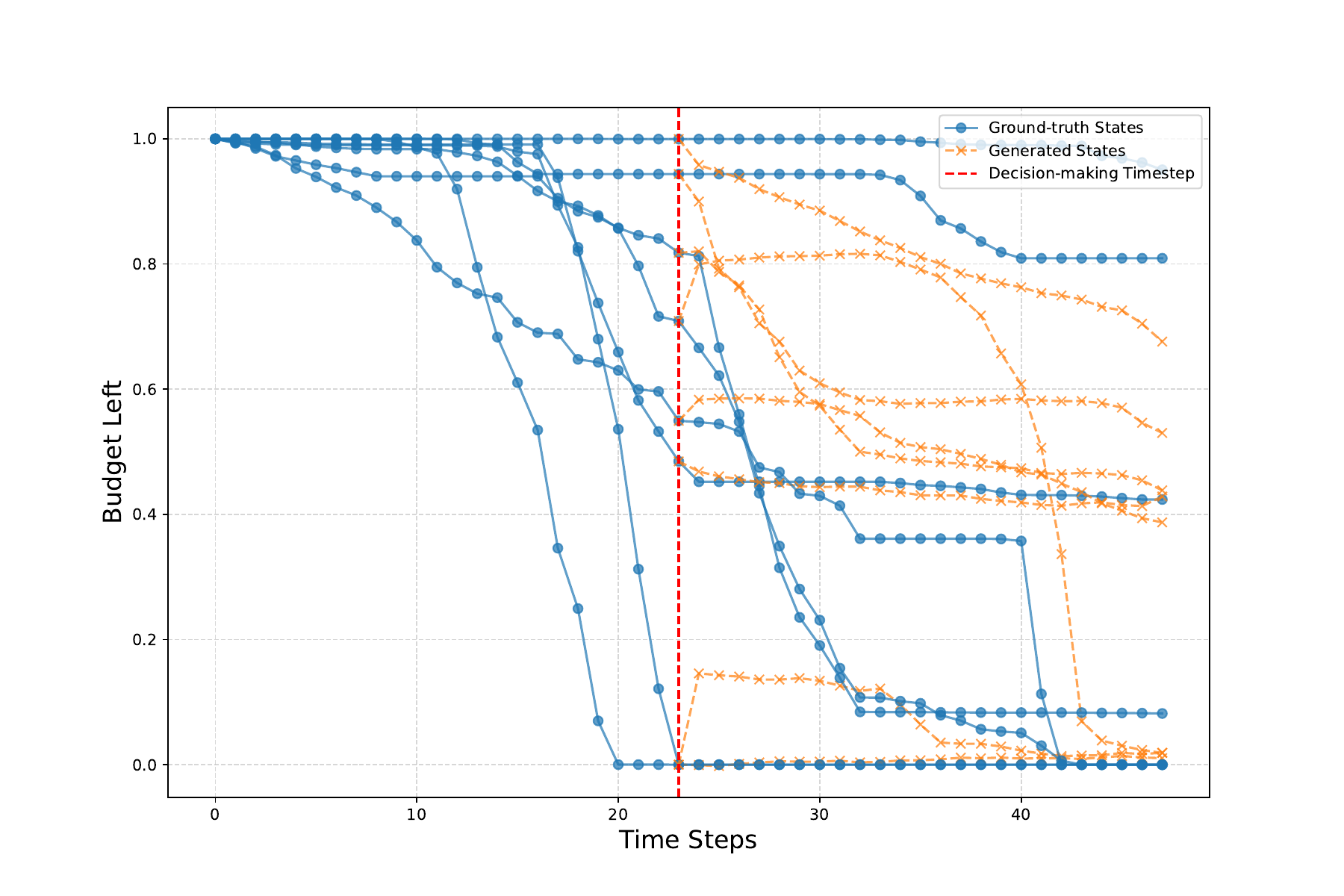}
    }
        \subfigure{
        \includegraphics[width=0.31\linewidth]{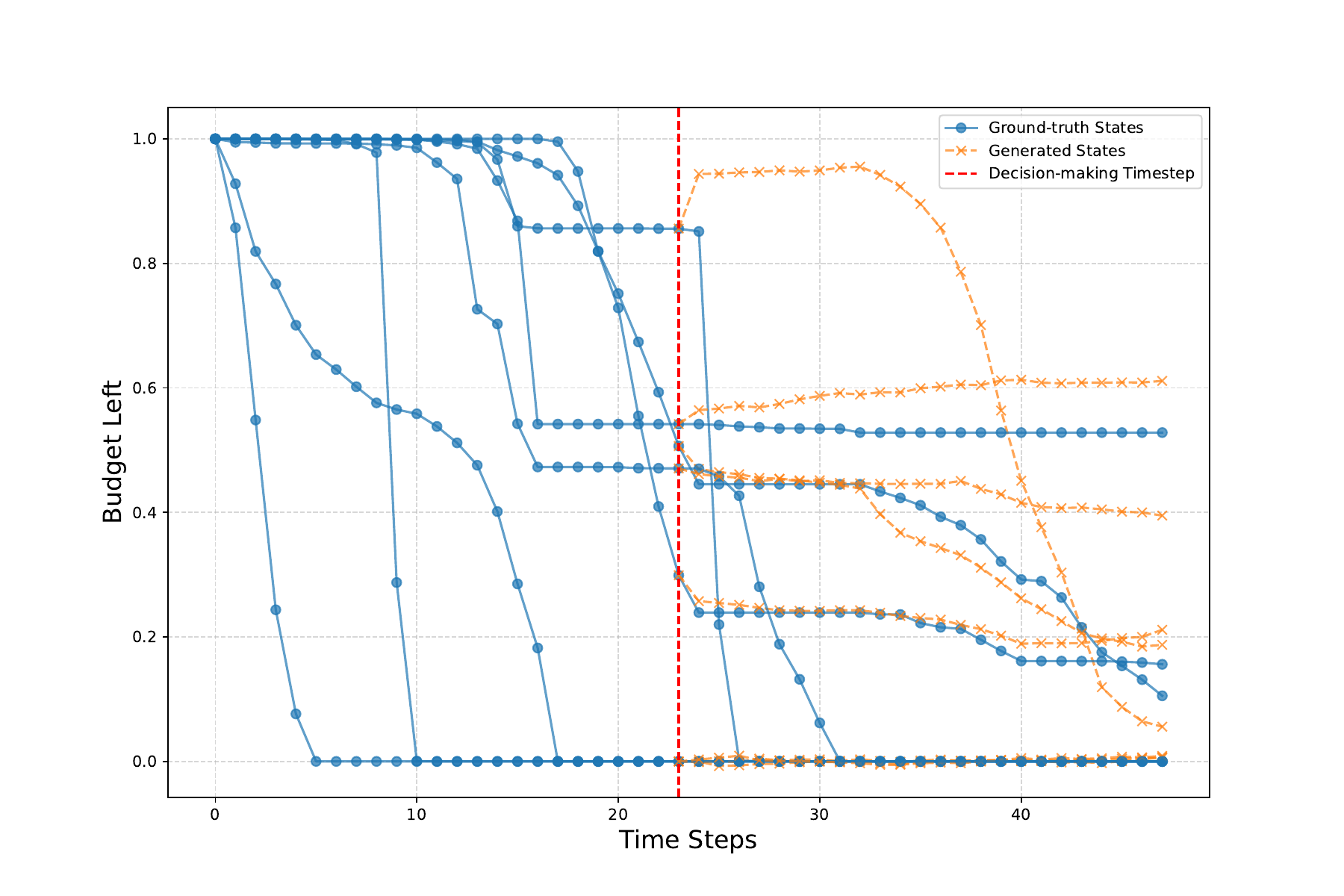}
    }
    \subfigure{
        \includegraphics[width=0.31\linewidth]{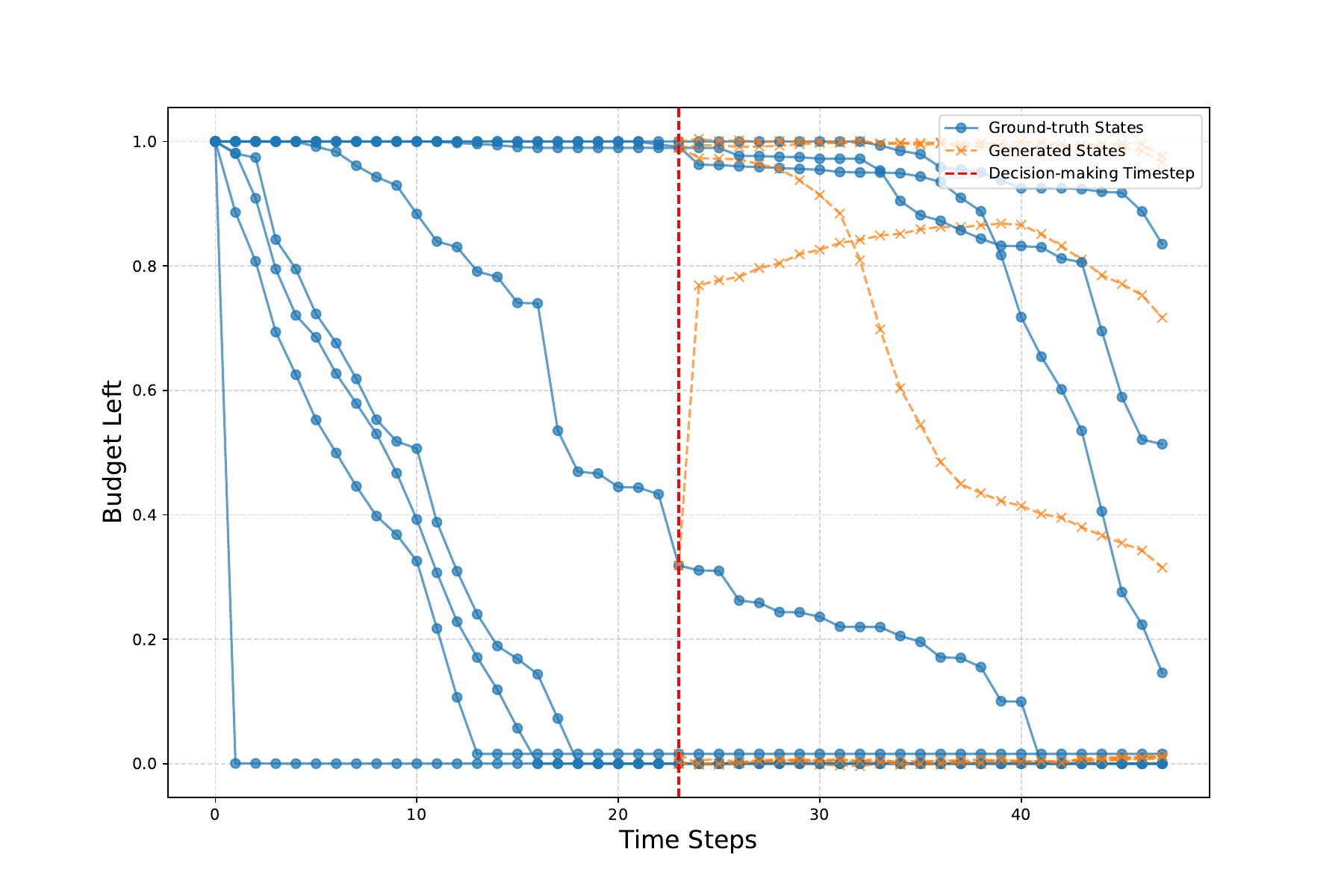}
    }
    \subfigure{
        \includegraphics[width=0.31\linewidth]{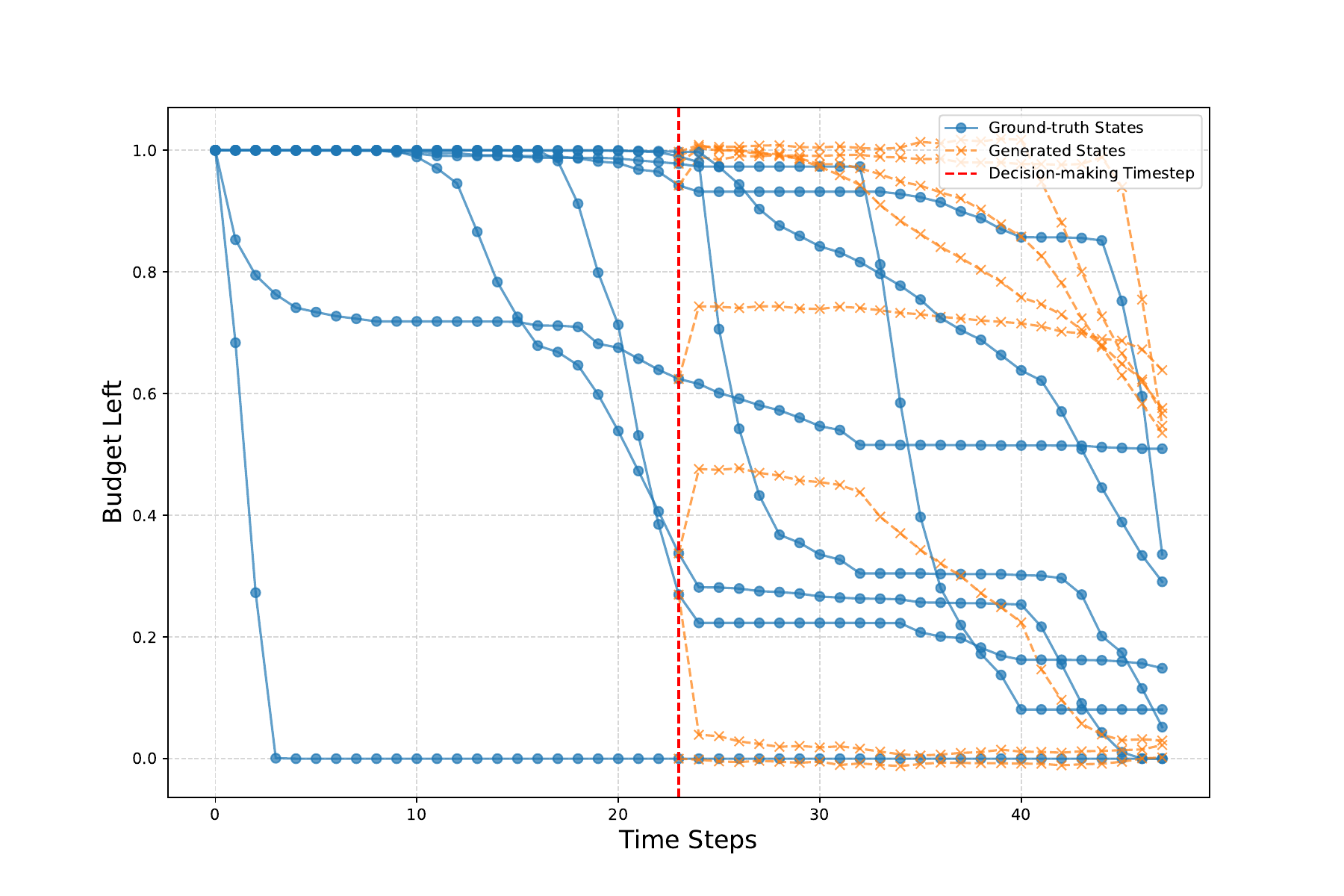}
    }
        \subfigure{
        \includegraphics[width=0.31\linewidth]{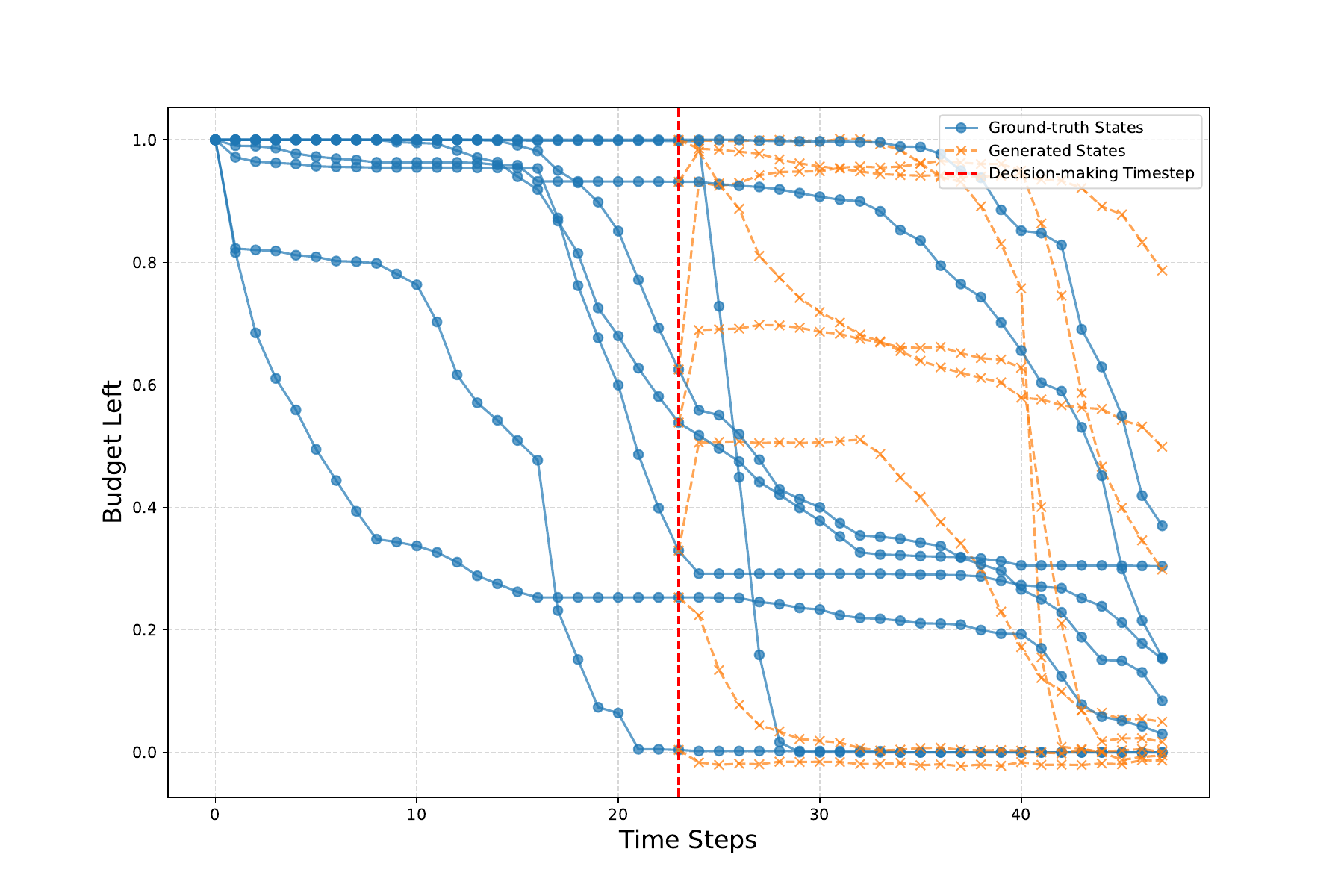}
    }
    \subfigure{
        \includegraphics[width=0.31\linewidth]{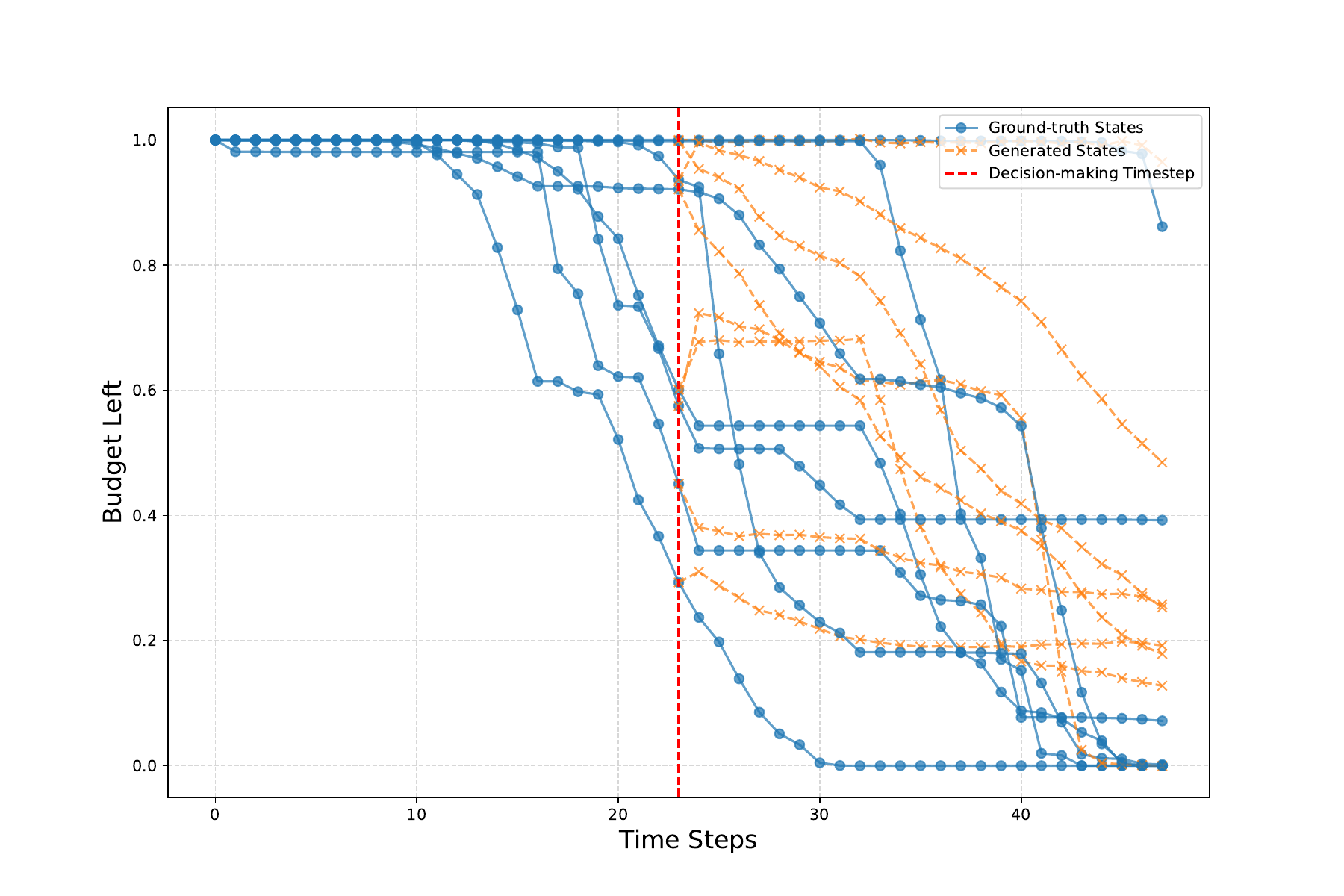}
    }
    \subfigure{
        \includegraphics[width=0.31\linewidth]{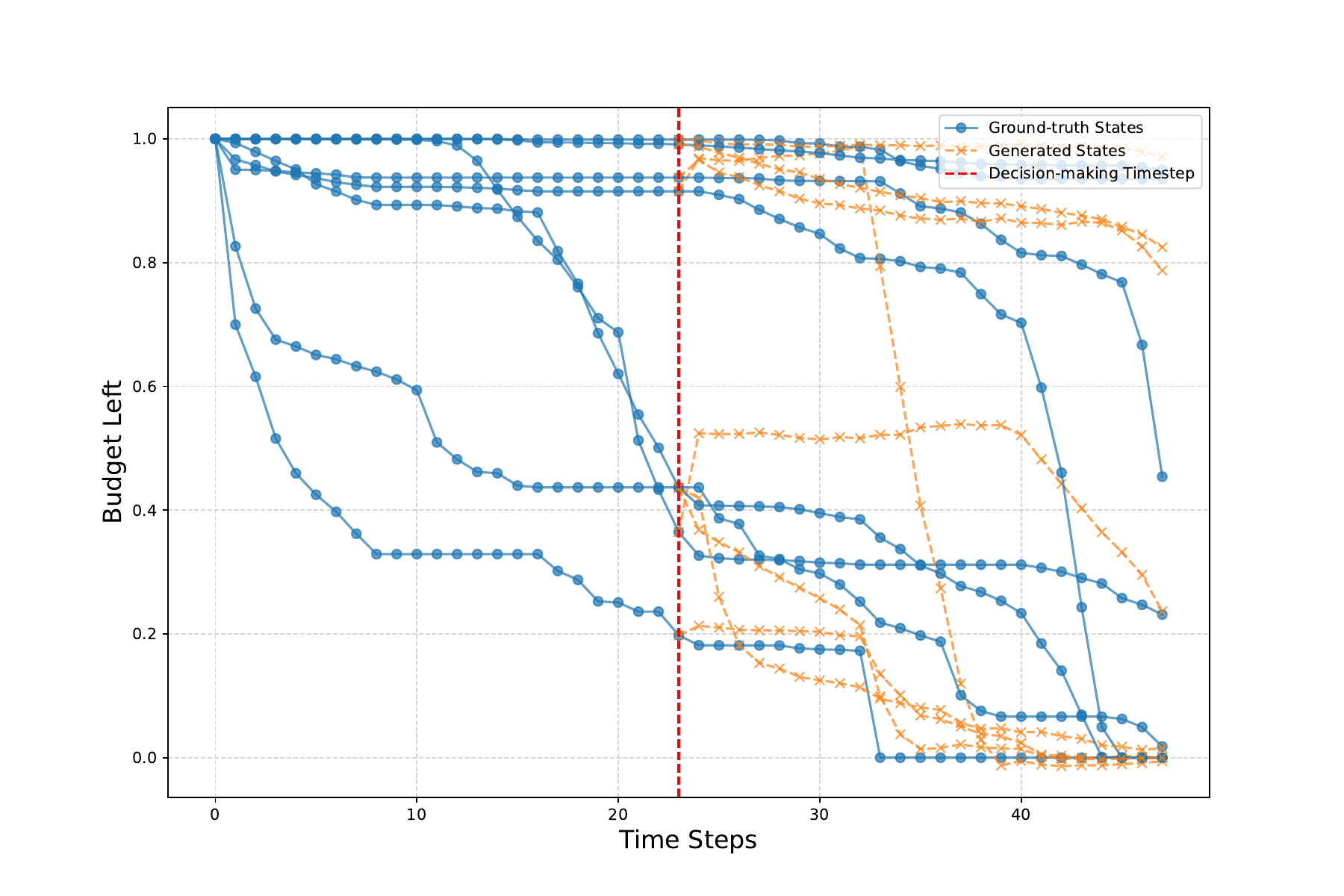}
    }
        \subfigure{
        \includegraphics[width=0.31\linewidth]{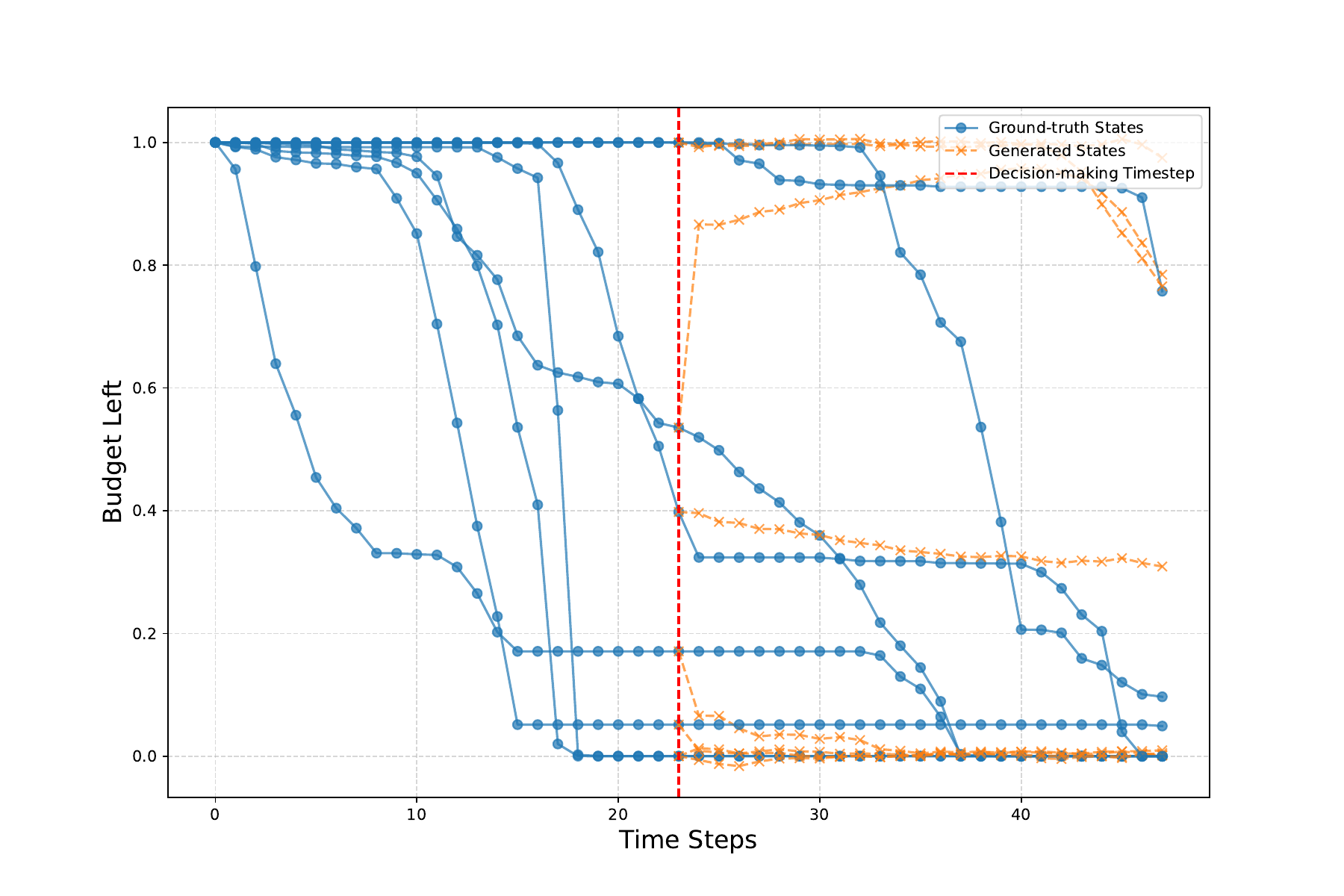}
    }
    \subfigure{
        \includegraphics[width=0.31\linewidth]{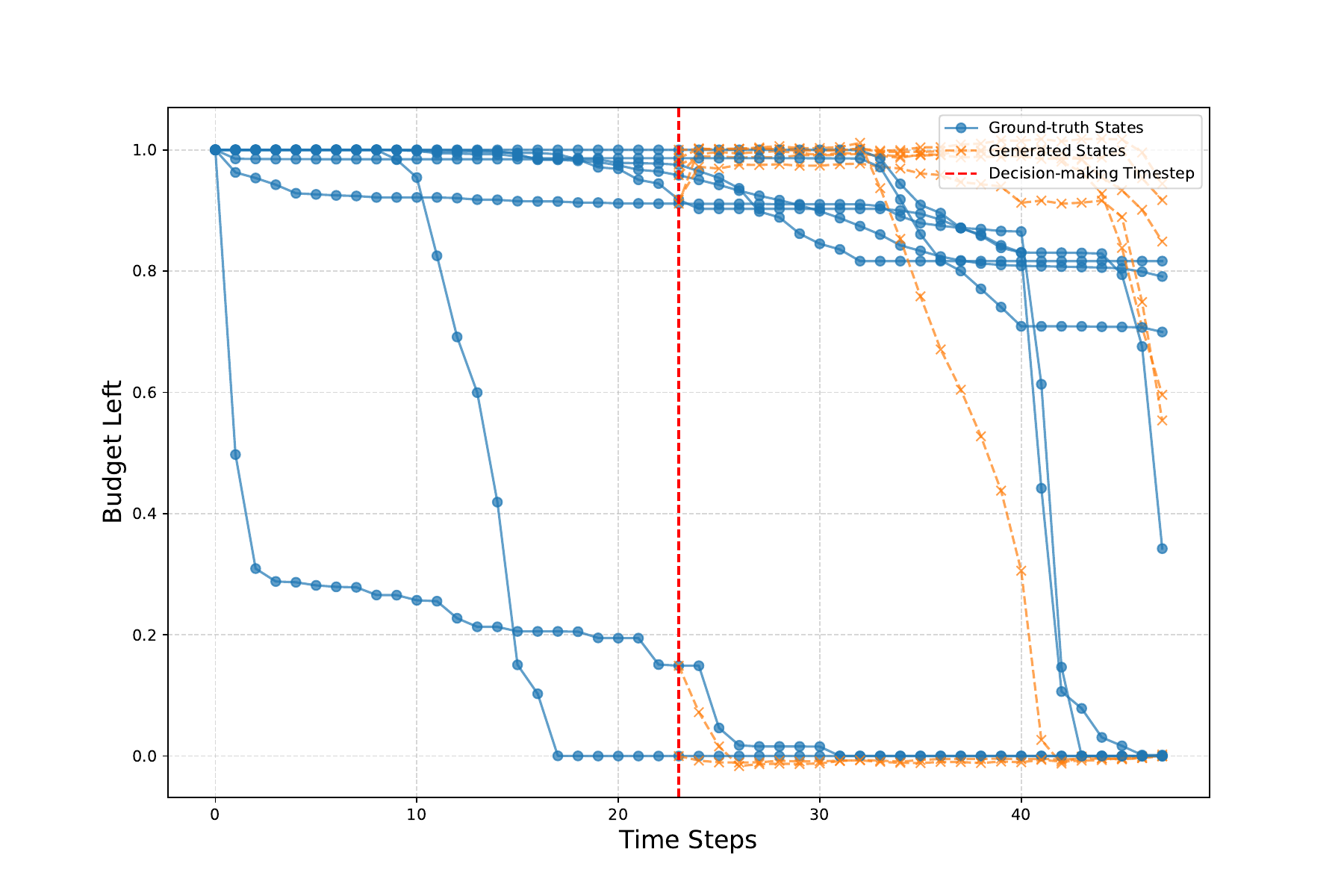}
    }
    \subfigure{
        \includegraphics[width=0.31\linewidth]{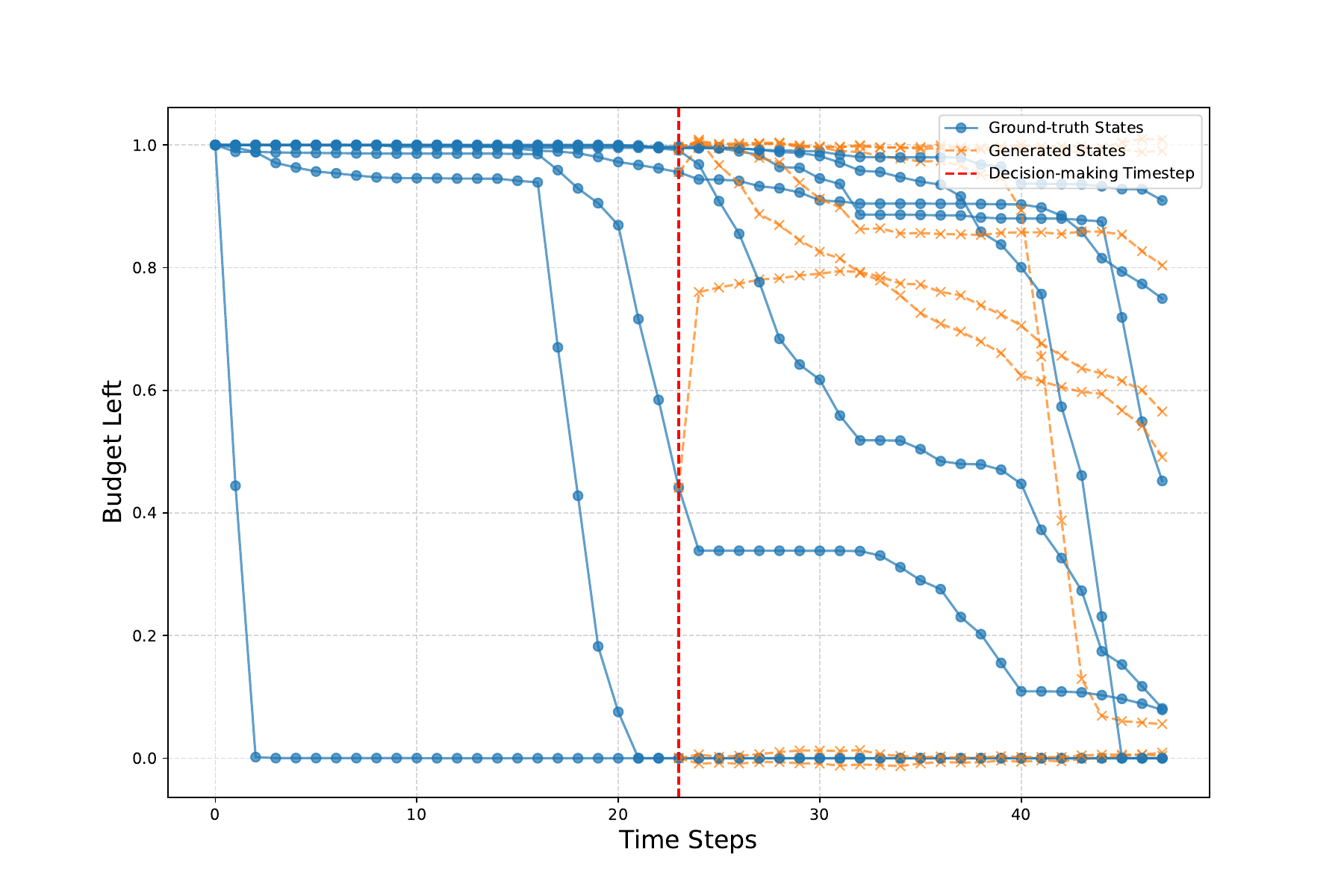}
    }
        \subfigure{
        \includegraphics[width=0.31\linewidth]{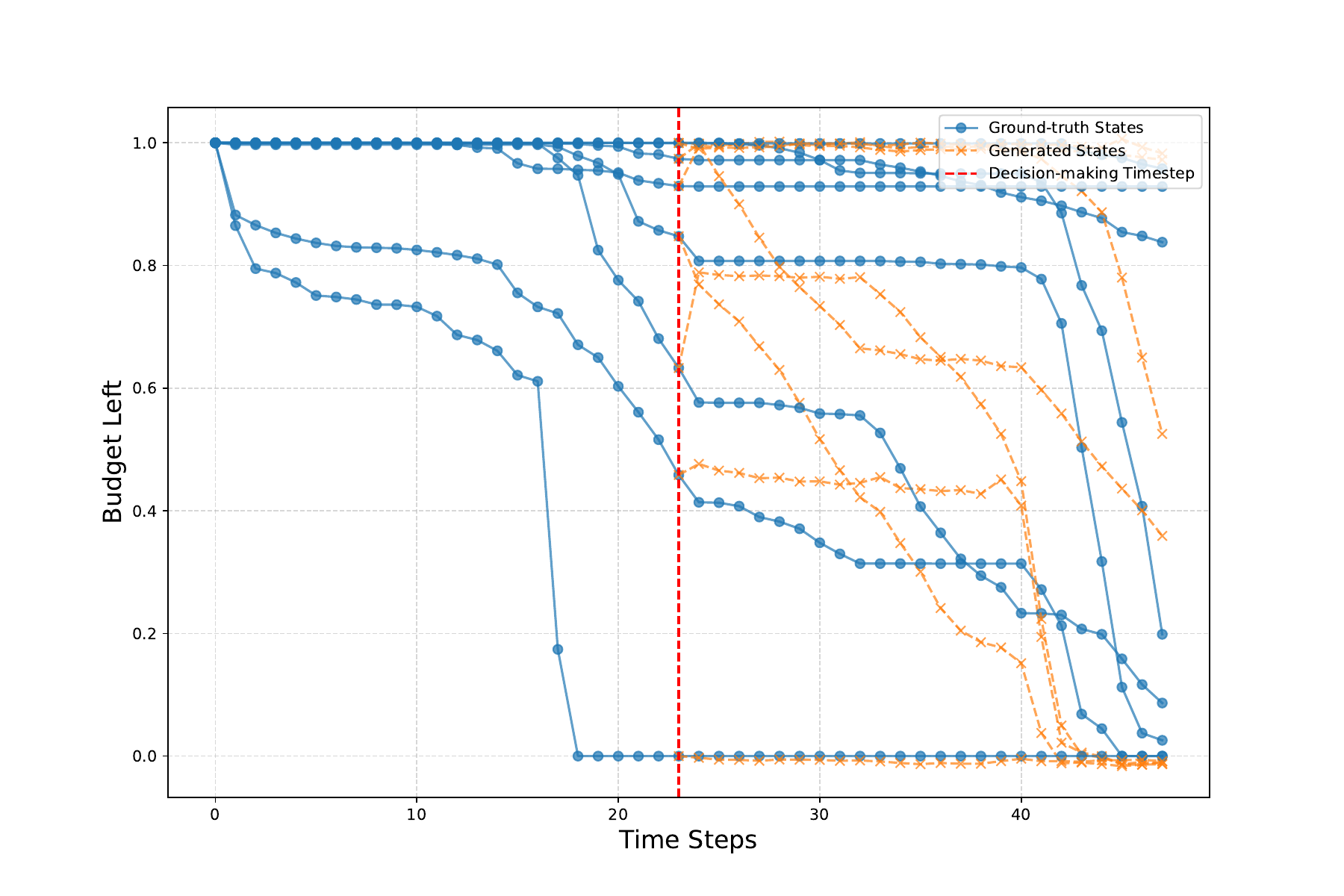}
    }
    \subfigure{
        \includegraphics[width=0.31\linewidth]{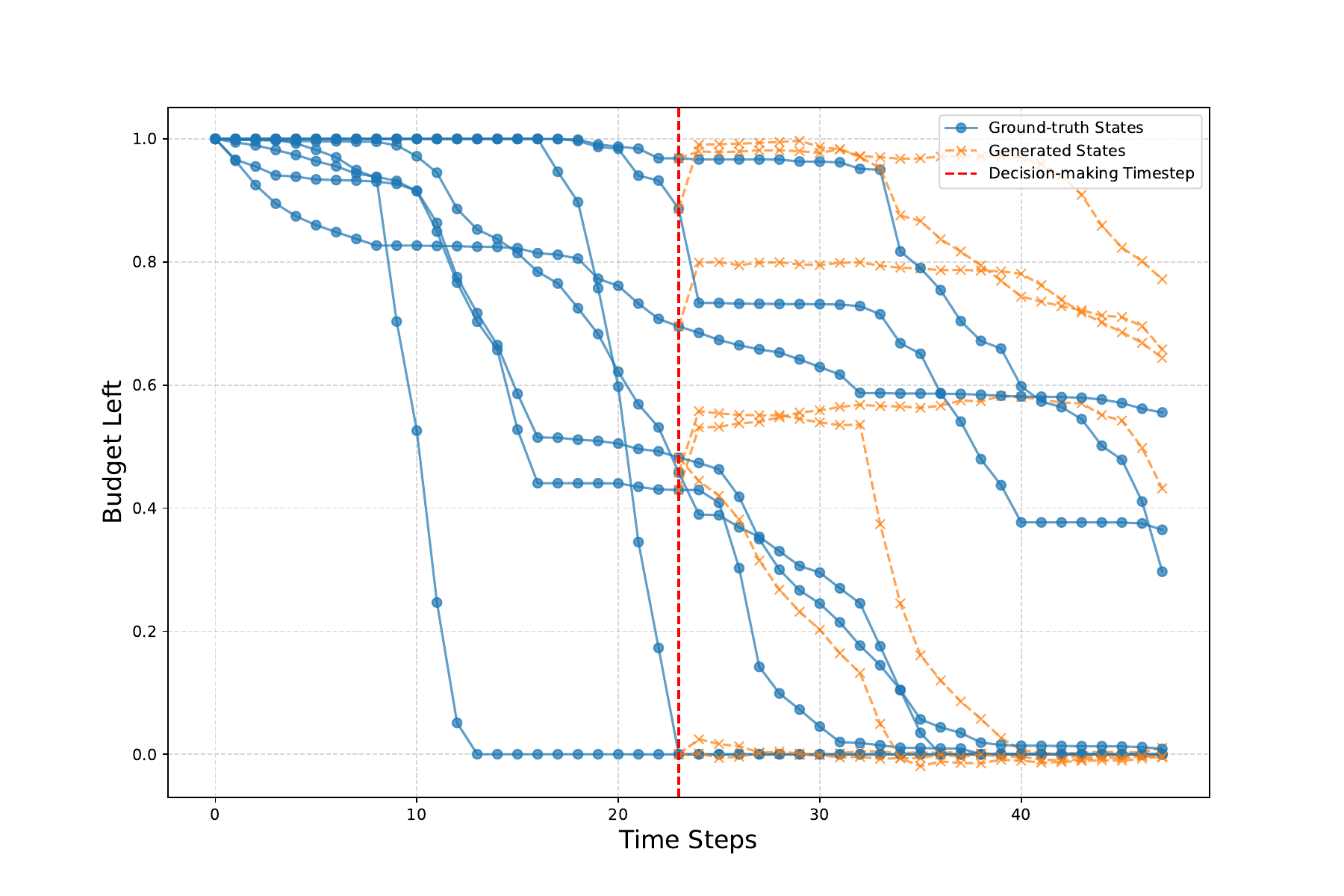}
    }
    \subfigure{
        \includegraphics[width=0.31\linewidth]{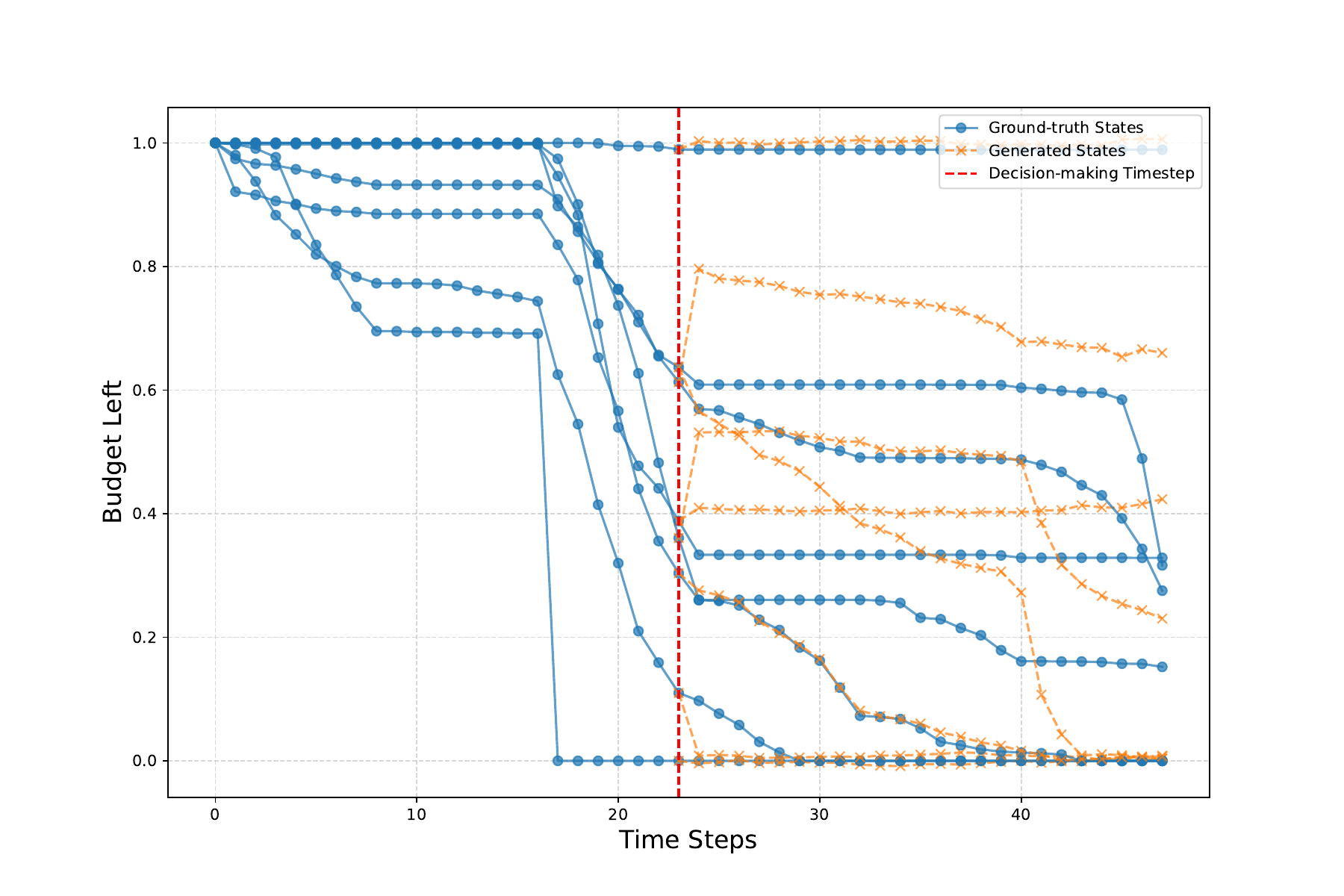}
    }
    \caption{Generated trajectories of DiffBid.}
    \label{fig:diffbid_gen}
\end{figure}

\begin{figure}[htbp]
    \centering
    \subfigure{
        \includegraphics[width=0.31\linewidth]{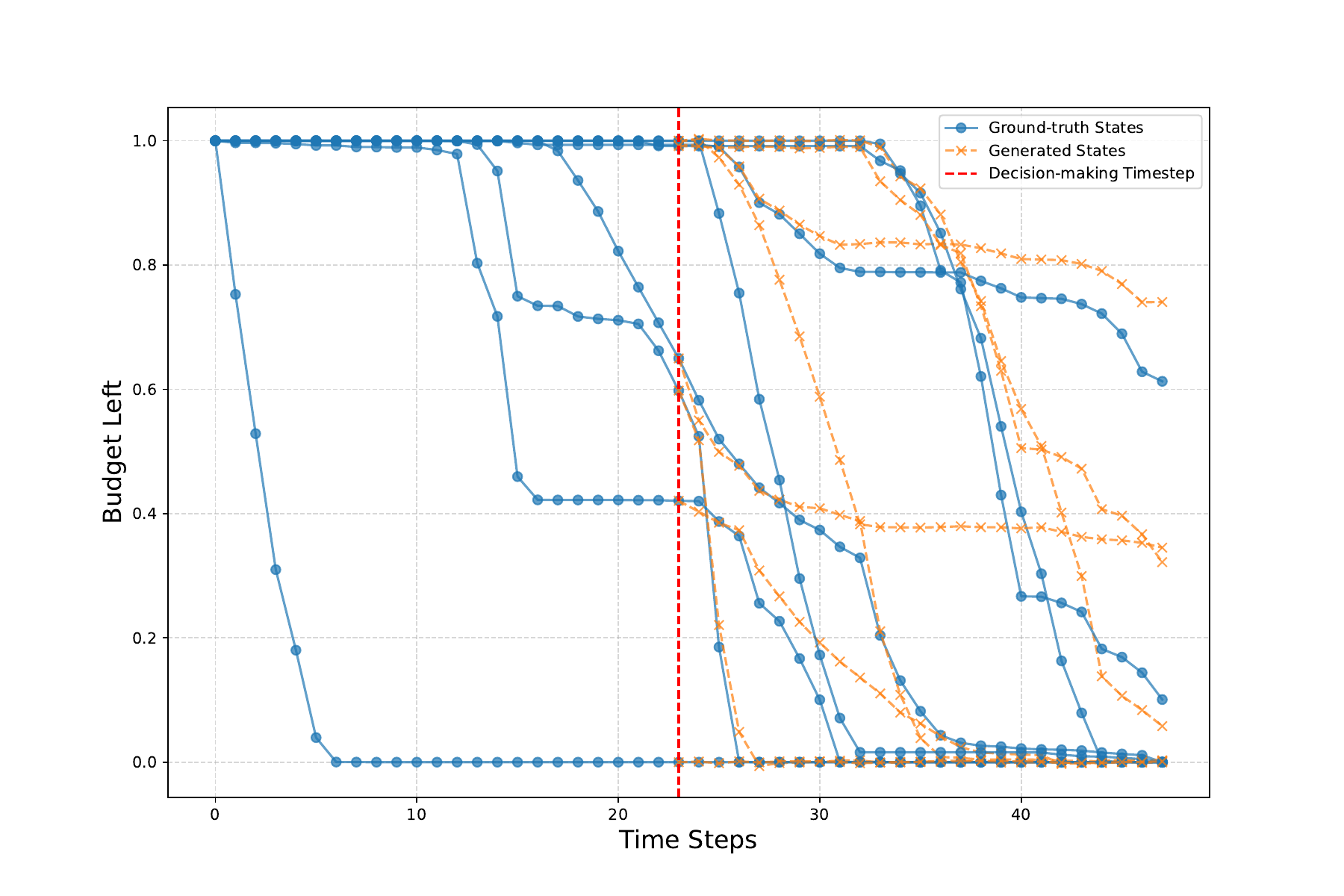}
    }
    \subfigure{
        \includegraphics[width=0.31\linewidth]{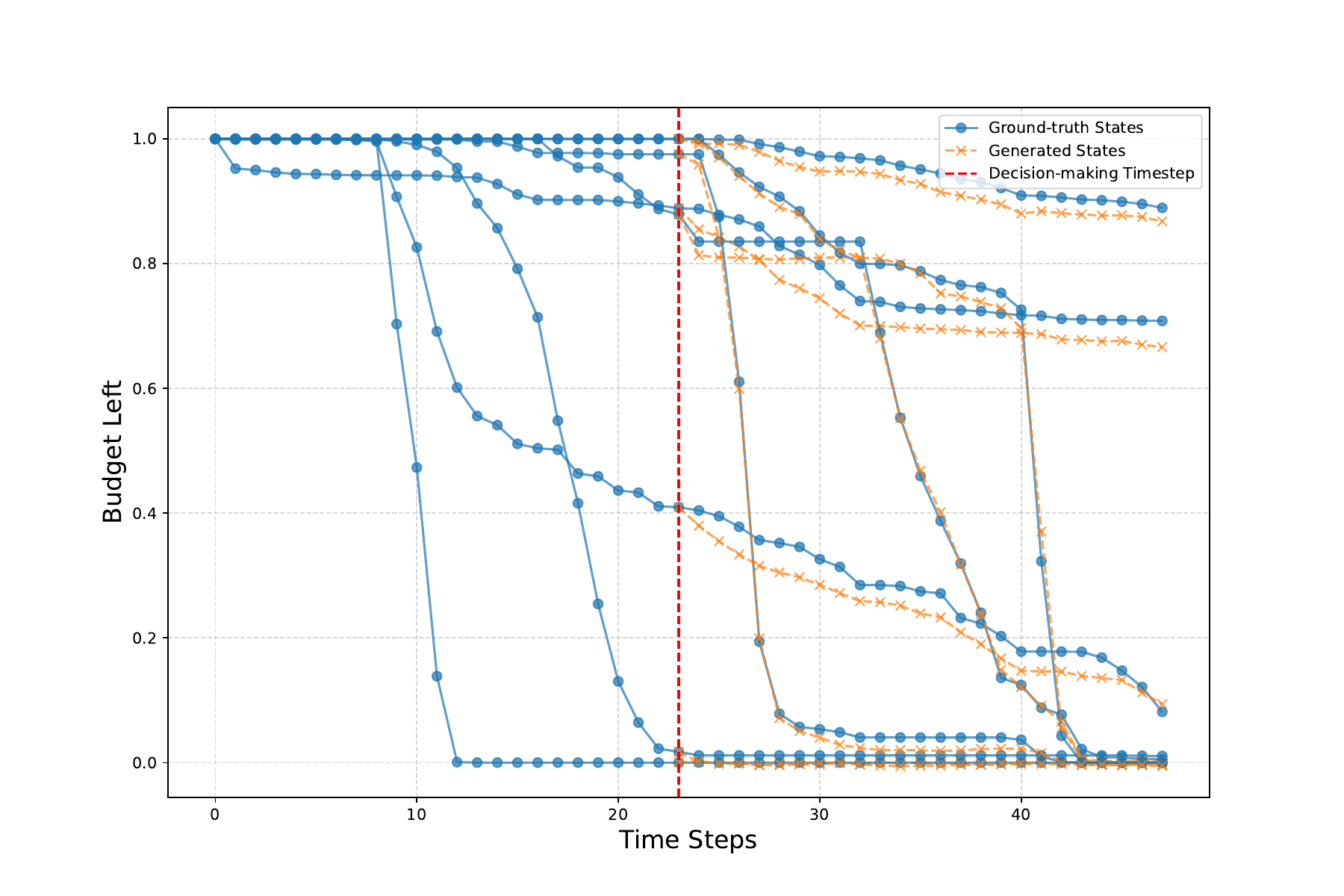}
    }
    \subfigure{
        \includegraphics[width=0.31\linewidth]{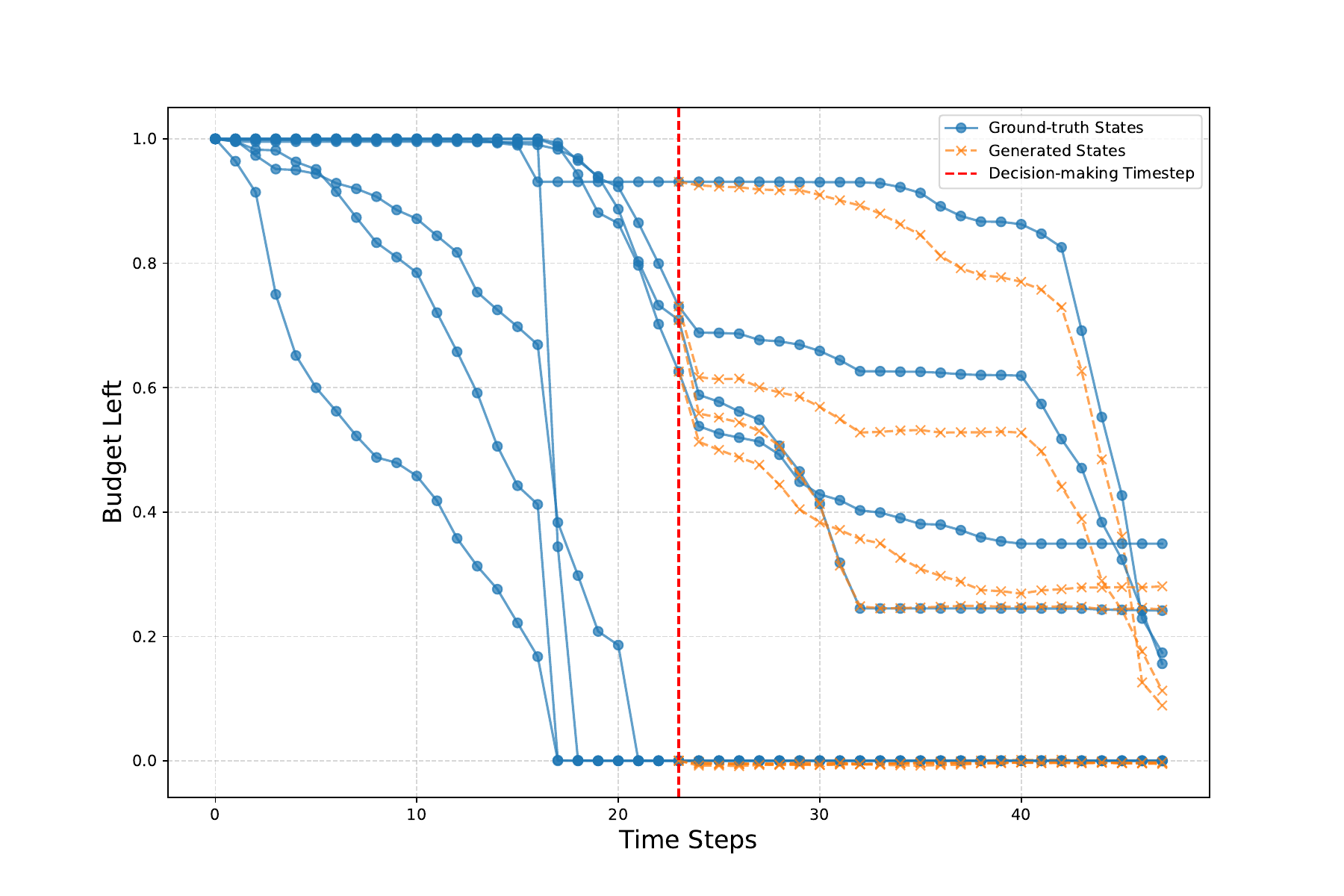}
    }
    \subfigure{
        \includegraphics[width=0.31\linewidth]{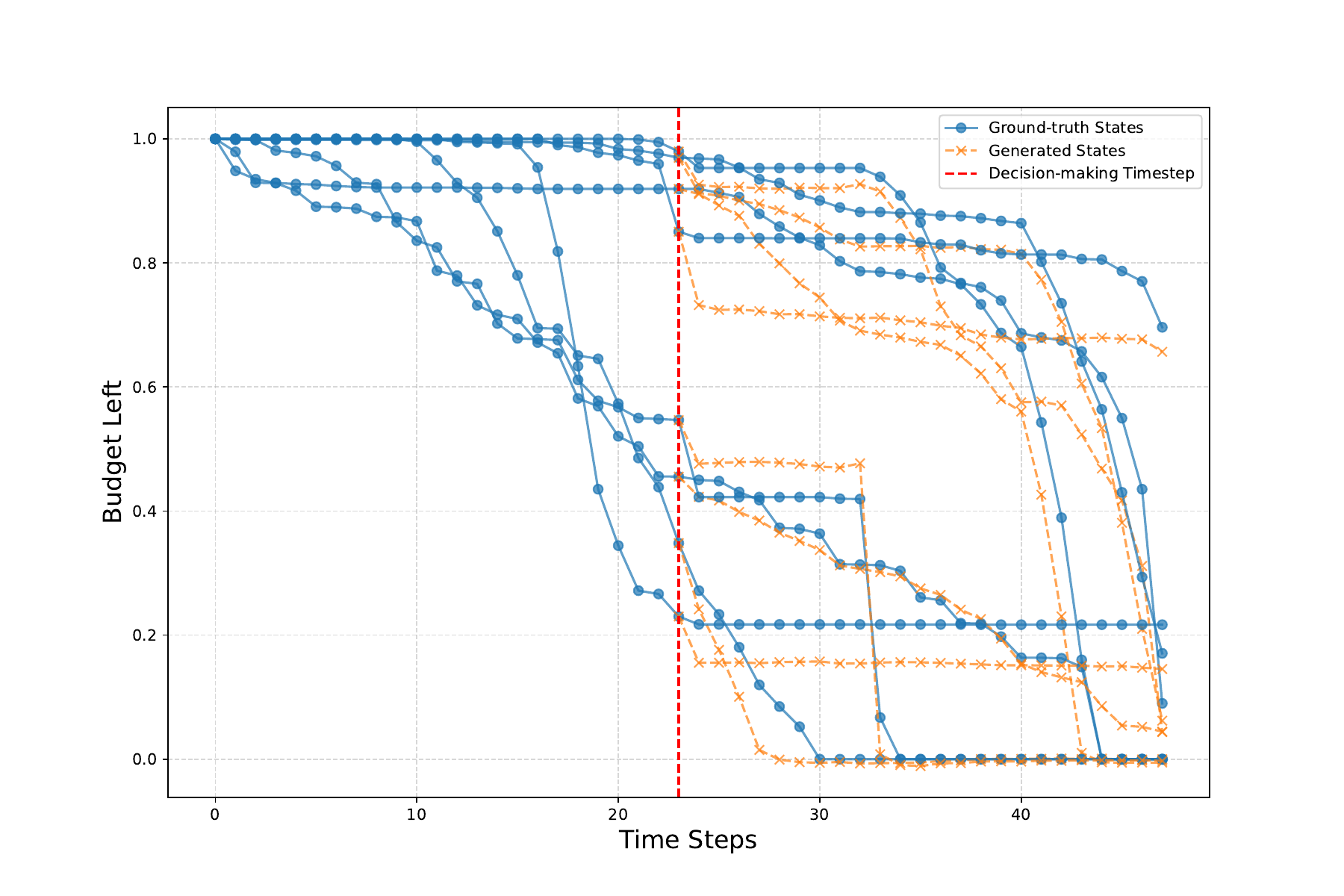}
    }
    \subfigure{
        \includegraphics[width=0.31\linewidth]{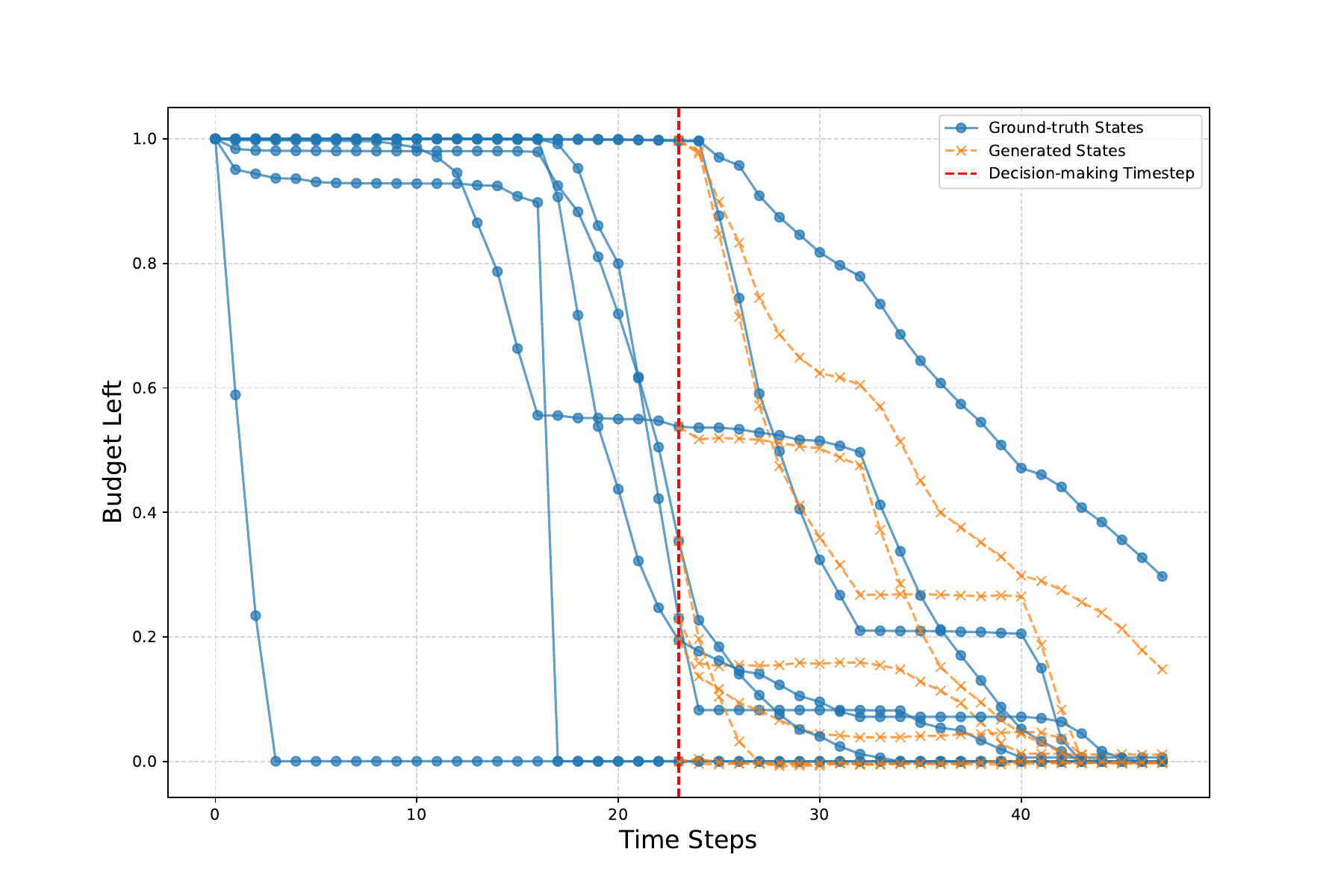}
    }
    \subfigure{
        \includegraphics[width=0.31\linewidth]{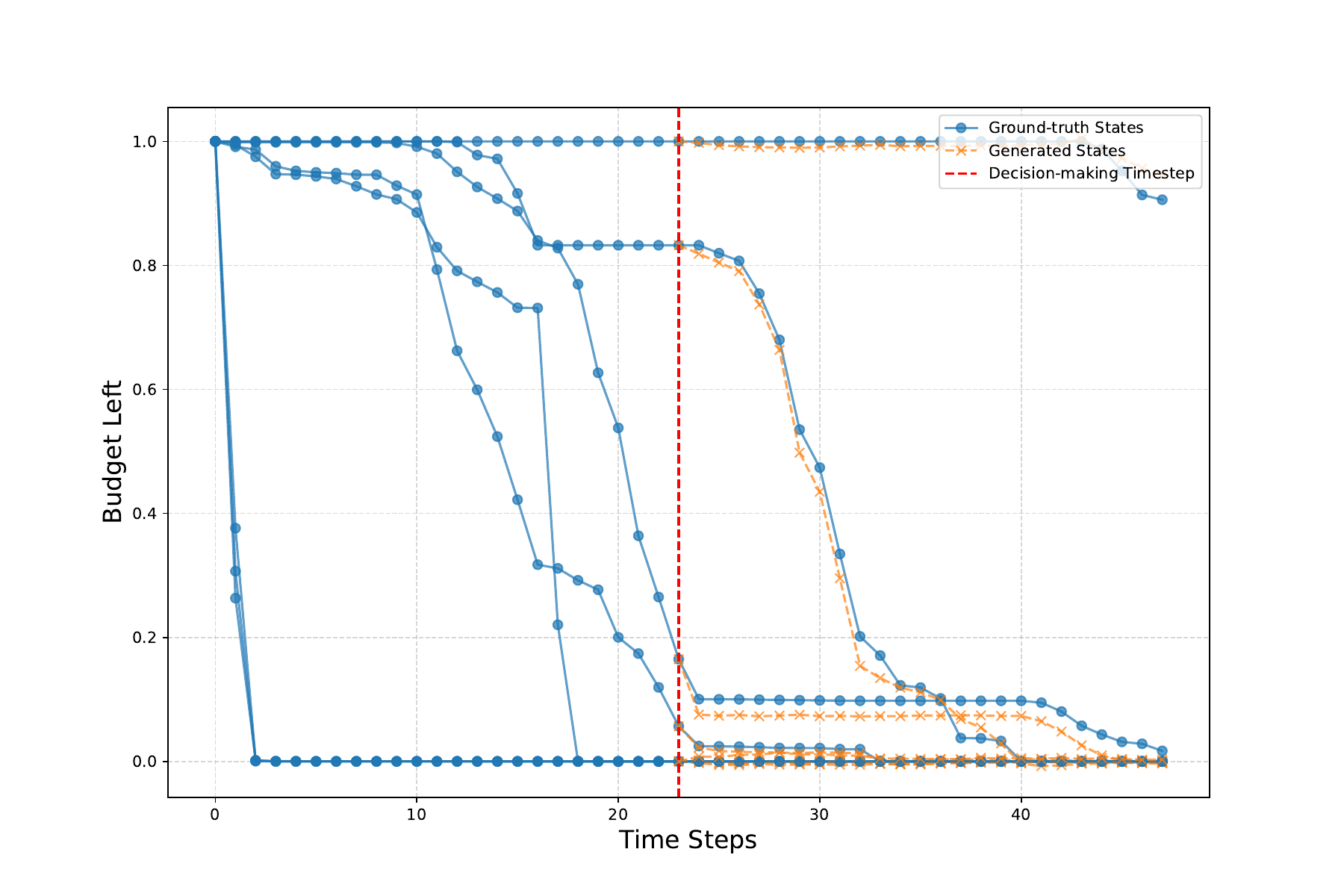}
    }
        \subfigure{
        \includegraphics[width=0.31\linewidth]{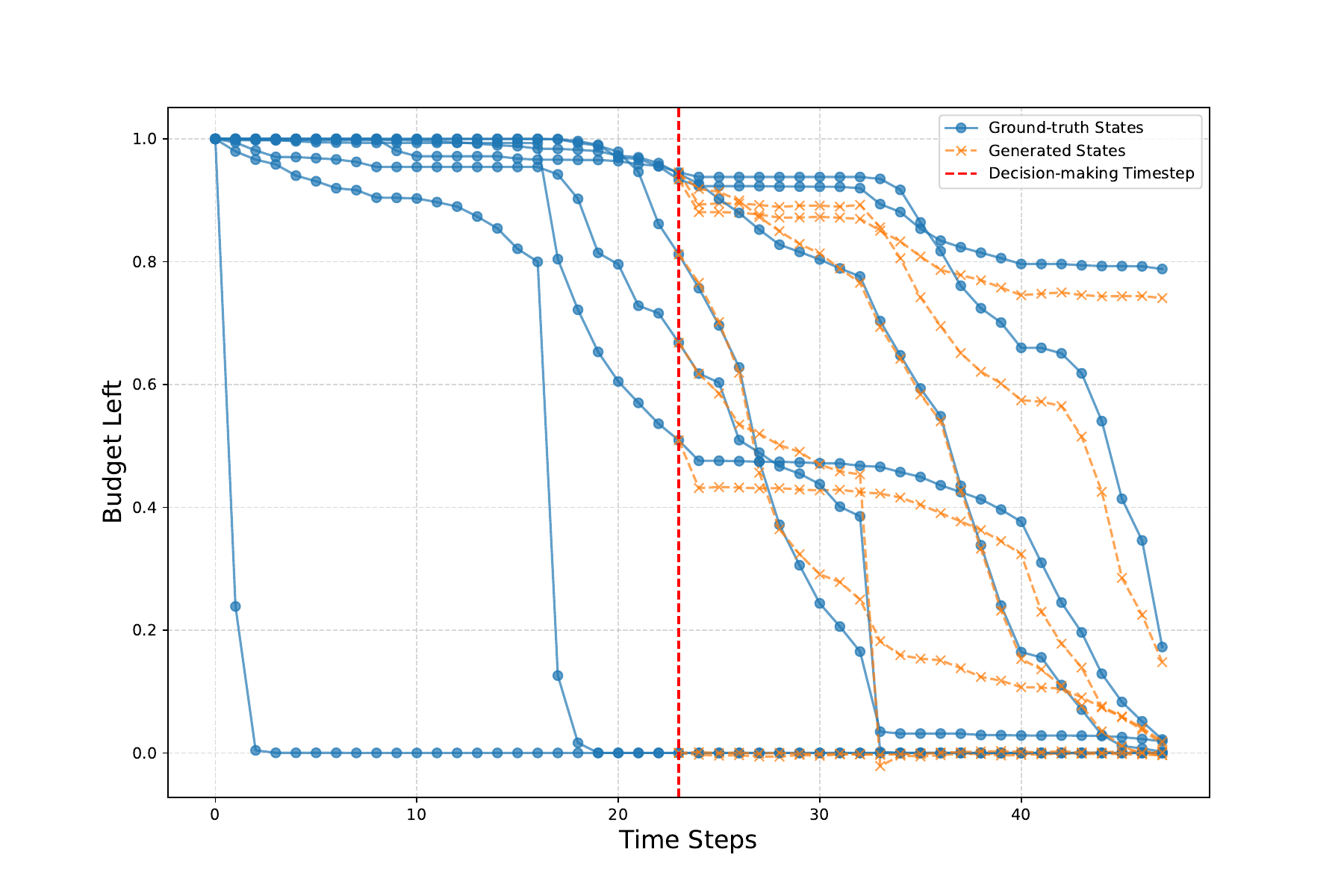}
    }
    \subfigure{
        \includegraphics[width=0.31\linewidth]{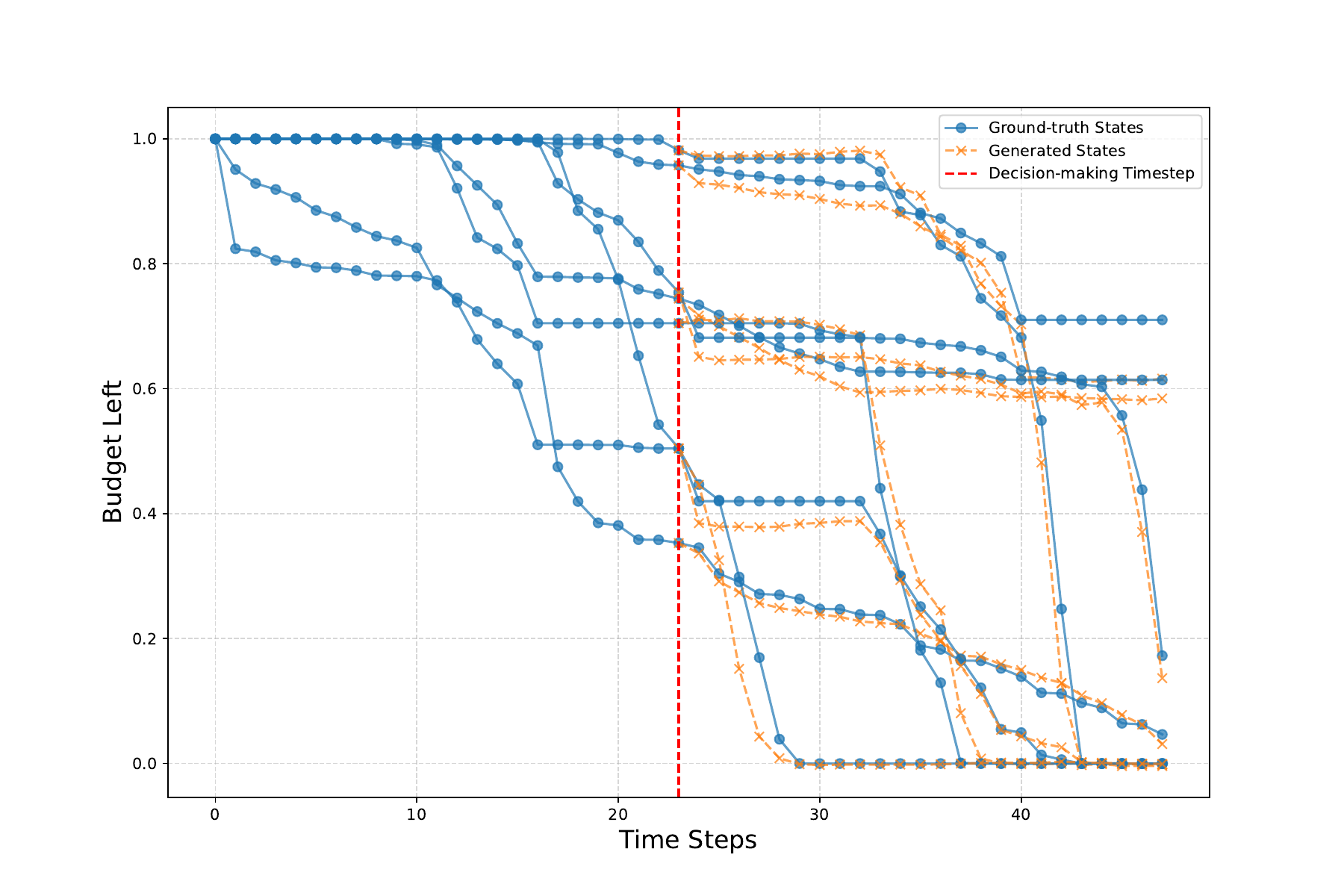}
    }
    \subfigure{
        \includegraphics[width=0.31\linewidth]{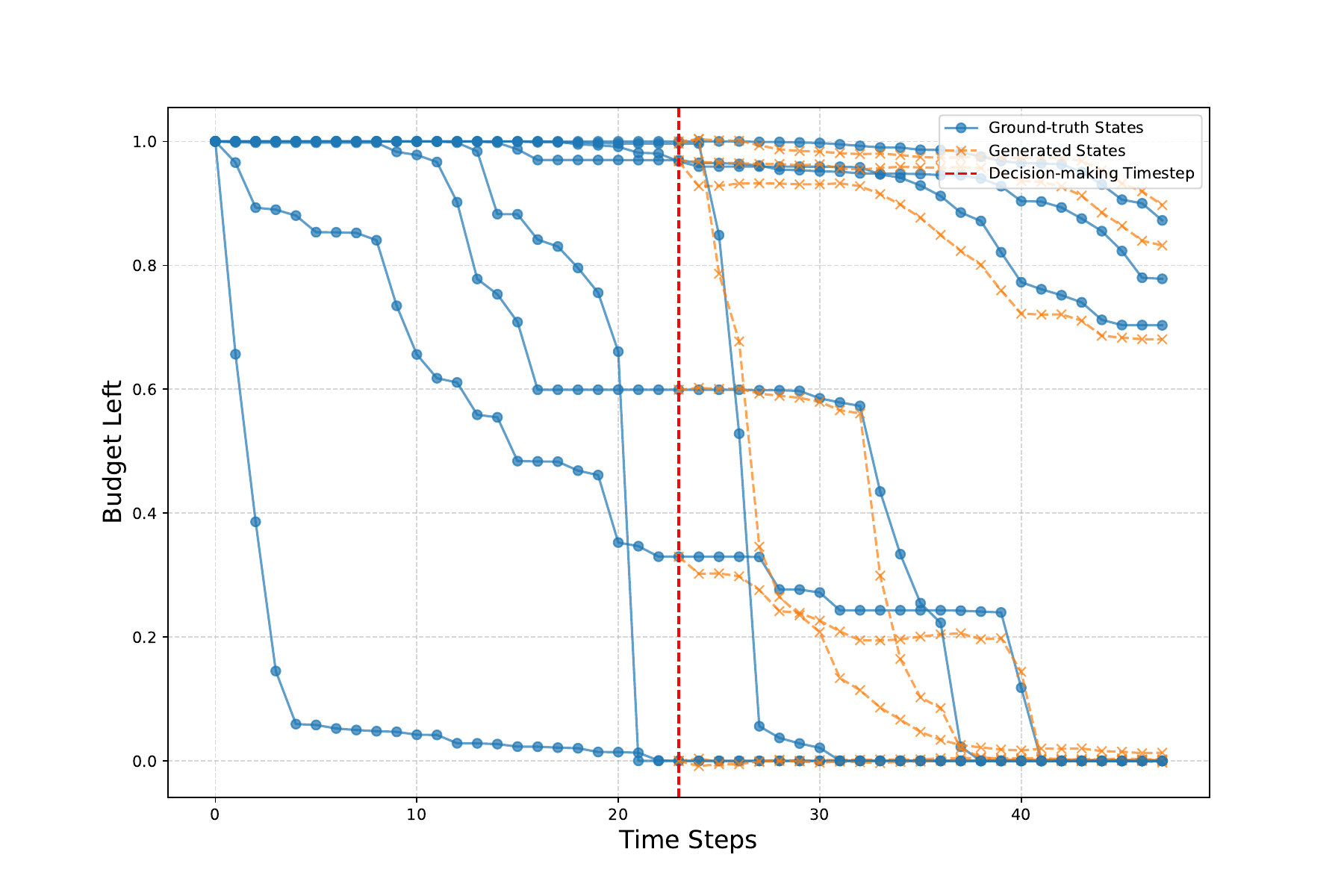}
    }
        \subfigure{
        \includegraphics[width=0.31\linewidth]{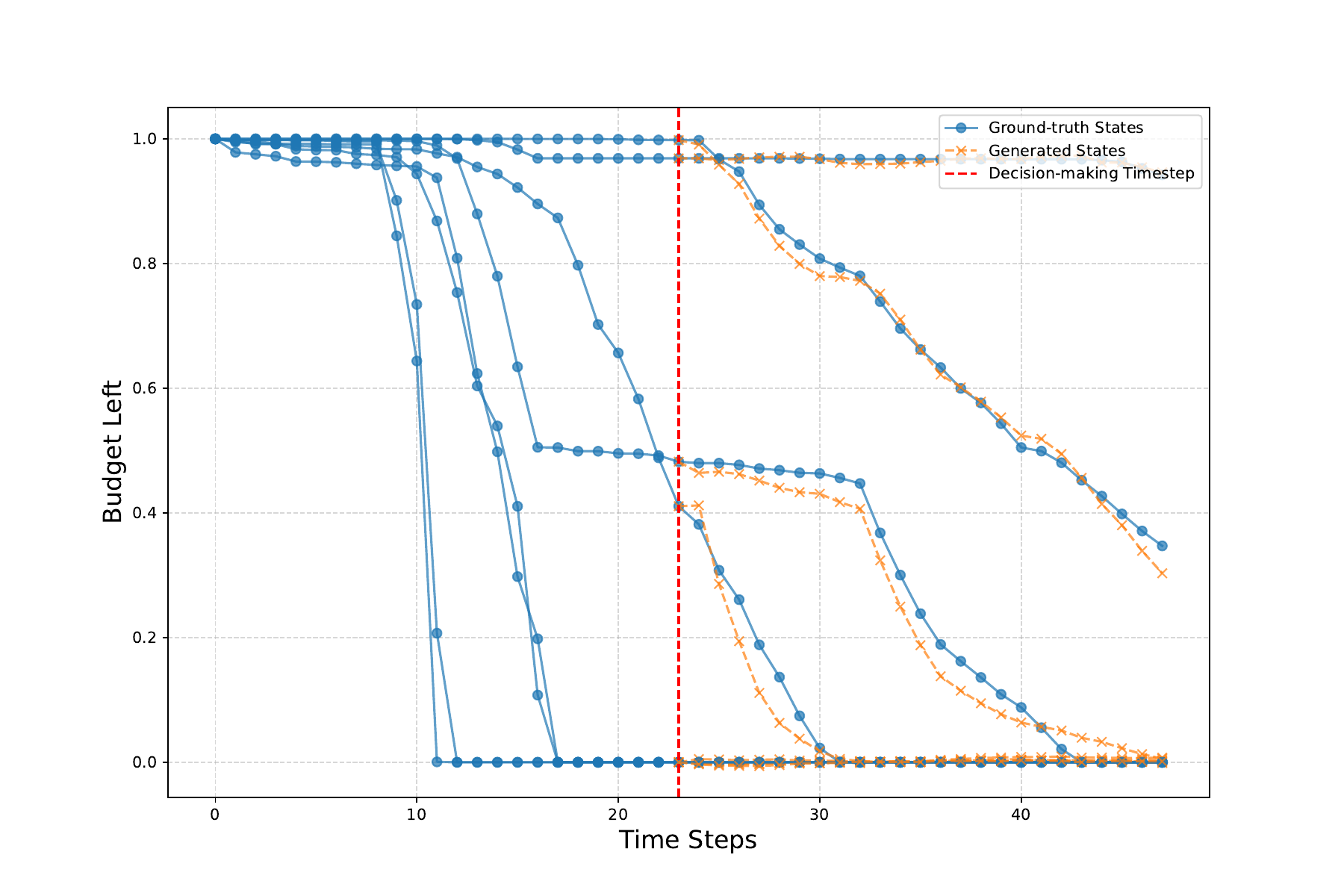}
    }
    \subfigure{
        \includegraphics[width=0.31\linewidth]{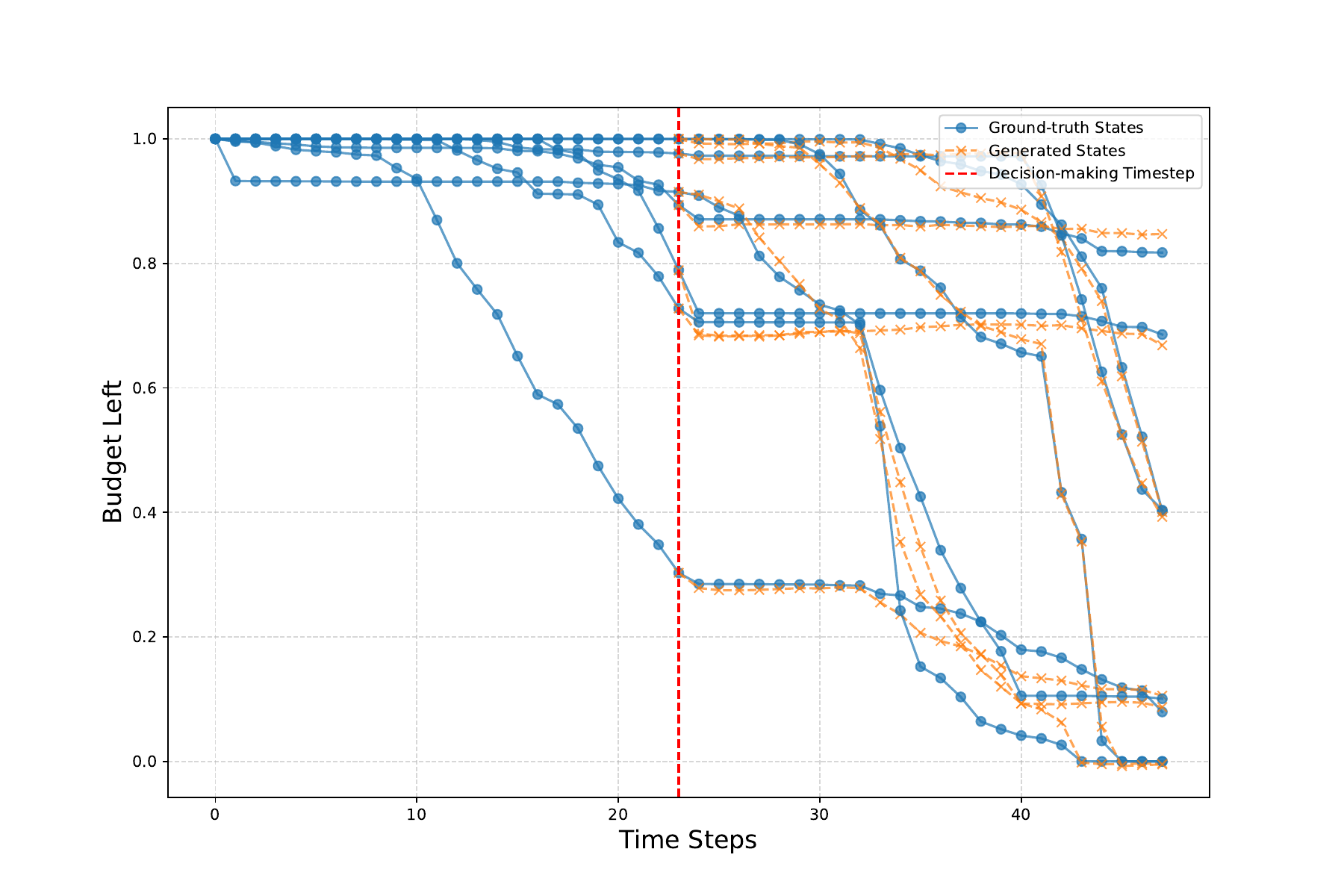}
    }
    \subfigure{
        \includegraphics[width=0.31\linewidth]{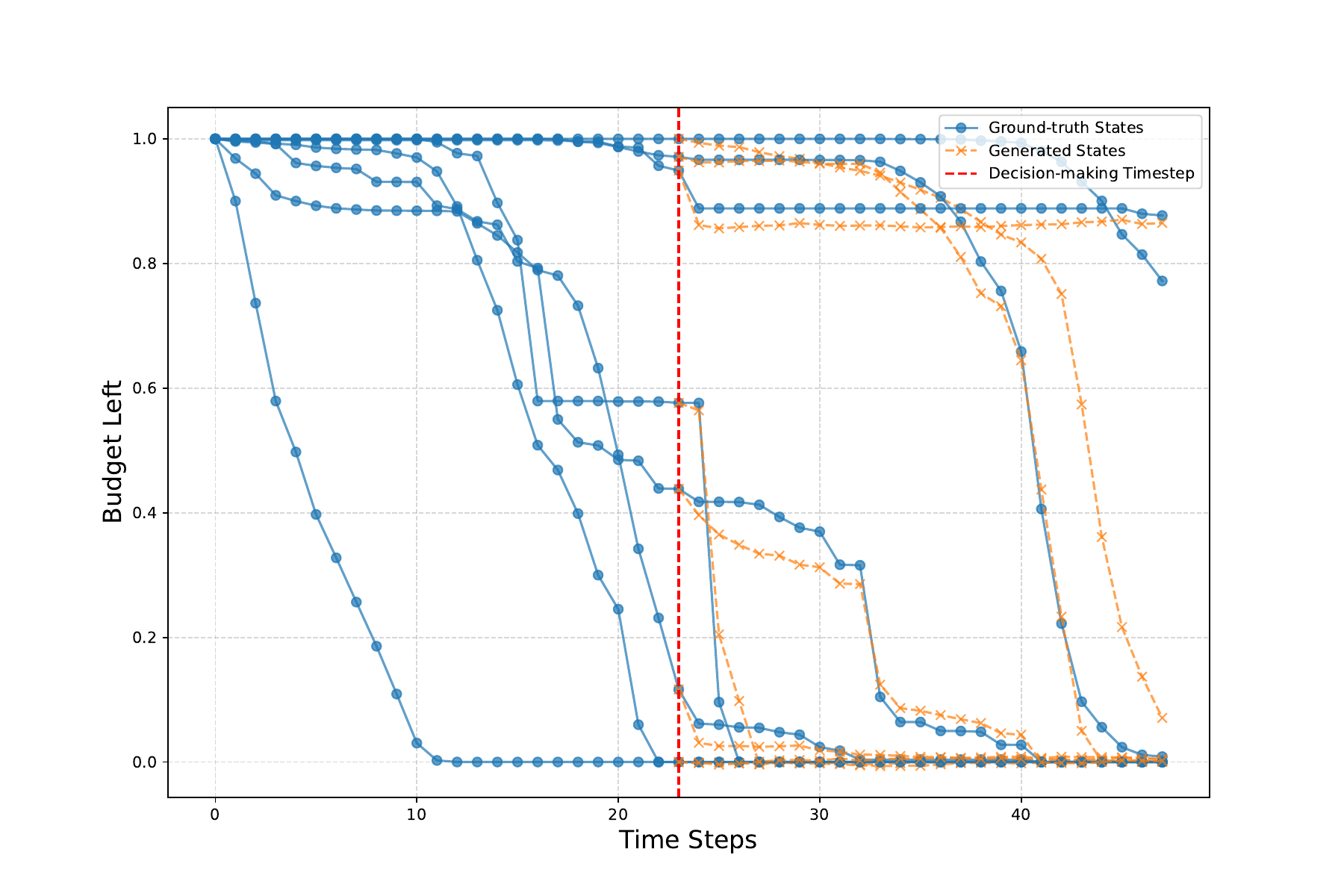}
    }
        \subfigure{
        \includegraphics[width=0.31\linewidth]{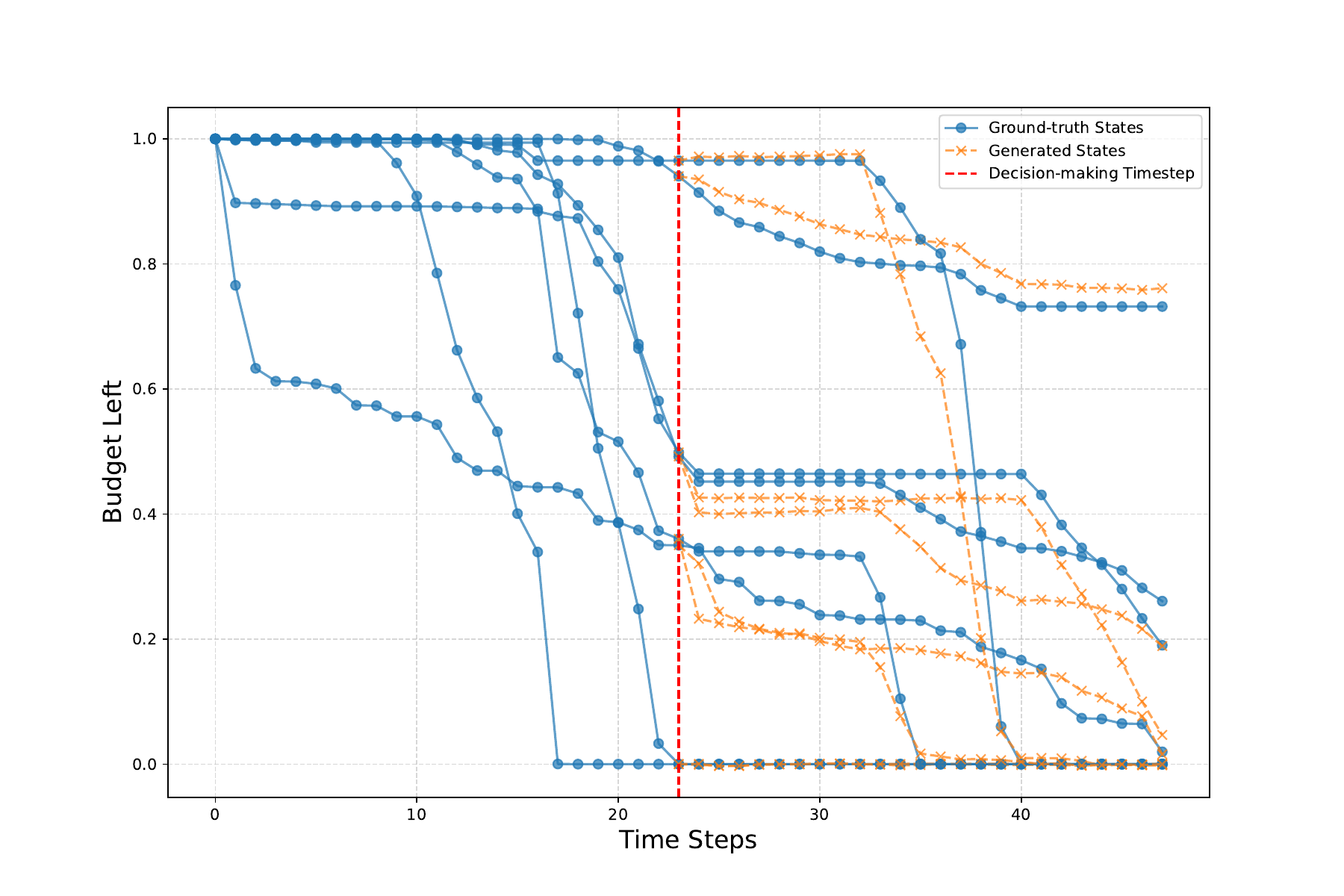}
    }
    \subfigure{
        \includegraphics[width=0.31\linewidth]{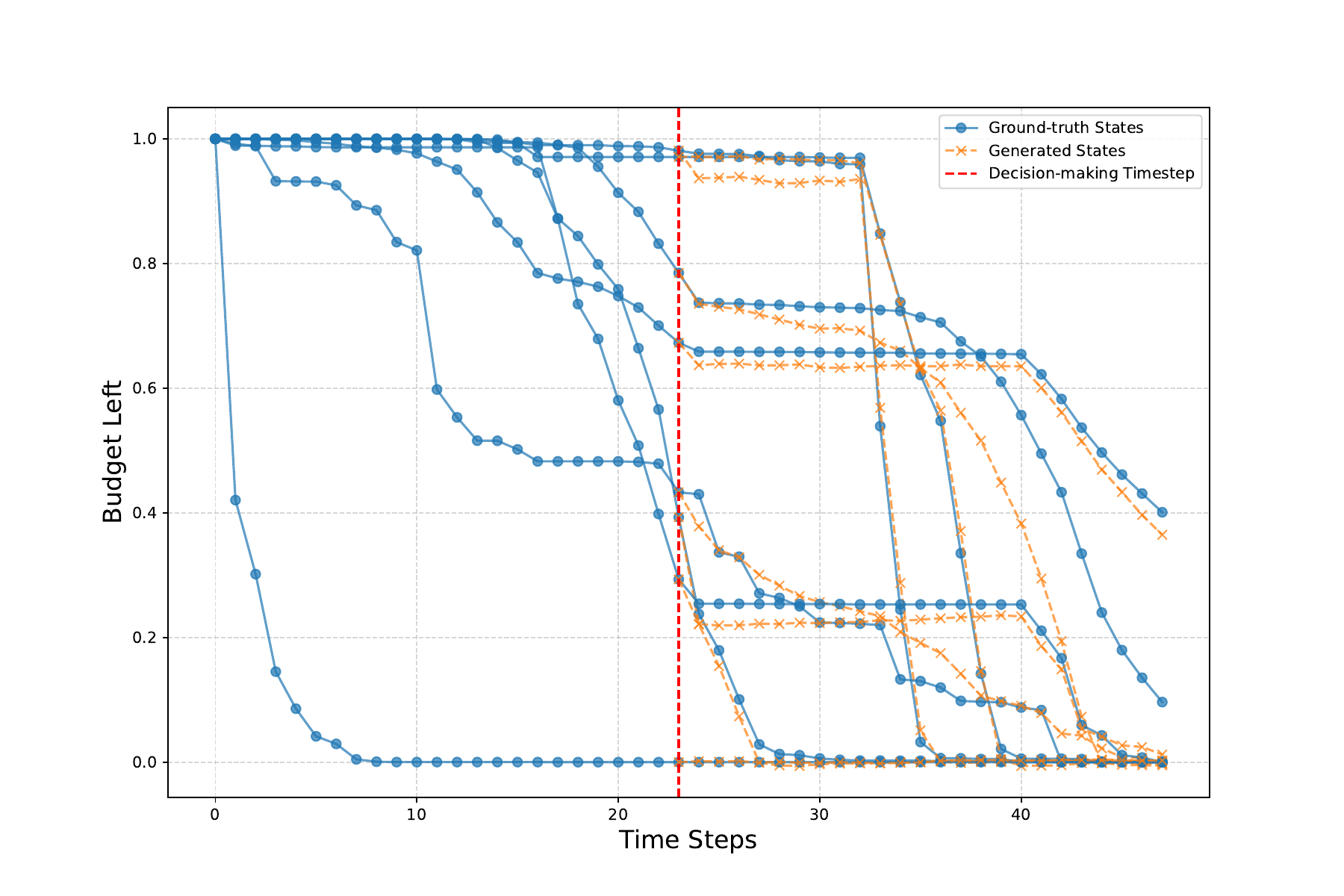}
    }
    \subfigure{
        \includegraphics[width=0.31\linewidth]{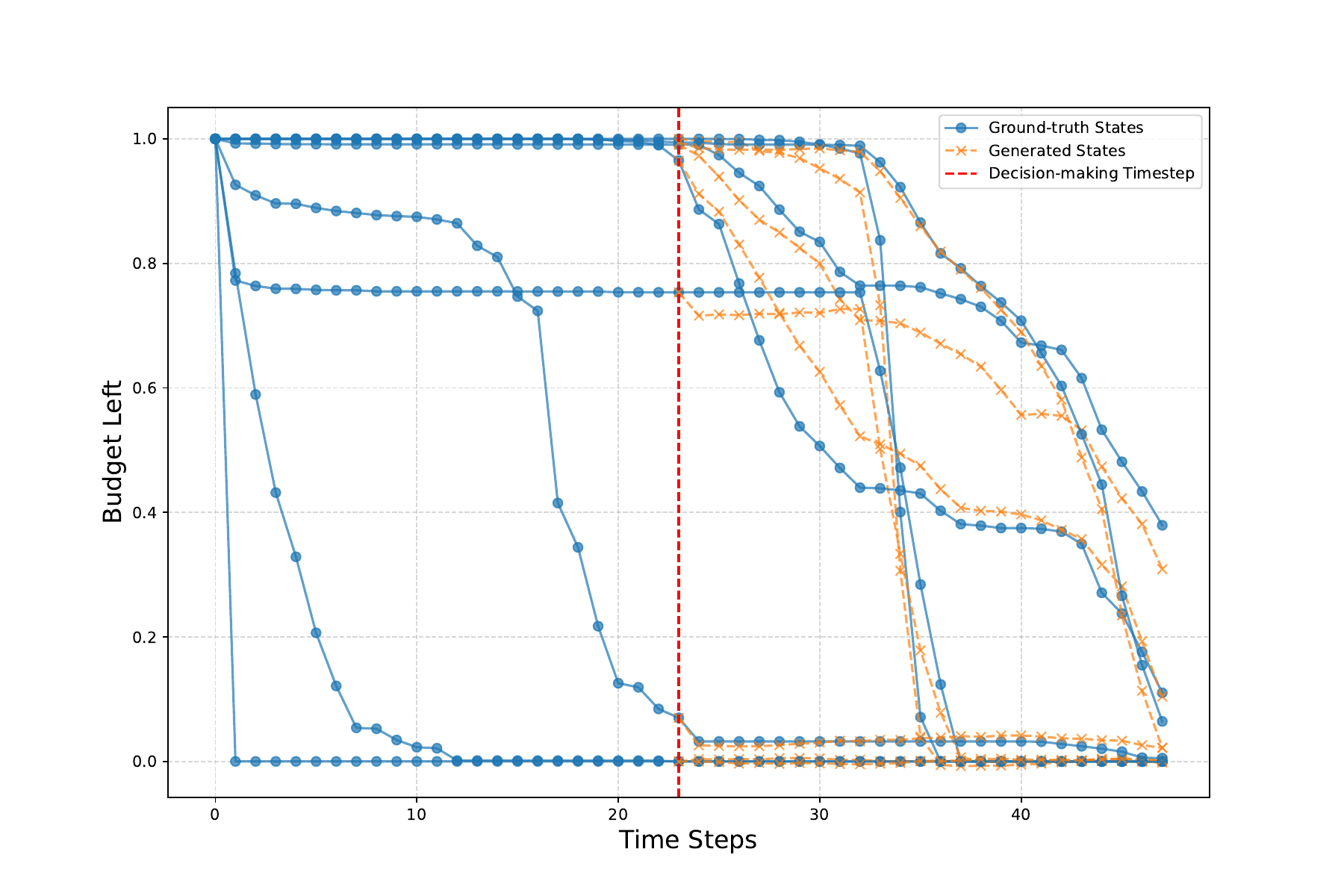}
    }
        \subfigure{
        \includegraphics[width=0.31\linewidth]{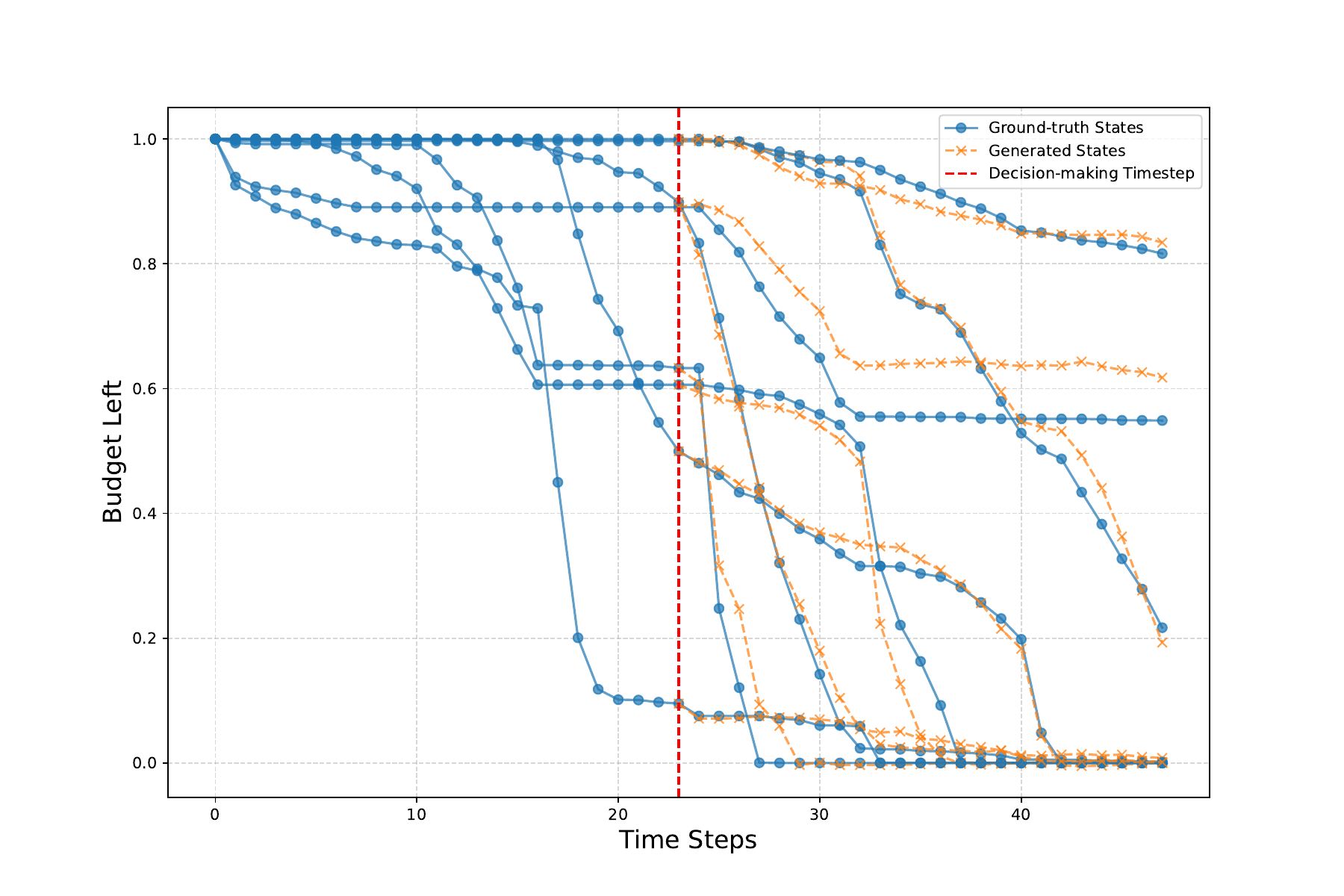}
    }
    \subfigure{
        \includegraphics[width=0.31\linewidth]{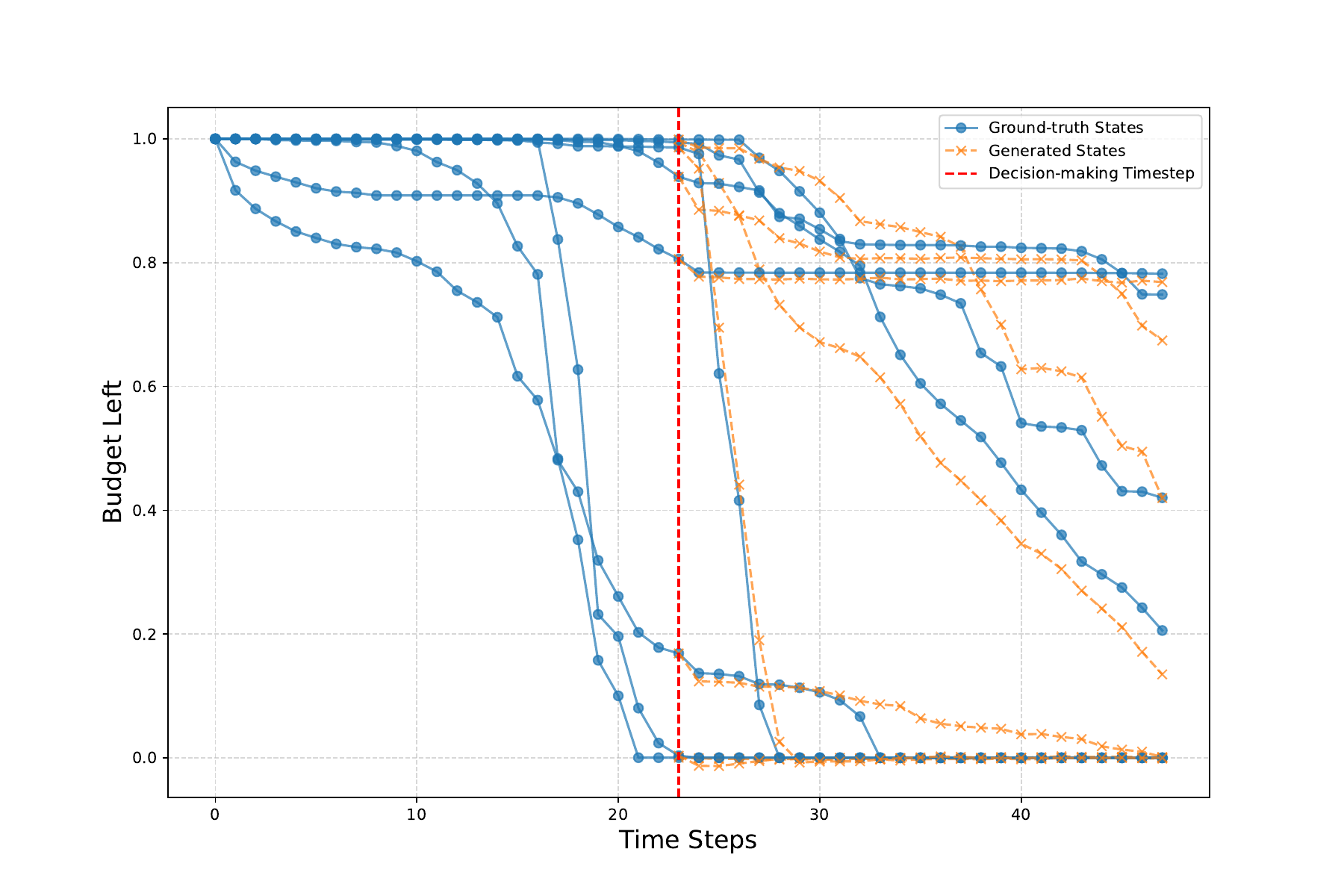}
    }
    \subfigure{
        \includegraphics[width=0.31\linewidth]{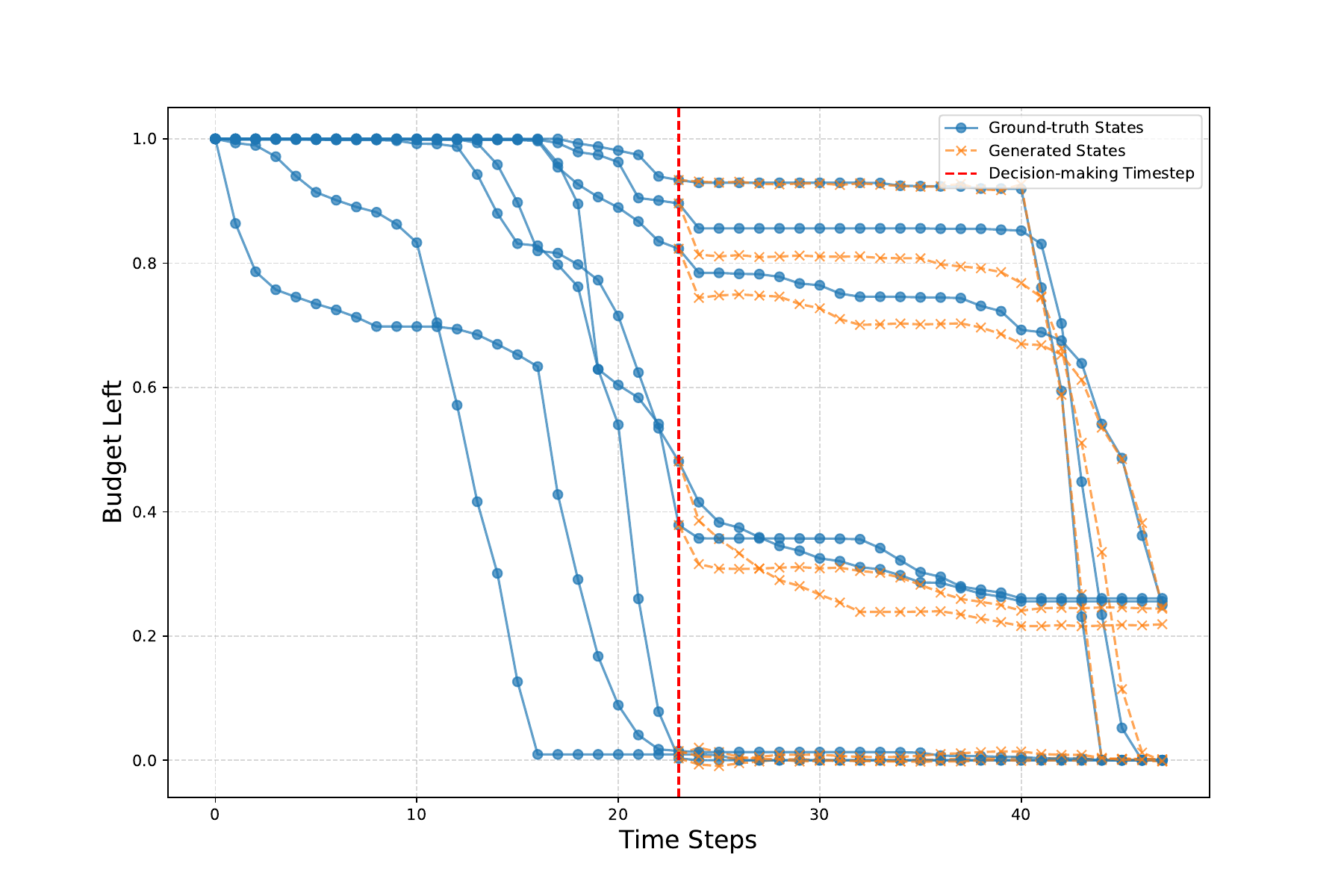}
    }
        \subfigure{
        \includegraphics[width=0.31\linewidth]{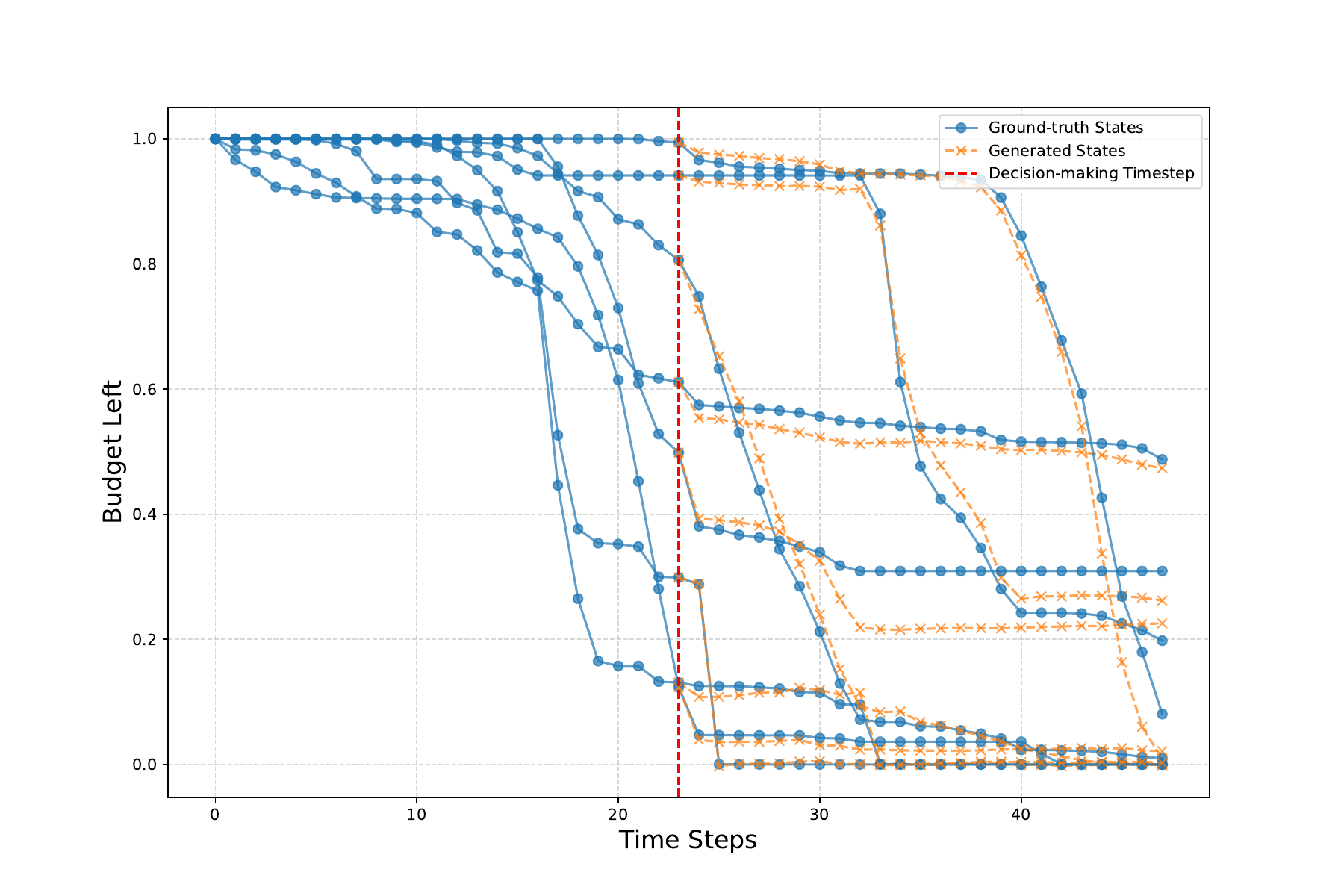}
    }
    \subfigure{
        \includegraphics[width=0.31\linewidth]{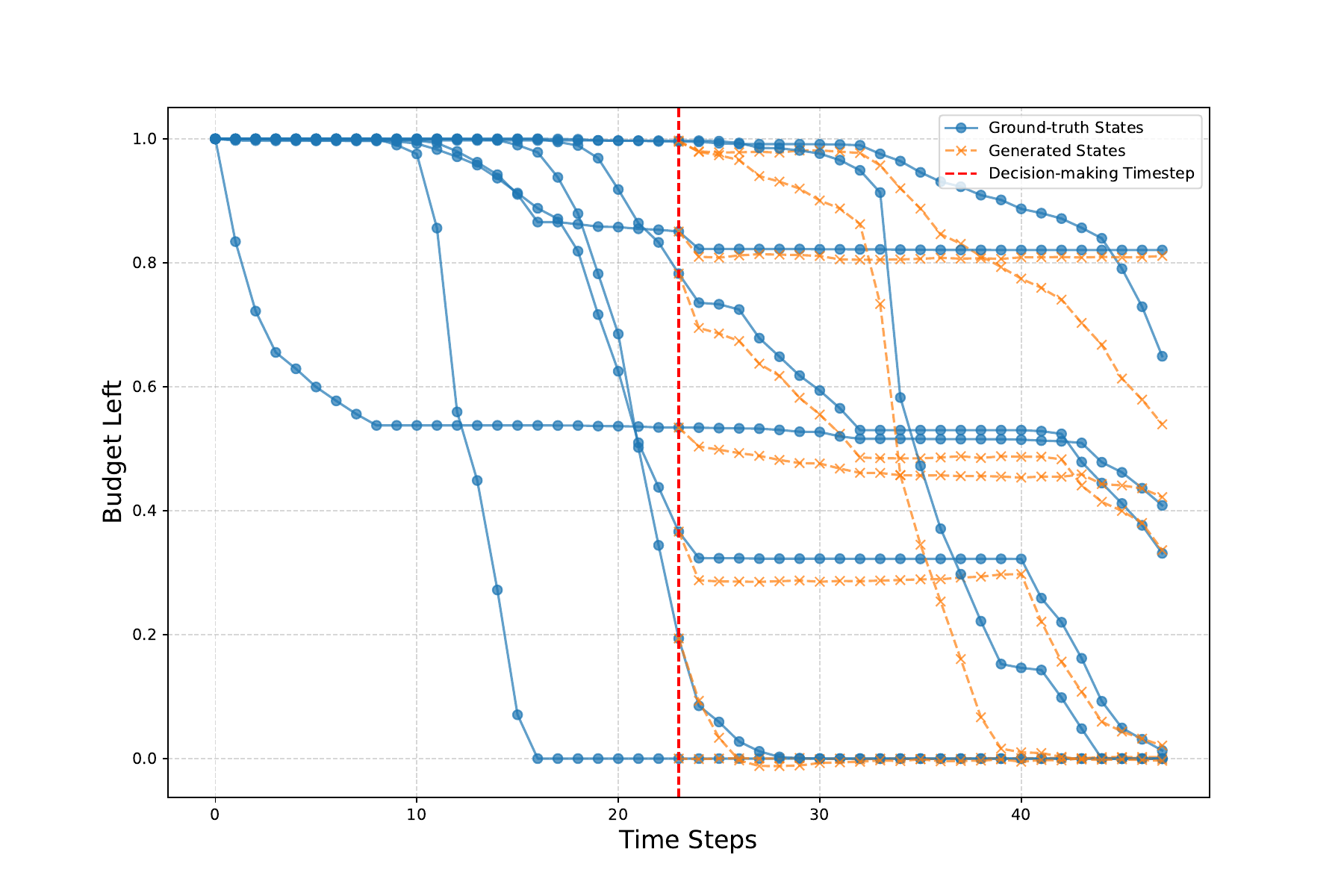}
    }
    \subfigure{
        \includegraphics[width=0.31\linewidth]{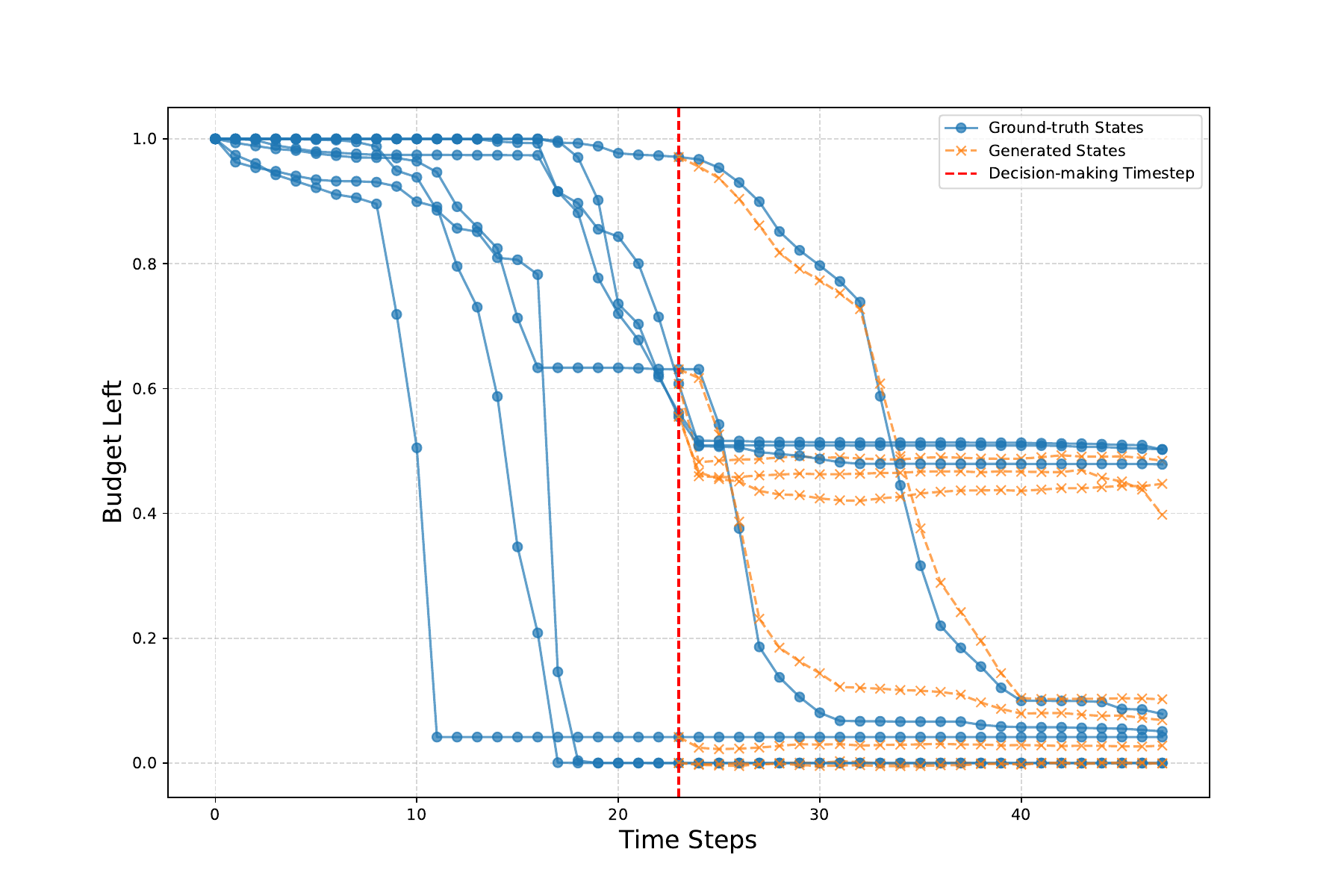}
    }
    \caption{Generated trajectories of CBD.}
    \label{fig:cbd_gen}
\end{figure}

We show more generated trajectories for visualization of the generation uncertainty issue in Fig. \ref{fig:diffbid_gen} and Fig. \ref{fig:cbd_gen}.
The blue lines represent the actual trajectories found in the dataset, while the yellow lines depict the generated trajectories, which incorporate observations from the trajectory spanning from $t=0$ to 23. A red line highlights the specific timestep at which decision-making occurs.
As the results demonstrate, our CBD method effectively alleviates the issue of generation uncertainty by producing dynamic legitimate and well-aligned generations.

% \subsection{Property Alignment Performance}
% \begin{figure}[t!]
%     \centering
%     \subfigure[Selection on Smoothness]{
%     \includegraphics[width=0.26\linewidth]{figs/smooth_over_time.pdf}
%     % \caption{}
%     \label{fig:smooth_over_time}
%     }
%     \subfigure[Selection on Early-Spend]{
%     \includegraphics[width=0.26\linewidth]{figs/early_spend_over_time.pdf}
%     % \caption{}
%     \label{fig:early_over_time}
%     }
%     \subfigure[Smoothness Dist.]{
%     \includegraphics[width=0.205\linewidth]{figs/smooth.pdf}
%     % \caption{}
%     \label{fig:smooth}
%     }
%     \subfigure[Early-Spend Dist.]{
%     \includegraphics[width=0.205\linewidth]{figs/early_spend.pdf}
%     % \caption{}
%     \label{fig:early}
%     }
%     \vspace{-4mm}
%     \caption{Results for property alignment. (a-b) show the generated trajectories with different properties, including smoothness and early spending; (c-d) illustrate the property distribution after aligning the generated trajectories to the desired properties at each bidding timestep.}
%     \label{fig:control}
%     \vspace{-5mm}
% \end{figure}

\end{document}